\documentclass[prd,showpacs,amsmath,amssymb,nobibnotes,nofootinbib,11pt]{revtex4-1}
\usepackage{appendix}
\usepackage{hyperref}
\usepackage{graphicx}
\usepackage{amsmath}
\usepackage{txfonts}
\usepackage{subfigure}
\usepackage{amssymb}
\usepackage{subfigure}
\usepackage{multirow}
\usepackage{youngtab}
\usepackage{colortbl}
\usepackage{slashbox}

\usepackage{hhline}
\newtheorem{theorem}{Theorem}[section]
\newtheorem{definition}[theorem]{Definition}

\newenvironment{defn*}{\begin{definition}}{\end{definition}}

\def\be{\begin{equation}} \def\ee{\end{equation}} 
\def\bea{\begin{eqnarray}} \def\eea{\end{eqnarray}}

\def\bs{\begin{split}}
\def\es{\end{split}}

\begin{document}
\title{3+1 Orthogonal and Conformal Decomposition of the Einstein Equation and the ADM Formalism for General Relativity}

\author{Suat Dengiz}
\email{e169656@metu.edu.tr}
\affiliation{Department of Physics, Middle East Technical University,
Ankara, Turkey}

\begin{abstract}
In this work, two particular orthogonal and conformal decompositions  of the 3+1 dimensional Einstein equation and Arnowitt-Deser-Misner (ADM) formalism for general relativity are obtained. In order to do these, the 3+1 foliation of the four-dimensional spacetime, the fundamental conformal transformations and the Hamiltonian form of general relativity that leads to the ADM formalism, defined for the conserved quantities of the hypersurfaces of the globally-hyperbolic asymptotically flat spacetimes, are reconstructed. All the calculations up to chapter 7 are just a review.

We propose a method in chapter 7 which gives an interesting relation between the Cotton (Conformal) soliton and the static vacuum solutions. The formulation that we introduce can be extended to find the gradient Cotton soliton and the solutions of Topologically Massive Gravity (TMG) as well as the gradient Ricci soliton.
\end{abstract}

\maketitle
\tableofcontents

\newpage

\section{INTRODUCTION}
\label{sec:introduction}

In this MS thesis, we have followed Eric Gourgoulhon's lecture notes titled \emph{3+1 Formalism and Bases of Numerical Relativity (arXiv:gr-qc/0703035v1} \cite{Gourgoulhon:2007ue} and E. Poisson's book : \emph{A Relativist's Toolkit, The Mathematics of Black-Hole Mechanics, Cambridge University Press, Cambridge (2004)} \cite{Poisson:2004ue}.

Historically, the 3+1 approach has been put forward G. Darmaois (1927)\cite{Darmois:1927vy}, A. Lichnerowicz (1930-40) \cite{Lichnerowicz:1939vy}, \cite{Lichnerowicz:1982} and Y. Choquet-Bruhat (1952)\cite{Bruhat:1952}. During 1958, 3+1 formalism started to be used to construct the Hamiltonian form of general relativity by P. A. M. Dirac \cite{Dirac:1958sc}, \cite{Dirac:1958sq} and later by R. Arnowitt, S. Deser and C. W. Misner (1962) \cite{Arnowitt:1959ah}. The 3+1 formalism became popular in the numerical relativity community during 1970 \cite{Gourgoulhon:2007ue}.

\emph{The 3+1 formalism is used to rewrite the Einstein equation as an initial value problem and construct the Hamiltonian form of the general relativity}. This method is based on the concept of the hypersurface, $ \Sigma_t $, which is independent of whether the given spacetime is a solution of the Einstein equation or not. In this formalism, we consider that there is an embedding mapping $ \Phi $ which maps the points of a hypersurface into the corresponding points of the four-dimensional manifold $ \mathcal{M} $ such that $ \mathcal{M} $ is covered by the continuous set of hypersurfaces $ (\Sigma_t)_{t \in R} $. Furthermore, the well-known Gauss-Codazzi relations and the 3+1 decomposition of the spacetime Ricci scalar curvature are the fundamental equations of the 3+1 decompositions of the spacetime $ (\mathcal{M},\textbf{g}) $. And they play a crucial role in the 3+1 decompositions of the Einstein equation. The Gauss-Codazzi relations are defined on a single hypersurface. On the other hand, the 3+1 decomposition of the spacetime Ricci scalar is obtained from the flow of the hypersurfaces. Moreover, the foliation is valid for any spacetime with a Lorentzian metric so we have to restrict our selves to the globally-hyperbolic spacetimes. And the foliation kinematics of the globally-hyperbolic spacetimes allow us to construct the Ricci equation whose contraction with respect to the induced 3-metric gives the last fundamental equation of the 3+1 formalism (i.e the 3+1 expression of the spacetime scalar curvature) \cite{Gourgoulhon:2007ue}.

The 3+1 decomposition of the Einstein equation is obtained by using the Gauss-Codazzi relations, the 3+1 decomposition of spacetime Ricci scalar and the 3+1 decomposition of stress-energy tensor. Basically, the four-dimensional Einstein equation decomposes into three main equations which are known as: \emph{the dynamical Einstein equation}, \emph{Hamiltonian constraint} and \emph{Momentum constraint}. The dynamical Einstein equation is obtained from the full projection of the Einstein equation onto the hypersurface and has 6 independent components, the Hamiltonian constraint is obtained from the full projection of Einstein equation along the normal vector and has 1 independent component and the Momentum constraint is obtained from the mixed projection of the Einstein equation and it has 3 independent components. Therefore, as we expect, the total number of independent components are 10 which is exactly the number of independent components of the Einstein equation in four-dimensional spacetime \cite{Gourgoulhon:2007ue}. 

The 3+1 dimensional Einstein system is modified to the Cauchy problem (or initial-value problem) by rewriting it as a set of PDEs (Partial Differential Equations) and specifying with the help of particular choices of the lapse function N and shift vector $ \beta $. Choosing a scalar field $ N $, a vector field $ \beta $ and a spatial coordinate system $ (x^i) $ on an initial hypersurface allows us to define a unique coordinate system $ (x^\alpha) $ within a neighborhood of $ \Sigma_0 $ such that $ x^0 = 0 $ corresponds $ \Sigma_t $. That is , N and $ \beta $ are depend on the coordinate systems. And also, the lapse function N at each point of $ \Sigma_0 $ leads us to define a unique vector \textbf{m}($ = $ N \textbf{n}) which is used to construct the neighboring hypersurface $ \Sigma_{\delta t} $ by Lie dragging each point of  $ \Sigma_0 $ along \textbf{m}. Therefore, the 3+1 dimensional Einstein system can be turned into as a PDEs system by using tensor components which are expanded with respect to the coordinates $ (x^\alpha)=(t,x^i) $ adapted to the foliation. The PDEs form of the 3+1 dimensional Einstein system contains only tensor fields of $ \Sigma_t $ and their time derivatives which implies that they can be taken as a \emph{time evolving tensor fields on a given} $ \Sigma_t $. The PDEs form of the 3+1 dimensional Einstein system is a system of second-order, non-linear PDEs for unknown ($ \gamma_{ij}\,,\, K_{ij}\,,\,N\,,\, \beta^i $ ) when the matter source terms ($ E\,,\,p_i\,,\,S_{ij} $ ) are given. Here $ \gamma_{ij} $ is the metric of the hypersurface, $ K_{ij} $ is the extrinsic curvature, $ E $ is the energy density, $ p_i $ is the momentum density and $ S_{ij} $ is the stress tensor. The crucial point is that the PDEs form of the 3+1 Einstein system contains neither the time derivative of N nor of $ \beta $. This means that \emph{they are not dynamical variables} rather they are just quantities associated with the coordinates $ (x^\alpha)=(t,x^i) $ (that is, Lagrange multipliers). Therefore, PDEs form of the 3+1 Einstein system can be converted into the initial value problem by choosing particular N and $ \beta $ \cite{Gourgoulhon:2007ue}, \cite{Poisson:2004ue}.

Beside the orthogonal decomposition that is used for 3+1 formalism, the \emph{conformal decomposition} is also used to define the flow of the hypersurfaces by continuously mapping an initial well-defined conformal background metric $ \tilde{\gamma}_{ij} $ into the induced 3-metric of $ \Sigma_t $. Lichnerowicz \cite{Lichnerowicz:1939vy} proposed that by mean of the particular conformal decomposition of the extrinsic curvature, one can arrange the constraint equations which allows us to define initial data for the Cauchy problem. In addition to this, York has shown that the conformal decompositions can be used for the time evolution \cite{York:1981bg}. That's, he has proved that \emph{the two degrees of freedom of the gravitational field are carried by the conformal equivalence classes of the induced 3-metric} \cite{York:1981bg}. The Weyl tensor is used to check whether a given spacetime, whose dimension is greater than 3, is conformally flat or not.And it disappears for lower-dimensional manifolds. In this case, the Cotton-York tensor \{\cite{York:1979cf}, \cite{York:1981bg}, \cite{Cotton:1899hj}\}, $ C^{ij} $, does \emph{the same task of the Weyl tensor in higher dimensional spacetime}. Furthermore, the Cotton-York \cite{York:1979cf}, \cite{York:1981bg}, \cite{Cotton:1899hj} tensor of weight $ 5/6 $, $ C^{ij}_*= \gamma^{5/6} C^{ij} $ , is \emph{conformally invariant}.      

The Hamiltonian model approaches a physical state at a \emph{certain time} and gives the evolution of the state as time varies. This model is being transformed into the gravitational theory as a state on \emph{a particular spacelike hypersurface} \cite{Dirac:1958sc}, \cite{Dirac:1958sq}. Now, the gravitational theory is a covariant theory and locally has Lorentz symmetry. In order to write the Einstein equations into the Hamiltonian form, people started to give up the main covariance property of the gravitational theory by choosing a family of particular coordinate systems such that `` $ x^0 = $ constant``  corresponds a spacelike hypersurface. Instead of the set ($ \gamma_{ij}\,,\, K_{ij}\,,\,N\,,\, \beta^i $ ) in the PDEs form of 3+1 Einstein system, Arnowitt, Deser and Misner have proposed the ADM formalism of the general relativity in which \emph{conjugate momentum of the induced three-metric $ \gamma_{ij} $}, $ \pi^{ij}=\sqrt{\gamma}(K\gamma^{ij}-K^{ij}) $, is used. In the ADM formalism \cite{Arnowitt:1959ah}, $ \pi^{ij} $ and $ \gamma_{ij} $ are the dynamical variables and the Lapse function N and the shift vector $ \beta $ are taken as Lagrange multipliers \cite{Arnowitt:1959ah}. In the chapter 6, we will see this in detail \cite{Gourgoulhon:2007ue}, \cite{Poisson:2004ue}.

The action for the General Relativity (when the boundary term is different than zero) contains Einstein-Hilbert part and Matter part. The infinitesimal four-dimensional volume element is taken as the union of two spacelike hypersurfaces $ \Sigma_{t_1} $, $ \Sigma_{t_2} $ which are at the upper and lower boundaries and a timelike hypersurface $ \mathcal{B} $ between $ \Sigma_{t_1} $, $ \Sigma_{t_2} $. Now, 3+1 decomposition of $ \mathcal{M} $ and 2+1 decomposition of the timelike hypersurface with proper choice of vectors lead us to the conserved quantities of the ADM mass, ADM linear momentum and ADM angular (by using rotational Killing vectors) of a given hypersurface. However, due to the fact that global quantities of mass, linear momentum and angular momentum are defined only for \emph{asymptotically flat spacetimes}, the ADM formulas are valid just for \emph{the spacetimes which asymptotically converge to well-defined spacetimes such as Minkowski spacetime} \cite{Gourgoulhon:2007ue}, \cite{Poisson:2004ue}, \cite{Arnowitt:1959ah}.

Finally, R. Bartnik and P. Tod introduce \cite{Bartnik:2005hh} the conditions on the intrinsic quantities of the $ \Sigma_t $. They ensure whether the $ \Sigma_t $ is a hypersurface of a spacetime which is a solution of the four-dimensional static vacuum field ( with $ \Lambda =0 $) or not. In addition to \cite{Bartnik:2005hh}, we introduce the equations  (\ref{dkt3}), (\ref{dkt4}) (for $ \Lambda= 0 $ case) and (\ref{dkt1}), (\ref{dkt2}) (for $ \Lambda \ne 0 $ case). These equations can be used to find which solutions of the gradient Cotton (Conformal) soliton \cite{Kisisel:2008jx} are also the solutions of the static vacuum fields equations. However,  We have not also been able to solve the constraint equations and have not found explicit metric. Moreover, we think that our method can be extended to the Ricci soliton \cite{Hamilton:1982} and Topologically Massive Gravity (TMG) \cite{Deser:1981wh}, \cite{Deser:1982vy}, \cite{Chow:2009km}.

\section{GEOMETRY OF HYPERSURFACES}
\label{sec:Geometry of hypersurfaces}

Since 3+1 decomposition of the spacetime is constructed by slicing the spacetime with a continuous set of the hypersurfaces, $ (\Sigma_t)_{t \in R} $, we will deal with the geometrical fundamentals of the hypersurface. The geometrical results that we will obtain in this chapter are fully independent of whether the given spacetime is a solution of the Einstein equation or not. The only constraint is that the spacetime must have Lorentzian metric \cite{Gourgoulhon:2007ue}, \cite{Carroll:2004fv}, \cite{Straumann:2004} .

\subsection{Notations and Basic Geometrical Tools}

\subsubsection{Spacetime and Tensor Fields}

 We assume a real, smooth $ (i.e. \, C^\infty) $ four-dimensional manifold  $ \mathcal{M} $ which endowed a Lorentzian metric of signature $ (-,+,+,+) $ and a connection  $ \nabla $. In general, no one can define a global vector space on manifolds. Therefore, it is considered that at each point of the manifolds there is a space of vectors  $ \mathcal{T}_p (\mathcal{M}) $ (titled as \emph{tangent space at the point p}) and corresponding space of linear forms  $\mathcal{T}^*_p(\mathcal{M}) $ (titled as \emph{dual space} or \emph{cotangent space at the point p} ). Furthermore, we suppose that all the Greek letters $ \{\alpha,\beta,\gamma,...\} $ run in $ \{0,1,2,3\} $ are free indices , $ \{\mu,\nu,\rho,...\} $ are dummy indices and all Latin letters $ \{i,j,k,...\} $ run in $ \{1,2,3\} $ and $ \{a,b,c,...\} $ run in $ \{2,3\} $.  

Since the $ \mathcal{T}_p (\mathcal{M}) $ and  $\mathcal{T}^*_p(\mathcal{M}) $ are vector spaces, we consider that there is a set of basis $ (e^\alpha) $ which spans  $ \mathcal{T}_p (\mathcal{M}) $ and the associated dual set of basis $ (e_\alpha) $ which spans  $\mathcal{T}^*_p(\mathcal{M}) $ such that $ e^\alpha(e_\beta)=\delta^\alpha{_\beta} $. Therefore, any tensor field $ \mathcal{\textbf{T}} $ of type $ \binom {p}{q} $ can be expanded with respect to these bases as
\begin{equation}
\mathcal{\textbf{T}} =T^ {\alpha_1...\alpha_p}{_{\beta_1...\beta_q}} e_{\alpha_1} \otimes...\otimes e_{\alpha_p} \otimes e^{\beta_1} \otimes...\otimes e^{\beta_q} \, .
\label{exp-ten}
\end{equation}
Here $ T^ {\alpha_1...\alpha_p}{_{\beta_1...\beta_q}} $ are the related components of $ \mathcal{\textbf{T}} $ relative to the bases $ (e^\alpha) $ and $ (e_\alpha) $. A tensor field \textbf{T} with rank $ \binom{p}{q} $ turns into another tensor field $ \nabla \textbf{T} $ with the rank $ \binom{p}{q+1} $ when the covariant derivative acts on it. Therefore, the expansion of $ \nabla \mathcal{\textbf{T}} $ in these bases is
\begin{equation}
 \nabla \mathcal{\textbf{T}} = T^ {\alpha_1...\alpha_p}{_{\beta_1...\beta_q;\gamma}} e_{\alpha_1} \otimes...\otimes e_{\alpha_p} \otimes e^{\beta_1} \otimes...\otimes e^{\beta_q} \otimes e^\gamma \, .
\label{exp_tens}
\end{equation}
The contraction of covariant derivative of the tensor field $ \mathcal{\textbf{T}} $ with an arbitrary vector field \textbf{u} gives us the \emph{covariant derivative of  $ \mathcal{\textbf{T}} $ along the vector field} \textbf{u} which \emph{does not change the rank of tensor fields} $ \mathcal{\textbf{T}} $
\begin{equation}
\begin{aligned}
 \nabla_{\textbf{u}} \mathcal{\textbf{T}}  &=\nabla \mathcal{\textbf{T}} (\underbrace{...,...,} \textbf{u}) \, ,\\
 & \qquad \quad p+q \, slots
\end{aligned}
\end{equation}
where $ u^\mu \nabla_\mu T^ {\alpha_1...\alpha_p}{_{\beta_1...\beta_q}} $ are the components of $ \nabla_{\textbf{u}}\mathcal{\textbf{T}} $ with respect to $ (e^\alpha) $ and $ (e_\alpha) $. 

\subsubsection{Scalar Products and Metric Duality}

In general, we do not know how to relate the elements of $\mathcal{T}_p(\mathcal{M}) $ ( or of  $\mathcal{T}^*_p(\mathcal{M}) $ ). The concept of metric is introduced to do this task. Now, \emph{the scalar product} of two vectors is taken place by mean of the related metric \textbf{g} of the manifold $ \mathcal{M} $
\begin{equation*}
 \forall (\textbf{u},\textbf{v}) \in \mathcal{T}_p(\mathcal{M})  \otimes \mathcal{T}_p(\mathcal{M}) \, , 
\end{equation*}
\begin{equation*}
 \textbf{u}.\textbf{v}= g(u^\mu e_\mu,v^\nu e_\nu ) =u^\mu v^\nu g(e_\mu,e_\nu)=g_{\mu\nu}u^\mu v^\nu =u^\mu v_\mu \, .
\end{equation*}
Here the metric \textbf{g} is taken as if it has two slots for inputting vectors. Alternatively, the same job is done by bracket notation : $ \forall (\tilde{\textbf{w}},\textbf{v}) \in \mathcal{T}^*_p(\mathcal{M})  \otimes \mathcal{T}_p(\mathcal{M})  $, 
\begin{equation*}
\begin{aligned}
 <\tilde{\textbf{w}},\textbf{v}>&=<w_\mu e^\mu ,v^\nu e_\nu> \\
&=w_\mu v^\nu <e^\mu,e_\nu> \\
&=w_\mu v^\nu e^\mu(e_\nu) \\
&=w_\mu v^\nu \delta^\mu{_\nu} \\
&=w_\mu v^\mu \, .
\end{aligned}
\end{equation*}
As we see in equation (\ref{exp_tens}), '' $ \nabla_\beta w_\alpha e^\alpha \otimes e^\beta $ '' are the components of the 2-form $ \nabla \tilde{\textbf{w}} $ relative to the bases $ e^\alpha \otimes e^\beta $ of $ \mathcal{T}^*(\mathcal{M})  \otimes \mathcal{T}^*(\mathcal{M})  $. Then, \emph{the directional covariant derivative of a 1-form $ \tilde{\textbf{w}} $ along a vector field $ \textbf{u} $,$ \nabla_{\textbf{u}}\tilde{\textbf{w}} $,} is a 1-form 
\begin{equation*}
\begin{aligned}
\nabla_{\textbf{u}}\tilde{\textbf{w}} &= \nabla \tilde{\textbf{w}}(\textbf{u}) \\
&=\Big [ \nabla_\gamma w_\beta e^\beta e^\gamma \Big ](u^\mu e_\nu ) \\
&=u^\mu \nabla_\gamma w_\beta e^\beta e^\gamma(e_\mu) \\
 &= u^\mu \nabla_\gamma w_\beta e^\beta \delta^\gamma{_\mu} \\
& =u^\mu \nabla_\mu w_\beta e^\beta \, .
\end{aligned}
\end{equation*}
 Since the directional derivative, $ \nabla_{\textbf{u}}\tilde{\textbf{w}} $, is a 1-form, we use the bracket notation to get a scalar from it: $ \forall (\tilde{\textbf{w}},\textbf{u},\textbf{v}) \in \mathcal{T}^*(\mathcal{M})  \otimes \mathcal{T}(\mathcal{M})  \otimes \mathcal{T}(\mathcal{M}) $ , 
\begin{equation*}
\begin{aligned}
 \nabla \tilde{\textbf{w}}(\textbf{u},\textbf{v}) &= <\nabla_\textbf{u} \tilde{\textbf{w}},\textbf{v}> \\
&= < u^\mu \nabla_\mu w_\beta e^\beta,v^\nu e_\nu > \\
&= u^\mu \nabla_\mu w_\beta v^\nu <e^\beta,e_\nu> \\
&= u^\mu \nabla_\mu w_\beta v^\nu \delta^\beta{_\nu} \\
&=u^\mu v^\nu \nabla_\mu w_\nu \, .
\end {aligned}
\end{equation*}
Any element of $\mathcal{T}_p(\mathcal{M}) $ (or $\mathcal{T}^*_p(\mathcal{M}) $) can be mapped into $\mathcal{T}^*_p(\mathcal{M}) $ (or $\mathcal{T}_p(\mathcal{M}) $) by mean of the 2-form \textbf{g}. That's, the metric \textbf{g} induces an isometry between $\mathcal{T}_p(\mathcal{M}) $ and $\mathcal{T}^*_p(\mathcal{M}) $. Some of crucial properties of this isometry are
\begin{enumerate}
 \item The dual of any vector \textbf{u} ($ \in  \mathcal{T}_p(\mathcal{M}) $) is a unique linear form of $\mathcal{T}^*_p(\mathcal{M}) $  and denoted by $ \tilde{\textbf{u}} $  such that the scalar product is defined as
\begin{equation}
 \forall \, \textbf{v} \in \mathcal{T}_p(\mathcal{M}) \,,\, \, < \tilde{\textbf{u}},\textbf{v}>= \textbf{g}(\textbf{u},\textbf{v}) \, .
\end{equation}
\item The dual of any linear form  $ \tilde{\textbf{w}} $  ( $ \in  \mathcal{T}^*_p(\mathcal{M})  $), $ \tilde{\tilde{\textbf{w}}} $, is a unique vector \textbf{w} $ \in  \mathcal{T}_p(\mathcal{M}) $ such that 
\begin{equation*}
 \forall \textbf{v} \in  \mathcal{T}_p(\mathcal{M})  \,,\,\, \textbf{g}(\tilde{\tilde{\textbf{w}}},\textbf{v})= <\textbf{w},\textbf{v}> \, .
\end{equation*}
\item $ \textbf{T}:\mathcal{T}_p(\mathcal{M}) \otimes \mathcal{T}_p(\mathcal{M}) \rightarrow \mathcal{R} $ (i.e any rank $ \binom{0}{2} $ tensor \textbf{T} maps 2-vectors of tangent spaces at the point p into the space of scalars). An \emph{endomorphism} $ \overset{\rightarrow}{\textbf{T}} $ is induced from \textbf{T} such that $ \overset{\rightarrow}{\textbf{T}} : \mathcal{T}(\mathcal{M}) \rightarrow \mathcal{T}(\mathcal{M}) $ and it satisfies 
\begin{equation}
\begin{aligned}
 \textbf{T}(\textbf{u},\textbf{v})&= u^\alpha v^\beta T_{\gamma\sigma} \delta_\alpha{^\gamma} \delta_\beta{^\sigma} \\
&=u^\alpha v^\sigma T_{\gamma\sigma}\delta_\alpha{^\gamma} \\
&=u^\alpha v^\sigma T_{\gamma\sigma} e_\alpha(e^\gamma) \\
&=u^\alpha v^\sigma T_{\gamma\sigma}<e^\gamma, e_\alpha> \\
&=u^\alpha v^\sigma T^{\gamma}{_\sigma} g(e_\gamma,e_\alpha) \\
&=g_{\gamma\alpha}u^\alpha v^\sigma T^{\gamma}{_\sigma} \\
&=\textbf{u}.\overset{\rightarrow}{\textbf{T}}(\textbf{v}) \, .
\end{aligned}
\end{equation}
As we see in the equation (\ref{exp-ten}) and because the endomorphism $ \overset{\rightarrow}{\textbf{T}} $ is a 1-form, $ T^\alpha{_\beta} $ are the components of $ \overset{\rightarrow}{\textbf{T}} $ relative to $ (e_\alpha) $.
\end{enumerate}
\subsubsection{Curvature Tensor}

According to gravity, the matter curves the geometry and the geometry determines the motion of the matter. The rank $ \binom{1}{3} $ Riemann curvature tensor  measures how much the spacetime is curved. Basically, it is a map which sends a 1-form and 3 vectors into the real, smooth space of scalar fields $ C^\infty (\mathcal{M},\mathcal{R}) $ 
\begin{equation*}
 {^4}\textbf{R} :\mathcal{T}^*(\mathcal{M}) \otimes \mathcal{T}(\mathcal{M})^3 \longrightarrow C^\infty (\mathcal{M},\mathcal{R}) \, , \qquad \qquad \qquad \qquad \qquad \quad \qquad \qquad 
\end{equation*}
\begin{equation}
 \Big (\tilde{\textbf{w}},\textbf{w},\textbf{u},\textbf{v} \Big ) \longrightarrow < \tilde{\textbf{w}},\nabla_\textbf{u} \nabla_\textbf{v} \textbf{w} 
-\nabla_\textbf{v}\nabla_\textbf{u} \textbf{w} -\nabla_{[\textbf{u},\textbf{v}]} \textbf{w}> \, .
\label{Riem}
\end{equation}
The Riemann tensor $ {^4}\textbf{R} $ is assumed to be machine which has 1 slot for 1-form and 3 slots for vectors. The relation (\ref{Riem}) is nothing but a tensor field on $ \mathcal{M} $. Furthermore, $ {^4}R^\gamma{_{\delta\alpha\beta}} $ is the components of $ {^4}\textbf{R} $ with respect to a proper set of basis $ (e_\alpha) $ and $ (e^\alpha) $ of $ \mathcal{T}_p(\mathcal{M}) $ and $ \mathcal{T}^*_p(\mathcal{M}) $. Now, the crucial properties of $ {^4}\textbf{R} $ are 
\begin{enumerate}
 \item $ {^4}R_{\alpha\beta\gamma\delta} $ ( $ =g_{\alpha\gamma} {^4}R^\gamma{_{\beta\gamma\delta}} \, . $ ) is anti-symmetric between the $ 1^{st} $ two terms $ \alpha $ and $ \beta $  and between the $ 2^{nd} $ two terms $ \gamma $ and $ \delta $.
\item $ {^4}R_{\alpha\beta\gamma\delta} $ satisfies the cyclic property between the last three indices which is known as second Bianchi identity
\begin{equation}
{^4}R_{\alpha\beta\gamma\delta}+{^4}R_{\alpha\gamma\delta\beta}+{^4}R_{\alpha\delta\beta\gamma}=0 \, .
\end{equation}
\item For the torsion-free spacetime, the well-known relation of the Ricci identity is 
\begin{equation}
 [\nabla_\alpha,\nabla_\beta]w^\gamma = {^4}R^\gamma{_{\mu\alpha\beta}}w^\mu \, .
\end{equation}
\item One-times contraction of $ {^4}R_{\gamma\alpha\delta\beta} $ , i.e.  $ \delta \rightarrow \beta $, leads us to a symmetric, bilinear-form Ricci tensor $ {^4}\textbf{R} $. The $ {^4}\textbf{R} $ is considered to be a machine that has 2 slots for vectors 
\begin{equation}
\begin{aligned}
 {^4}\textbf{R}:\mathcal{T}(\mathcal{M}) \otimes \mathcal{T}(\mathcal{M}) &\longrightarrow C^\infty (\mathcal{M},\mathcal{R}) \, ,\\
(\textbf{u},\textbf{v}) &\longrightarrow {^4}R(e^\mu,\textbf{u},e_\mu,\textbf{v}) \, .
\label{Ricci}
\end{aligned}
\end{equation}
Also, the trace of the Ricci tensor relative to the dual of $ \textbf{g} $ results in the spacetime Scalar curvature $ R $. \item The \emph{traceless} part of the spacetime Riemann tensor titled as \textbf{Weyl tensor}, ${^4}\textbf{C} $ which gives \emph{\textbf{whether a given spacetime is conformally flat or not}} is obtained by subtracting all the \emph{trace} part (i.e. the Ricci tensor) and the \emph{trace-trace} part (i.e. the Ricci scalar $ {^4}R =g^{\mu\nu}R_{\mu\nu} \, . $ ) of the spacetime Riemann curvature tensor from itself
\begin{equation}
\begin{aligned}
 {^4}C^\gamma{_{\delta\alpha\beta}} =&{^4}R^\gamma{_{\delta\alpha\beta}}-\frac{1}{2}\Big ({^4}R^\gamma{_\alpha}g_{\delta\beta}
-{^4}R^\gamma{_\beta}g_{\delta\alpha}+{^4}R_{\delta\beta} \delta^\gamma{_\alpha}-{^4}R_{\delta\alpha} \delta^\gamma{_\beta} \Big ) \\
\qquad & +\frac{1}{6}{^4}R \Big ( g_{\delta\alpha} \delta^\gamma{_\beta}-g_{\delta\beta} \delta^\gamma{_\alpha} \Big ) \, .
\end{aligned}
\end{equation}
We need to emphasize that \emph{the Weyl tensor ,  ${^4}\textbf{C} $, vanishes for spacetime whose dimension is lower than 4. Thus, in the lower dimensional geometry the spacetime Riemann tensor can be written in terms of the Ricci tensor, the metric and the scalar curvature tensor.}
\end{enumerate}

\subsubsection{Hypersurface Embedded in Spacetime}

\begin{figure}[h]
\centering
\includegraphics[width=0.8\textwidth]{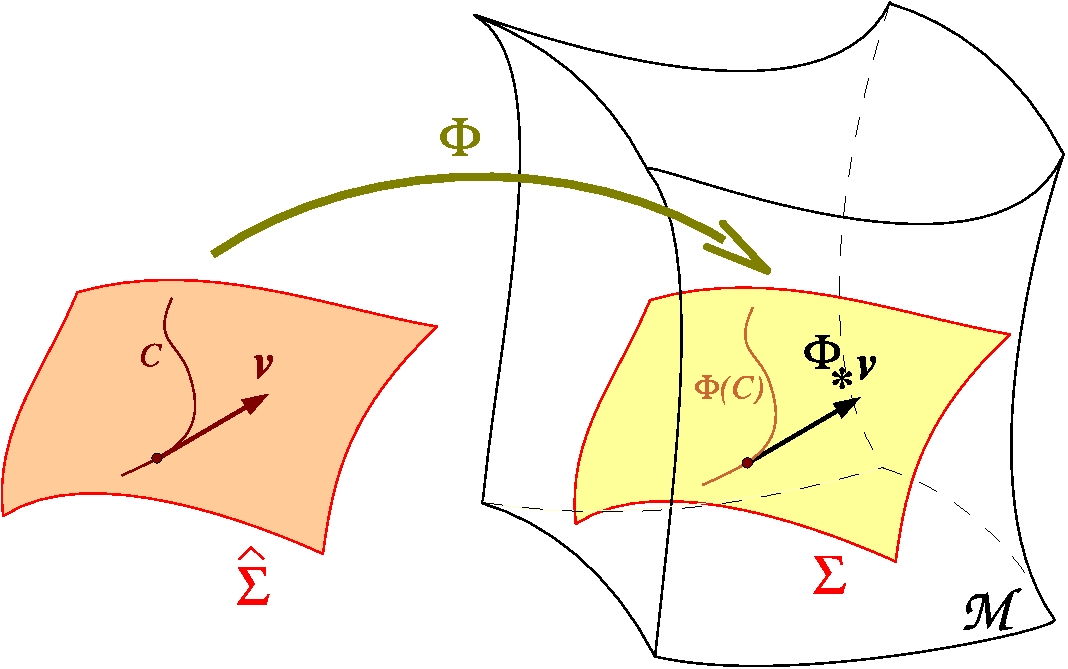}
\caption{The embedding of $ \hat{\Sigma} $ into $ \mathcal{M} $}
\label{hyp_embed}
\end{figure}

As we see in the figure (\ref{hyp_embed}), the set of points, $ \forall p \in \mathcal{M} $, at which the scalar field is constant corresponds a hypersurface $ \Sigma $ of the four-dimensional manifold $ \mathcal{M} $ which is an image of a three-dimensional manifold $ \hat{\Sigma} $ under the homeomorphism  $ \Phi $. Since the three-dimensional manifold $ \hat{\Sigma} $ is something like to be embedded into the four-dimensional manifold $ \mathcal{M} $, we say that the mapping $ \Phi $ is an embedding mapping
\begin{equation}
 \Phi:  \hat{\Sigma}  \rightarrow \mathcal{M} \, .
\end{equation}
Furthermore, ''\emph{one-to-one character of the embedding mapping $ \Phi $ ensures that the hypersurfaces do not intersect}''.

The embedding mapping $ \Phi $ induces two well-known mappings of \emph{the push-forward mapping}, $ \Phi_* $ , and \emph{the pull-back mapping}, $ \Phi^* $. First, $ \Phi_* $ maps the vectors of the tangent space of the three-dimensional manifold $ \hat{\Sigma} $, $ \mathcal{T}_p(\hat{\Sigma}) $, into the corresponding vectors of $ \mathcal{T}_p(\mathcal{M}) $
\begin{equation}
\begin{aligned}
 \Phi_* :\mathcal{T}_p(\hat{\Sigma}) &\longrightarrow \mathcal{T}_p(\mathcal{M}) \, ,\\
\textbf{v}=(v^x,v^y,v^z) &\longrightarrow \Phi_* \textbf{v}=(0,v^x,v^y,v^z) \, .
\label{push_forw}
\end{aligned}
\end{equation}
Here $ v^i=(v^x,v^y,v^z) $ is the components of the vector $ \textbf{v} $ with respect to the natural basis $ \partial/ \partial x^i $of $ \mathcal{T}_p(\mathcal{M}) $ associated with the coordinates $ (x^i) $. On the other hand, $ \Phi^* $ maps the linear forms of $ \mathcal{T}^*_p(\mathcal{M}) $ into the corresponding linear forms of $ \mathcal{T}^*_p(\hat{\Sigma}) $
\begin{equation}
\begin{aligned}
 \Phi^*:\mathcal{T}^*_p(\mathcal{M}) &\longrightarrow \mathcal{T}^*_p(\hat{\Sigma}) \, ,\\
\tilde {\textbf{w}} &\longrightarrow \Phi^*\tilde{\textbf{w}}:\mathcal{T}_p(\hat{\Sigma}) \longrightarrow \mathcal{R} \, ,\\
& \qquad \qquad \qquad \textbf{v} \longrightarrow <\tilde{\textbf{w}},\Phi_*\textbf{v} > \, .
\label{pull_back}
\end{aligned}
\end{equation}
Further insight, the mapping $ \Phi^* $ acts on the multilinear forms of $ \mathcal{T}_p(\mathcal{M}) $, too
\begin{equation}
 \forall (\textbf{v}_1,...,\textbf{v}_n) \in \mathcal{T}_p(\Sigma)^n\,,\, (\Phi^*T)(\textbf{v}_1,...,\textbf{v}_n)=T(\Phi_* \textbf{v}_1,...,\Phi_* \textbf{v}_n ) \, ,
\end{equation}
where \emph{T} is an n-form. Especially, the pull-back of the 2-form spacetime metric \textbf{g} takes a great attention. The pull-back of \textbf{g} is called the induced metric, $ \gamma $, of the hypersurface, $ \Sigma_t $ and known as \textbf{the first fundamental form of} $ \Sigma_t $  
\begin{equation}
 \gamma=\Phi^*\textbf{g} \, .
\end{equation}
Moreover, the scalar product between any two vectors of the tangent space of the hypersurface is same either by using \textbf{g} or $ \gamma $: $ \forall (\textbf{u},\textbf{v}) \in \mathcal{T}_p(\Sigma) \times \mathcal{T}_p(\Sigma) $ ,
\begin{equation}
\textbf{u}.\textbf{v}=\textbf{g}(\textbf{u},\textbf{v})=\gamma(\textbf{u},\textbf{v}) \,.
\end{equation}

\subsubsection{Normal Vector}

We consider there is a scalar field $ t $  on $ \mathcal{M} $ such that each ''$ t $ =constant'' corresponds the hypersurface $ \Sigma_t $ of $ \mathcal{M} $ and the vector $ \nabla t $ is normal to $ \Sigma $. Then, the dual of the vector $ \nabla t $ is the gradient 1-form $ dt $ such that the relation between them is
\begin{equation}
 \nabla^\alpha t =g^{\alpha\mu} \nabla_\mu t = g^{\alpha\mu} (\textbf{d} t)_\mu \, .
\end{equation}
and $ \forall \, \textbf{v} \in \Sigma $, the scalar product vanishes
\begin{equation}
 < \textbf{d}t,\textbf{v}>=0 \, .
\end{equation}
Furthermore, the type of the normal vector  $ \nabla t $ is determined by the type of the hypersurface $ \Sigma $: That's, if the signature of induced metric $ \gamma $ of $ \Sigma $ is (+,+,+) then $ \Sigma $ is \emph{spacelike} and the corresponding normal vector  $ \nabla t $ is \emph{timelike}, contrary, if the signature of induced is (-,+,+) then $ \Sigma $ is \emph{timelike} and the corresponding normal vector  $ \nabla t $ is \emph{spacelike} and, finally, if the induced metric is degenerate , i.e. has signature of (0,+,+) then either $ \Sigma_t $ or $ \nabla t $ are \emph{null}. 

Although $ \nabla t $ is a unique normal vector to $ \Sigma_t $, it is not a unit normal vector. Therefore, in the not-null case, we normalize it to get a unit normal vector of \textbf{n}
\begin{equation}
 \hat{\textbf{n}}=\frac{\nabla t}{\sqrt{ \pm \nabla t.\nabla t}} \, ,
\end{equation}
where the positive sign (+) is used for a timelike hypersurface and the negative sign (-) is used for a spacelike hypersurface. Thus, the norm of the unit normal vector is
\[
 \hat{\textbf{n}} . \hat{\textbf{n}}=\pm \frac{\nabla t.\nabla t}{\nabla t.\nabla t}=
\begin{cases}
 -1 & if\,\, \Sigma\,\, is\,\, spacelike \\
+1 &  if\,\, \Sigma\,\, is\,\, timelike
\end{cases}
\]

\subsubsection{Intrinsic Curvature}

For the not-null case, one can always propose a \emph{unique Levi-Civita connection $ D $} which is still torsion-free and metric compatible on a hypersurface $ \Sigma $. Moreover, the intrinsic covariant derivative $ D $ is defined by using the induced metric $ \gamma $. Now, as we see in the relation (\ref{Riem}), the Riemann curvature tensor measures the curvature by using the spacetime connection $ \nabla $. However, in order to measure the curvature of the hypersurface $ \Sigma $ (i.e. \emph{the intrinsic curvature of the $ \Sigma $}), we replace the spacetime connection $ \nabla $ with the intrinsic connection $ D $.
\begin{equation}
 \forall \, \textbf{v} \in \mathcal{T}(\Sigma)\,,\,\,[D_i,D_j]v^k = R^k{_{lij}} v^l \, .
\label{intr_ricc_ident} 
\end{equation}
That's, the \emph{intrinsic curvature of a given hypersurface} is nothing but the curvature which is measured (or felt) by anybody \emph{moving on the hypersurface}. Also, as we did in the equation ( \ref{Ricci}), one-times contraction on the intrinsic curvature tensor gives us the Ricci tensor of the hypersurface $ \Sigma $
\begin{equation}
 R_{ij}=R^k{_{ikj}} \, .
\end{equation}
Finally, the contraction of the intrinsic Ricci tensor results into the \emph{intrinsic scalar curvature} (or \emph{Gaussian curvature}) of $ \Sigma $.  
\begin{equation}
 R=\gamma^{ij}R_{ij} \, .
\end{equation}

\subsubsection{Extrinsic Curvature}

In 3+1 formalism, the global manifold is assumed to be constructed by a family of embedded hypersurfaces. Naturally, we expect that there must be a machine (or a tensor field) which will measure \emph{how much the hypersurfaces are bending within the global manifold}. Fortunately, there is one which is known as the \emph{extrinsic curvature} \textbf{K}. To find the explicit form of \textbf{K} let us first image a physical case: when a drop of ink is being released orthogonally onto the surface of water, it spreads over the surface. Therefore, the orthogonal release of the drop is something related to the spread of it over the surface. By taking this approach as a reference, we see that we need a vector which is related to the spread of the unit normal vector $ \hat{\textbf{n}} $ over the hypersurface $ \Sigma $. Basically, this is done by \emph{endomorphism} \emph{Weingarten map} (or \emph{shape operator}) $ \chi $ of $ \mathcal{T}_p (\Sigma) $    
\begin{equation*}
\begin{aligned}
 \chi : \mathcal{T}_p(\Sigma) &\longrightarrow \mathcal{T}_p(\Sigma) \, ,\\
\textbf{v} &\longrightarrow \nabla_\textbf{v} \hat{\textbf{n}} \, .
\end{aligned}
\end{equation*}
In words, \emph{the machine $ \chi $ inserts the unit normal vector $ \hat{\textbf{n}} $ into its slot and migrates the directional derivative of $ \hat{\textbf{n}} $ which is an element of  $ \mathcal{T}_p (\Sigma) $}
\begin{equation*}
 \hat{\textbf{n}} . \chi(\textbf{v})= \hat{\textbf{n}}.\nabla_\textbf{v} \hat{\textbf{n}}=\frac{1}{2} \nabla_\textbf{v}[ \hat{\textbf{n}}. \hat{\textbf{n}}]=0 \, . 
\end{equation*}
Now, let us deduce the crucial property ''\emph{self-adjointness with respect to the induced metric $ \gamma $} '' of $ \chi $:
$  \forall (\textbf{u},\textbf{v}) \in \mathcal{T}_p(\Sigma) \otimes \mathcal{T}_p(\Sigma) $,
\begin{equation}
 \textbf{u}.\chi({\textbf{v}})=\textbf{u}.\nabla_\textbf{v} \hat{\textbf{n}}=\nabla_\textbf{v}[\textbf{u}.\hat{\textbf{n}}]
-\hat{\textbf{n}}.\nabla_\textbf{v} \textbf{u}=-\hat{\textbf{n}}.\nabla_\textbf{v} \textbf{u} \, ,
\label{Tors}
\end{equation}
We assume that the \emph{torsion} tensor is zero \Big ($ \nabla_\textbf{v} \textbf{u}-\nabla_\textbf{u} \textbf{v}-[\textbf{u},\textbf{v}]=0 $\Big). Then, ( \ref{Tors} ) becomes
\begin{equation}
\begin{aligned} 
 \textbf{u}.\chi({\textbf{v}})&=-\hat{\textbf{n}}.\Big (\nabla_\textbf{u} \textbf{v}-[\textbf{u},\textbf{v}]\Big ) \\
&=-\hat{\textbf{n}}.\nabla_\textbf{u} \textbf{v}+\hat{\textbf{n}}.[\textbf{u},\textbf{v}] \\
&=-\nabla_\textbf{u}\Big (\hat{\textbf{n}}.\textbf{v} \Big )+\textbf{v}.\nabla_\textbf{u} \hat{\textbf{n}}+\hat{\textbf{n}}.[\textbf{u},\textbf{v}] \\
&=\textbf{v}.\nabla_\textbf{u} \hat{\textbf{n}}+\hat{\textbf{n}}.[\textbf{u},\textbf{v}] \, .  
\label{zero}
\end{aligned}
\end{equation}
For the sake of \emph{self-adjointness of $ \chi $}, we need to show that the last term of (\ref{zero}) disappears
\begin{equation}
\begin{aligned} 
\nabla t.[\textbf{u},\textbf{v}]&=<dt,[\textbf{u},\textbf{v}]> \\
&=<dt,(\nabla_\textbf{u} \textbf{v}-\nabla_\textbf{v} \textbf{u})> \\
&=<dt,\nabla_\textbf{u} \textbf{v}>-<dt,\nabla_\textbf{v} \textbf{u})> \\
&=<\nabla_\mu t\, e^\mu,u^\nu \nabla_\nu v^\delta e_\delta>-<\nabla_\mu t\, e^\mu,v^\nu \nabla_\nu u^\delta e_\delta> \\
&=\nabla_\mu t\, u^\nu \nabla_\nu v^\delta <e^\mu,e_\delta> -\nabla_\mu t\, v^\nu \nabla_\nu u^\delta < e^\mu,e_\delta> \\
&=\nabla_\mu t\, u^\nu \nabla_\nu v^\delta \delta^\mu{_\delta}-\nabla_\mu t\, v^\nu \nabla_\nu u^\delta \delta^\mu{_\delta} \\
&=\nabla_\mu t\, u^\nu \nabla_\nu v^\mu-\nabla_\mu t\, v^\nu \nabla_\nu u^\mu \\
&=u^\nu \Big [\nabla_\nu (v^\mu \nabla_\mu t )-v^\mu \nabla_\nu \nabla_\mu t \Big]-v^\nu \Big [\nabla_\nu (u^\mu \nabla_\mu t )-u^\mu \nabla_\nu \nabla_\mu t \Big] \\
&=0 \, ,
\end{aligned}
\end{equation}
where we used the fact that $ v^\mu $ is orthogonal to $ \nabla_\mu t $. Thus, we proved that $ \chi $ is really \emph{self-adjoint}
\begin{equation}
 \forall (\textbf{u},\textbf{v}) \in \mathcal{T}_p(\Sigma) \otimes \mathcal{T}_p(\Sigma)\,,\, \textbf{u}.\chi(\textbf{v})=\chi(\textbf{u}).\textbf{v} \, ,
\end{equation}
Since the Weingarten map $ \chi $ is self-adjoint, its \emph{eigenvalues} are taken as \textbf{\emph{the principal curvatures}}, $ \kappa_i $, of the hypersurface $ \Sigma $  and \emph{the corresponding eigenvectors} are taken as \textbf{\emph{the principal directions}} of the hypersurface such that the mean of the $ \kappa_i $ is known as \textbf{\emph{the mean curvature}},$ \mathcal{H} $, of $ \Sigma $
\begin{equation}
  \mathcal{H}=\frac{1}{3} \sum_{i=1}^{3} \kappa_i \, .
\end{equation}
Contrary to the intrinsic curvature, $ \kappa_i $ and $ \mathcal{H} $ are depend on how the hypersurface is embedded into $ \mathcal{M} $ so they are taken as extrinsic character of $ \Sigma $. Now, we are ready to construct the explicit mapping of 2-form of the extrinsic curvature $ \textbf{K} $ of $ \Sigma $ by using the Weingarten map $ \chi $. $ \textbf{K} $ is assumed to be a machine which has two slots for the vectors of $ \mathcal{T}_p(\Sigma) $ and whose output is a scalar  
\begin{equation}
\begin{aligned}
 \textbf{K}:\mathcal{T}_p(\Sigma) \otimes \mathcal{T}_p(\Sigma) &\longrightarrow \mathcal{R} \, , \\
(\textbf{u},\textbf{v}) &\longrightarrow - \textbf{u}.\chi(\textbf{v}) \, \, .
\label{extr}
\end{aligned}
\end{equation}
This is the well-known relation of \textbf{\emph{the second fundamental form}} (or \textbf{\emph{the extrinsic curvature tensor}} ) of the hypersurface $ \Sigma $. Moreover, the relation between the contraction of $ \textbf{K} $ with respect to the induced metric $ \gamma $ and the mean curvature, $ \mathcal{H} $  of $ \Sigma $ is
\begin{equation}
 K=\gamma^{ij}K_{ij}=-3 \mathcal{H} \, .
\end{equation}

Up to now, we dealt with timelike and spacelike hypersurfaces. Now, we will restrict ourself to spacelike hypersurface in which the signature  of the induced metric $ \gamma $ is (+,+,+) and we will define the fundamental geometrical tools for it:

\subsubsection{The Orthogonal Projector}

The tangent space of $ \mathcal{M} $ at a point p, $ \mathcal{T}_p(\mathcal{M}) $, can be orthogonally decomposed into the corresponding tangent space of the hypersurface $ \Sigma $ at the point p, $ \mathcal{T}_p(\Sigma) $, and a one-dimensional vector space of $ \hat{\textbf{n}} $, $ \mathcal{V}$ect$ (\hat{\textbf{n}}) $
\begin{equation}
\mathcal{T}_p(\mathcal{M})=\mathcal{T}_p(\Sigma) \oplus \mathcal{V}ect(\hat{\textbf{n}}) \, .
\label{orth_dec}
\end{equation}
where $ \mathcal{V}$ect$ (\hat{\textbf{n}}) $ is a 1-dimensional vector space for $ \hat{\textbf{n}} $. Because in the null case $ \mathcal{V}$ect$ (\hat{\textbf{n}})  \subset \mathcal{T}_p(\Sigma) $, the orthogonal decomposition of vector space (\ref{orth_dec}) is \emph{valid only for spacelike and timelike hypersurfaces}. Now, the orthogonal decomposition (\ref{orth_dec}) of $ \mathcal{T}_p(\mathcal{M}) $ allows us to define an operator $ \overset{\rightarrow}{\gamma} $ which projects the elements of $ \mathcal{T}_p(\mathcal{M}) $ into of $ \mathcal{T}_p(\Sigma) $
\begin{equation}
 \begin{aligned}
\overset{\rightarrow}{\gamma}:\mathcal{T}_p(\mathcal{M}) &\longrightarrow \mathcal{T}_p(\Sigma) \, \\  
\textbf{v} &\longrightarrow \textbf{v}+ (\hat{\textbf{n}}.\textbf{v}) \hat{\textbf{n}} \, ,
\label{orth_proj}
\end{aligned}
\end{equation}
here $ \overset{\rightarrow}{\gamma} $ is known as \emph{the orthogonal projection operator}. It selects the components of the vector of $ \Sigma $ among of $ \mathcal{M} $. Therefore, the projection of the unit normal vector $ \hat{\textbf{n}} $ is equal to zero [i.e. since $ \hat{\textbf{n}}.\hat{\textbf{n}}=-1 $, then, $ \overset{\rightarrow}{\gamma}(\hat{\textbf{n}}) =\hat{\textbf{n}}+(\hat{\textbf{n}}.\hat{\textbf{n}})\hat{\textbf{n}}=0  \, $] and it acts as an identity operator for vectors of $ \mathcal{T}_p (\Sigma) $ $ [i.e. \, \forall\, \textbf{v} \in \mathcal{T}_p(\Sigma)\,,\,\overset{\rightarrow}{\gamma}(\textbf{v})=\textbf{v}+(\hat{\textbf{n}}.\textbf{v})\hat{\textbf{n}}=\textbf{v} ] $. Further insight, the orthogonal projection operator $ \overset{\rightarrow}{\gamma} $ can be expanded relative a set of bases $ (e_\alpha) $ of $ \mathcal{T}_p(\mathcal{M}) $ and the corresponding components are 
\begin{equation}
 \gamma^\alpha{_\beta}=\delta^\alpha{_\beta}+n^\alpha n_\beta \, .
\label{orth_proj_oper}
\end{equation}
We mentioned in (\ref{push_forw}) and (\ref{pull_back}) the embedding $ \Phi $ induces the push-forward mapping, $ \Phi_* $ and the pull-back mapping, $ \Phi^* $ in the given direction and does not imply in the reverse directions. On the other hand, as we illustrated in (\ref{orth_proj}) that $ \overset{\rightarrow}{\gamma} $ carries the elements from $ \mathcal{T}_p(M) $ and projects them into of $ \mathcal{T}_p(\Sigma) $. And, it induces another mapping $ \overset{\rightarrow}{\gamma} ^*_{\mathcal{M}} $ between the corresponding dual spaces [from $ \mathcal{T}^*_p(\Sigma) $ to $ \mathcal{T}^*_p(\mathcal{M}) $]
\begin{equation}
 \begin{aligned}
\tilde{\textbf{w}} \in \mathcal{T}^*_p(\Sigma) \,, \textbf{v} \in \mathcal{T}_p(\mathcal{M}) ,\, \, \, \,\, \,\overset{\rightarrow}{\gamma} ^*_{\mathcal{M}}: \mathcal{T}^*_p(\Sigma) &\longrightarrow \mathcal{T}^*_p(\mathcal{M}) \, , \\
\tilde{\textbf{w}} &\longrightarrow \overset{\rightarrow}{\gamma} ^*_{\mathcal{M}} \tilde{\textbf{w}}:\mathcal{T}_p(\mathcal{M}) \longrightarrow \mathcal{R} \, ,\\
& \quad \qquad \qquad \qquad \quad \textbf{v} \longrightarrow <\overset{\rightarrow}{\gamma} ^*_{\mathcal{M}} \tilde{\textbf{w}},\textbf{v}>  \\
& \quad \qquad \qquad \qquad \qquad \quad =\overset{\rightarrow}{\gamma} ^*_M \tilde{\textbf{w}}(\textbf{v}) \\
& \quad \qquad \qquad \qquad \qquad \quad = \tilde{\textbf{w}} \Big(\overset{\rightarrow}{\gamma}(\textbf{v}) \Big ) \, . 
 \end{aligned}
\end{equation}
Also, the induced mapping $ \overset{\rightarrow}{\gamma} ^*_{\mathcal{M}} $ can map arbitrary n-form  $ \mathcal{A} $   of $ \mathcal{T}^*_p(\Sigma) $ 
\begin{equation}
\begin{aligned}
 \forall \,\mathcal{A} \in \mathcal{T}^*_p(\Sigma)^n\,,\,\overset{\rightarrow}{\gamma} ^*_{\mathcal{M}}: \mathcal{A} \longrightarrow \overset{\rightarrow}{\gamma} ^*_{\mathcal{M}} \mathcal{A} : \mathcal{T}_p(\mathcal{M})^n &\longrightarrow \mathcal{R} \, ,\\
(\textbf{v}_1,...,\textbf{v}_n) &\longrightarrow \overset{\rightarrow}{\gamma} ^*_{\mathcal{M}} \mathcal{A}(\textbf{v}_1,...,\textbf{v}_n) \, ,\\
&\qquad=\mathcal{A}(\overset{\rightarrow}{\gamma}({\textbf{v}}_1),...,\overset{\rightarrow}{\gamma}({\textbf{v}}_n)) \, .
\end{aligned}
\end{equation}
Particularly, the extension of 2-form induced metric $ \gamma $ to $ \mathcal{M} $ will act on the vectors of $ \mathcal{T}_p(\mathcal{M}) $ .Then, we denote it with the same symbol, $ \gamma=\overset{\rightarrow}{\gamma} ^*_{\mathcal{M}} \gamma $. The relation between the extended induced metric $ \gamma $ and the spacetime metric \textbf{g} is
\begin{equation}
 \gamma = \textbf{g} + \tilde{\textbf{n}} \otimes \tilde{\textbf{n}} =\overset{\rightarrow}{\gamma} ^*_{\mathcal{M}} \gamma \, .  
\label{indu}
\end{equation}
where $ \tilde{\textbf{n}} $ is a 1-form. As we did before, $  \gamma_{\alpha\beta}=g_{\alpha\beta}+n_\alpha n_\beta $ are the components of the extended induced metric $ \gamma $ ($ =\overset{\rightarrow}{\gamma} ^*_{\mathcal{M}} \gamma $) relative to a proper family of basis $ (e^\alpha) $ of $ \mathcal{T}^*_p(\mathcal{M}) $. Let us take a look at the action of $ \gamma $ on the particular cases:
\begin{enumerate}
 \item $ \forall (\textbf{v},\textbf{u}) \in \Sigma $, then, the induced metric $ \gamma $ and the spacetime metric \textbf{g} will do the same job on these vectors
\begin{equation}
\begin{aligned}
 \overset{\rightarrow}{\gamma} ^*_{\mathcal{M}} \gamma(\textbf{u},\textbf{v}) &= \textbf{g}(\textbf{u},\textbf{v}) + <\tilde{\textbf{n}},\textbf{u}> <\tilde{\textbf{n}},\textbf{v}> \\
& = \textbf{g}(\textbf{u},\textbf{v}) \\
&=g_{\mu\nu} u^\mu v^\nu \, .
\end{aligned}
\end{equation}
\item On the other hand, if one of these vectors (consider $ \textbf{u}=\lambda \hat{\textbf{n}} $) is collinear with $ \hat{\textbf{n}} $, then, the action of $ \gamma $ will be zero. That's, for any $ \textbf{v} \in T_p(\mathcal{M}) $
\begin{equation}
\begin{aligned}
 \gamma(\textbf{u},\textbf{v})&=\overset{\rightarrow}{\gamma} ^*_{\mathcal{M}} \gamma(\textbf{u},\textbf{v}) \\
&=\lambda \textbf{g}(\hat{\textbf{n}},\textbf{v}) + \lambda <\tilde{\textbf{n}},\hat{\textbf{n}}> <\tilde{\textbf{n}},\hat{\textbf{v}}> \\
&=\lambda \Big \{\textbf{g}(\hat{\textbf{n}},\textbf{v})-<\tilde{\textbf{n}},\hat{\textbf{v}}> \Big \} \\
&=\lambda \Big \{<\tilde{\textbf{n}},\hat{\textbf{v}}>-<\tilde{\textbf{n}},\hat{\textbf{v}}> \Big \} \\
&=0 \, .
\end{aligned}
\end{equation}
\end{enumerate}
By observing the equation (\ref{orth_proj_oper}) and the components of the extended metric $ \gamma $ ($ =\overset{\rightarrow}{\gamma}^* \gamma $ ), we see that the orthogonal projection operator $ \overset{\rightarrow}{\gamma} $ is obtained from the extended metric by raising its $ 1^{st} $ index. Indeed, we use the same symbol for the extension of the extrinsic curvature \textbf{K} to $ \mathcal{M} $, too:
\begin{equation}
 K= \overset{\rightarrow}{\gamma} ^*_{\mathcal{M}} K \, .
\end{equation}
Finally, with the help of the orthogonal projection operator $\overset{\rightarrow}{\gamma} $ any rank-(p+q) tensor \textbf{T} [ $ \in \mathcal{T}(\mathcal{M})^p \otimes \mathcal{T}^*(\mathcal{M})^q $ ] can be converted into another tensor, $\overset{\rightarrow}{\gamma}^*_{\mathcal{M}} \textbf{T} $, of \emph{same type} which is still an element of $ \mathcal{T}(\mathcal{\mathcal{M}})^p \otimes \mathcal{T}^*(\mathcal{M})^q $. The transformation between their components is by
\begin{equation}
( \overset{\rightarrow}{\gamma} ^*_{\mathcal{M}} T)^{\alpha_1...\alpha_p}{_{\beta_1...\beta_q}}=
\gamma^{\alpha_1}{_{\mu_1}}...\gamma^{\alpha_p}{_{\mu_p}}\gamma^{\nu_1}{_{\beta_1}}...\gamma^{\nu_q}{_{\beta_q}} T^{\mu_1...\mu_p}{_{\nu_1...\nu_q}} \, .
 \end{equation}

 \subsubsection{Relation Between ''\textbf{K}'' and $ \nabla_{\tilde{\textbf{n}}} $}

Up to now we have not said anything about the diffusion of the unit normal vector  $ \hat{\textbf{n}} $ within the neighborhood of a point p of the hypersurface. We only assumed $ \hat{\textbf{n}} $ to be at points of the hypersurface. Basically, we will see that deviation of $ \hat{\textbf{n}} $ leads us to the relation between the extrinsic curvature \textbf{K} and the covariant derivative along $ \hat{\textbf{n}} $. Therefore, we need to define the acceleration vector \textbf{a} of
 $ \hat{\textbf{n}} $
\begin{equation}
 \textbf{a} = \nabla_{\hat{\textbf{n}}} \hat{\textbf{n}} \, .
\label{accel}
\end{equation}
If we assume that $ \hat{\textbf{n}} $  is a 4-velocity of an observer (since $ \hat{\textbf{n}} $ is a timelike vector), then, $ \hat{\textbf{n}} $ is 4-acceleration of the observer. Furthermore, this deviation is an element of $ \mathcal{T}_p(\Sigma) $,
\begin{equation}
 \hat{\textbf{n}}.\textbf{a} = \hat{\textbf{n}}.\nabla_{\hat{\textbf{n}}} \hat{\textbf{n}}=\frac{1}{2}\nabla_{\hat{\textbf{n}}} [ \hat{\textbf{n}}.\hat{\textbf{n}}] 
=0 \, .
\end{equation}
Now, we have emphasized that \textbf{K} can be taken as a machine which has two slots for inputting vectors. Therefore, let us extend it to $ \mathcal{M} $ and insert two vectors belonging the tangent space of $ \mathcal{M} $ at p: $  \forall (\textbf{u},\textbf{v}) \in \mathcal{T}_p(\mathcal{M}) $,
\begin{equation}
 \begin{aligned}
(\overset{\rightarrow}{\gamma}^*_{\mathcal{M}} K)(\textbf{u},\textbf{v}) &=K\Big(\overset{\rightarrow}{\gamma}(\textbf{u}),\overset{\rightarrow}{\gamma}(\textbf{v}) \Big) \\
&=-\overset{\rightarrow}{\gamma}(\textbf{u}).\nabla_{\overset{\rightarrow}{\gamma}(\textbf{v})} \hat{\textbf{n}} \\
&=-[\textbf{u}+(\hat{\textbf{n}}.\textbf{u})\hat{\textbf{n}}]. \nabla_{[\textbf{v}+(\hat{\textbf{n}}.\textbf{v})\hat{\textbf{n}}]} \hat{\textbf{n}} \\
&=-\textbf{u}.\nabla_{\textbf{v}}\hat{\textbf{n}}-(\hat{\textbf{n}}.\textbf{u})\hat{\textbf{n}}.\nabla_{\textbf{v}}\hat{\textbf{n}}-
(\hat{\textbf{n}}.\textbf{v})(\textbf{u}.\textbf{a})-(\hat{\textbf{n}}.\textbf{v})(\hat{\textbf{n}}.\textbf{u})(\hat{\textbf{n}}.\textbf{a}) \\
&=-\textbf{u}.\nabla_{\textbf{v}}\hat{\textbf{n}}-(\hat{\textbf{n}}.\textbf{v})(\textbf{u}.\textbf{a}) \\
&=-<\nabla \tilde{\textbf{n}}(...,\textbf{v)}),\textbf{u}>-<\tilde{\textbf{a}},\textbf{u}><\tilde{\textbf{n}},\textbf{v}> \, ,
\label{K}
\end{aligned}
\end{equation}
where we used $\textbf{u}.\textbf{v}=g(\textbf{u},\textbf{v})=g(u^\mu e_\mu,v^\nu e_\nu)=g_{\mu\nu}u^\mu v^\nu=u^\mu v_\mu $. We know that the equation(\ref{K}) is valid for any pair of tangent vectors of $ \mathcal{M} $ so we can drop the vectors to get the compact form as
\begin{equation}
 \textbf{K}=-\nabla \tilde{\textbf{n}}-\tilde{\textbf{a}} \otimes \tilde{\textbf{n}} 
\Longrightarrow \nabla \tilde{\textbf{n}} =-\textbf{K}-\tilde{\textbf{a}} \otimes \tilde{\textbf{n}} \, ,
\label{k}
\end{equation}
here the symbol ( $ \tilde{} $ ) means dual of the vector. The components of the tensorial equation (\ref{k}) with respect to a given basis of $ \mathcal{T}^*_p(\mathcal{M}) $ are 
\begin{equation}
 \nabla_\beta n_\alpha = -K_{\alpha\beta}-a_\alpha n_\beta \, .
\label{kk}
\end{equation}
The equation ( \ref{kk}) is defined onto the four-dimensional manifold $ \mathcal{M} $. Then, let us pull-back it to $ \mathcal{T}^*_p(\Sigma) $
\begin{equation}
 \gamma^\mu{_\alpha} \gamma^\nu{_\beta} \nabla_\nu n_\mu = - \gamma^\mu{_\alpha} \gamma^\nu{_\beta} K_{\mu\nu}- \gamma^\mu{_\alpha} \gamma^\nu{_\beta} a_\mu n_\nu \, ,
\label{kkk}
\end{equation}
notice that the last term on the right hand side of the equation (\ref{kkk}) vanishes because the projection of $ n_\nu $ onto $ \Sigma $ is zero so we get
\begin{equation}
\gamma^\mu{_\alpha} \gamma^\nu{_\beta} \nabla_\nu n_\mu = - \gamma^\mu{_\alpha} \gamma^\nu{_\beta} K_{\mu\nu} \, ,
\label{kkkk1}
\end{equation}
or in compact form
\begin{equation}
 \textbf{K}= \overset{\rightarrow}{\gamma}^* \textbf{K} =- \overset{\rightarrow}{\gamma}^* \nabla \tilde{\textbf{n}} \, .
\label{kkkk}
\end{equation}
As we see in the equation (\ref{kkkk}), \emph{the projected form of $ \nabla \tilde{\textbf{n}} $ onto $ \Sigma $ (i.e. $-\textbf{K} $) is symmetric though the four-dimensional $ \nabla \tilde{\textbf{n}} $ is not}. Now, it is easy to show that the trace of the equation (\ref{kk}) with respect to $ g^{\alpha\beta} $ is
\begin{equation}
 K=-\nabla_\beta n^\beta \, ,
\end{equation}
or in compact form
\begin{equation}
 K=-\nabla.\hat{\textbf{n}} \, \, .
\label{kk2}
\end{equation}
The equation (\ref{kk2}) gives the relation between the scalar extrinsic curvature and the divergence of the unit normal vector.

\subsubsection{Relation between Connections of the Spacetime and of the Hypersurface}

Consider a tensor field $ \textbf{T} $ onto $ \Sigma $. Both the tensor field  $ \textbf{T} $ and its covariant derivative \textbf{DT} are tensor fields of $ \mathcal{M} $, too. As we implies before, we are able to convert \textbf{DT} into another vector field of $ \mathcal{M} $ which is denoted by $ \overset{\rightarrow}{\gamma}^* \nabla \textbf{T} $
\begin{equation}
 D \textbf{T} = \overset{\rightarrow}{\gamma}^* \nabla \textbf{T} \, ,
\end{equation}
or more precisely,
\begin{equation}
 D\textbf{T} = \overset{\rightarrow}{\gamma}^* \nabla \overset{\rightarrow}{\gamma}^*_{\mathcal{M}} \textbf{T} \, .
\label{covr_derivs} 
\end{equation}
And the related transformation of the components is given by
\begin{equation}
 D_\rho T^{\alpha_1...\alpha_p}{_{\beta_1...\beta_q}}=\gamma^{\alpha_1}{_{\mu_1}}...\gamma^{\alpha_p}{_{\mu_p}}
\gamma^{\nu_1}{_{\beta_1}}...\gamma^{\nu_q}{_{\beta_q}}\gamma^\sigma{_\rho} \nabla_\sigma T^{\mu_1...\mu_p}{_{\nu_1...\nu_q}} \, . 
\label{trans}
\end{equation}
The crucial properties of the transformation of ( \ref{covr_derivs}) are
\begin{enumerate}
 \item $ \overset{\rightarrow}{\gamma}^* \nabla \overset{\rightarrow}{\gamma}^*_{\mathcal{M}} \textbf{T} $ is a torsion-free connection on $ \Sigma $, it satisfies all the defining properties of a connection  [linearity, reduction to gradient for a scalar field, commutes with contraction and Leibniz rule].
\item $ \overset{\rightarrow}{\gamma}^* \nabla \overset{\rightarrow}{\gamma}^*_{\mathcal{M}} \textbf{T} $ is metric compatible
\begin{equation}
\begin{aligned}
\Big( \overset{\rightarrow}{\gamma}^* \nabla \gamma \Big)_{\alpha\beta\gamma}
&=\gamma^\mu{_\alpha} \gamma^\nu{_\beta} \gamma^\rho{_\gamma} \nabla_\rho \gamma_{\mu\nu} \\
&=\gamma^\mu{_\alpha} \gamma^\nu{_\beta} \gamma^\rho{_\gamma} \nabla_\rho \Big [g_{\mu\nu} + n_\mu n_\nu \Big ]\\
&=\gamma^\mu{_\alpha} \gamma^\nu{_\beta} \gamma^\rho{_\gamma} \Big \{ \nabla_\rho g_{\mu\nu} + (\nabla_\rho n_\mu) n_\nu + n_\mu (\nabla_\rho n_\nu) \Big \} \\       
&=0 \, ,
\end{aligned}
\end{equation}
where we have used the metric compatibility of $ \textbf{g} $ and the fact that the orthogonal projection of $ \hat{\textbf{n}} $  onto $ \Sigma $ vanishes.
\end{enumerate}
Finally, let us derive the relation between the connections:
\begin{equation}
\begin{aligned}
\forall(\textbf{u},\textbf{v}) \in \mathcal{T}_p(\Sigma) \otimes \mathcal{T}_p(\Sigma) \,,\,\, \Big (D_{\textbf{u}} \textbf{v} \Big)^\alpha &=u^\sigma D_{\sigma} v^\alpha \\
&= u^{\sigma} \gamma^\nu{_\sigma} \gamma^\alpha{_\mu} \nabla_\nu v^\mu \\
&= u^\nu \gamma^\alpha{_\mu} \nabla_\nu v^\mu \\
&= u^\nu \Big \{ \delta^\alpha{_\mu} + n^\alpha n_\mu \Big \} \nabla_\nu v^\mu \\
&= u^\nu \delta^\alpha{_\mu} \nabla_\nu v^\mu + u^\nu n^\alpha \underbrace{n_\mu \nabla_\nu v^\mu} \\
& \qquad \qquad \qquad \qquad \quad -v^\mu \nabla_\nu n_\mu \\
&=u^\nu \nabla_\nu v^\alpha - u^\nu v^\mu n^\alpha \nabla_\nu n_\mu \, .
\label{s}
\end{aligned}
\end{equation}
Now, the last term of the equation (\ref{s}) is nothing but the definition of the extrinsic curvature (\ref{extr}). By using this fact, the equation ( \ref{s}) becomes
\begin{equation}
 D_\textbf{u} \textbf{v}=\nabla_\textbf{u} \textbf{v} + \textbf{K}(\textbf{u},\textbf{v}) \hat{\textbf{n}} \, .
\end{equation}
That's, \emph{the difference between the directional covariant derivatives $ D_\textbf{u} \textbf{v} $ and $ \nabla_\textbf{u} \textbf{v} $ is given by the extrinsic curvature \textbf{K(u,v)}. Furthermore, the difference is along the unit normal vector $ \hat{\textbf{n}} $}.

\subsection{Geometry of Foliation}

\begin{figure}[h]
\centering
\includegraphics[width=0.6\textwidth]{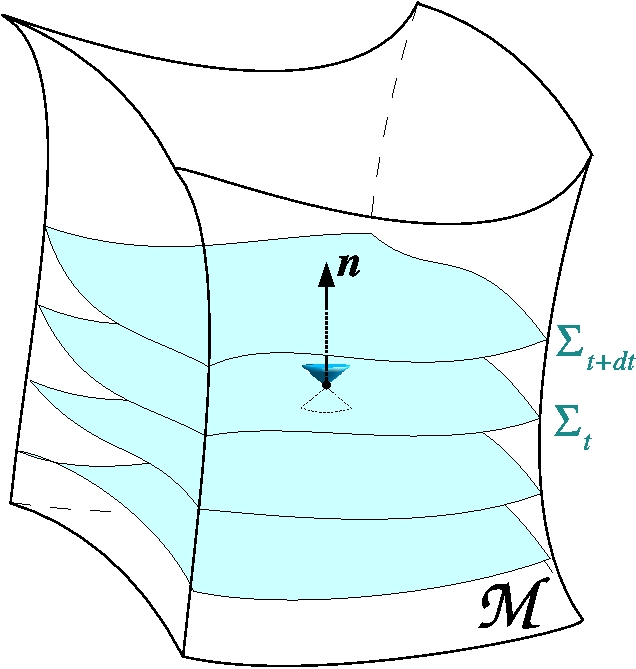}
\caption{The foliation of $ (\mathcal{M},\textbf{g} $)}
\label{fol_foliat}
\end{figure}

From now, as we see in the figure (\ref{fol_foliat}), we will deal with a continuous family of embedded hypersurfaces $ (\Sigma_t)_{t \in \mathcal{R}} $ which covers the spacetime $ (\mathcal{M},\textbf{g} $). However, the spacetime $ (\mathcal{M},\textbf{g} $) that we are going to foliate (i.e. to slice) is not any type of spacetime rather it is a \emph{globally-hyperbolic spacetime}. 

\subsubsection{Globally Hyperbolic Spacetimes and Foliation}

In order to define what the \emph{globally-hyperbolic spacetime} is, we need to first define the concept of \emph{Cauchy Surface}: \emph{Any spacelike hypersurface which is being intersected by causal curves (i.e. timelike and null) if and only if one time is called \textbf{A Cauchy Surface}}. Now, any spacetime on which there can be defined a continuous family of \emph{Cauchy surface} is called a \emph{globally-hyperbolic spacetime} [figure (\ref{fol_foliat})]. Also, it is obvious that the topology of the \emph{globally-hyperbolic spacetime} is essentially $ \Sigma \times \mathcal{R} $. In words, the foliation of the spacetime is that there is assumed to be a smooth scalar field $ \hat{t} $ on the four-dimensional manifold  $ \mathcal{M} $ such that the union of points at which the scalar field is identical construct the related hypersurface 
\begin{equation}
 \forall t \in \mathcal{R} \,,\, \Sigma_t = \Big \{p \in \mathcal{M} \,,\, \hat{t}(p)=t \Big \} \Longrightarrow \mathcal{M}= \underset{t \in R}{\bigcup} \Sigma_t \, ,
\end{equation}
where we assume that the gradient of the scalar field is always different than zero (i.e. regular) which ensures that the hypersurfaces never intersect         
\begin{equation}
 \Sigma_t \cap \Sigma_{t^{'}} = \emptyset\,\, for\,\, t \neq t^{'} \, .
\end{equation}

\subsubsection{Foliation Kinematics}

\begin{enumerate}
\item Lapse Function

For convention, let us use the symbol $ t $ as the scalar field on $ \mathcal{M} $. Then, the vector  $ \nabla t $ is essentially normal to the hypersurface $ \Sigma_t $ and not necessarily a unit normal vector. Therefore, we suppose that there exists a scalar field $ N \,(> 0 $ and called \emph{\textbf{lapse function}} \cite{Wheeler:1964og}) which is used to re-normalize $ \nabla t $ :
\begin{equation}
 \hat{\textbf{n}} = - N \nabla t = - \frac{\nabla t}{\sqrt{-{\nabla t.\nabla t}}} \,\Longrightarrow N = \frac{1}{\sqrt{-\nabla t . \nabla t}} \, ,
\label{laps}
\end{equation}
the minus sign in the equation (\ref{laps}) guarantees that  $ \hat{\textbf{n}} $ is \emph{future-oriented}. Because $ \nabla t $ is a vector, its dual is the gradient 1-form $ d t $. The crucial point is that the lapse function makes the $ d t $ a unit 1-form, too: Suppose that there is a scalar field $ S $ such that $ \tilde{\textbf{n}} = S\, dt $ so what is $ S $?,
\begin{equation}
\begin{aligned}
 <\tilde{\textbf{n}},\hat{\textbf{n}}> &=<S dt,-N \nabla t> \\
&=S N \Big \{- <dt,\nabla t > \Big \} \\
&= S.N \frac{1}{N^2}\\
&=-1 \, ,
\label{S}
\end{aligned}
\end{equation}
so $ S=-N $. Thus,
\begin{equation}
 \tilde{\textbf{n}}=-N dt \, .
\end{equation}

\begin{figure}[h]
\centering
\includegraphics[scale=0.4]{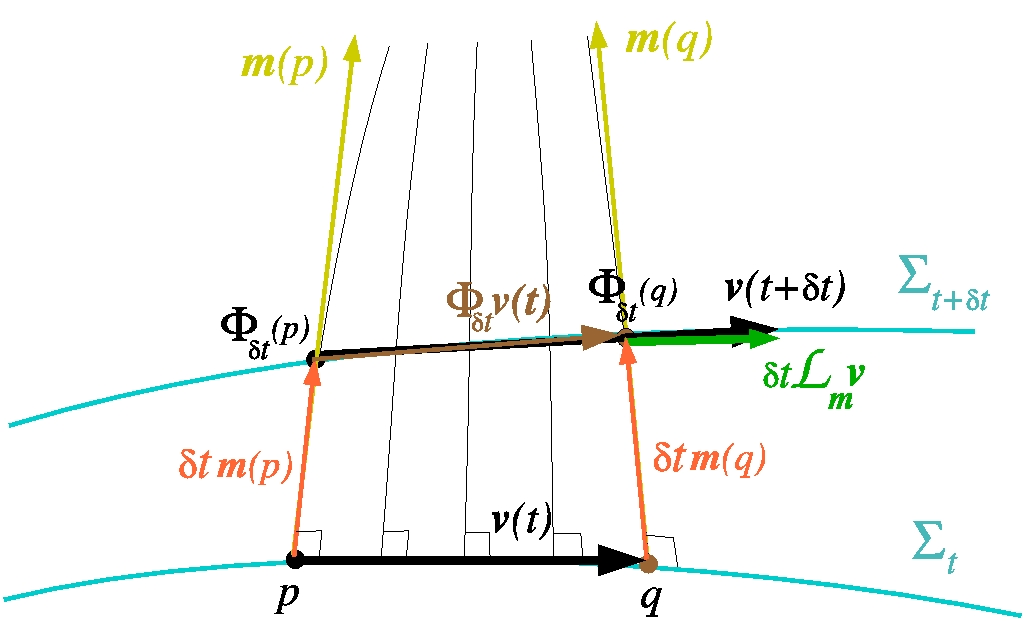}
\caption{The Lie Dragging of $ \textbf{v} \in  \mathcal{T}(\Sigma_t) $ along \textbf{m} such that  $ \mathcal{L}_\textbf{m} \textbf{v}  \in \mathcal{T}(\Sigma_t) $.Here
the diffeomorphism $ \Phi $ mappings the elements of $ \mathcal{T}_p(\Sigma_t) $ into  of $ \mathcal{T}_q(\Sigma_t) $ }
\label{fol_liem_vect}
\end{figure}

\newpage

\item Normal Evolution Vector

Now, when we evaluate the inner product between the vector $ \nabla t $ and the normal unit vector $ \hat{\textbf{n}} $, we will see that it is not equal to 1. This means that \emph{the normal unit vector $ \hat{\textbf{n}} $ can not follow the flow of the scalar field though it locally does}. Then, the modification of the $ \hat{\textbf{n}} $ to the evolution of the hypersurfaces seems as a primary condition. Therefore, we propose a new normal vector \textbf{m} (known as \emph{\textbf{normal evolution vector}})
\begin{equation}
 \textbf{m}=N \textbf{n} \Longrightarrow \textbf{m}.\textbf{m}=-N^2 \, ,
\label{N}
\end{equation}
\quad such that the normal evolution vector \textbf{m} is being adapted to the flow of the scalar field $ t $,
\begin{equation}
\begin{aligned}
 <dt,\textbf{m}> &= < dt,N \hat{\textbf{n}}> \\
&=N <dt,\hat{\textbf{n}}> \\
&= N <dt,-N \nabla t>  \\
&=-N^2 < dt,\nabla t> \\
&=-N^2.\frac{1}{N^2} \\
&=1 \, .
\end{aligned}
\end{equation}
This modification provides the evolution of the hypersurfaces. That's, all the points of the initial hypersurface  $ \Sigma_t $ are being carried along the vector $ \delta t \textbf{m} $ such that the union of carried points construct the neighbor hypersurface $\Sigma_{t+\delta t} \, $[see figure \ref{fol-liem-sigma}]. This evolution of the hypersurface is known as the \emph{\textbf{Lie dragging of hypersurfaces}} along \textbf{m}. As we see from the figure (\ref{fol_liem_vect}) the Lie dragging along \textbf{m} does not disturb the elements tangent to $ \Sigma_t $. That is , they are still the elements tangent to $ \Sigma_t $ after dragging along \textbf{m}:
\begin{equation}
 \forall \textbf{v} \in \mathcal{T}(\Sigma_t) \,,\,\, \mathcal{L}_\textbf{m} \textbf{v}\in \mathcal{T}(\Sigma_t) \, .
\end{equation}

 \begin{figure}[h]
\centering
\includegraphics[scale=0.4]{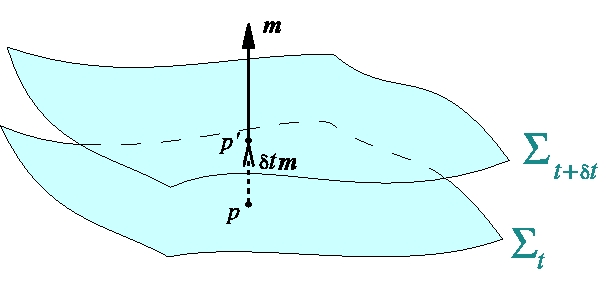}
\caption{The Lie Dragging of the Points}
\label{fol-liem-sigma}
\end{figure}

\item Eulerian Observers

Actually, the unit normal vector $ \hat{\textbf{n}} $ can be taken as the 4-velocity of \emph{\textbf{the Eulerian Observer}}. Then, the world line of \emph{Eulerian Observer} will intersect the hypersurface $ \Sigma_t $ only one time which says that all the simultaneous events at each constant scalar field (i.e. local) construct the hypersurface $ \Sigma_t $. Now, assume that there are two points (=events) p ($ \in \Sigma_t $) whose coordinate time is $ t $ and $ p^{'} $ ($ \in \Sigma_{t + \delta t } $) whose time coordinate is $ t +\delta t $ on the world-line of a \emph{\textbf{Eulerian Observer}}. Then, the elapsed time difference $ \delta \tau $ between two events with respect to the observer's clock is given by
\begin{equation}
\delta \tau=\sqrt{-\textbf{g}(\delta t \, \textbf{m},\delta t \, \textbf{m})}=\sqrt{\textbf{g}(-\textbf{\textbf{m}},\textbf{\textbf{m}})} \, \delta t \, ,
 \end{equation}
from the equation (\ref{N}), we get
\begin{equation}
 \delta \tau = N \, \delta t \, .
\end{equation}
As in the equation(\ref{accel}), the corresponding 4-acceleration of the \emph{the Eulerian Observer} is an element of $ \mathcal{T}_p(\Sigma_t) $
\begin{equation}
 \textbf{a}=\nabla_{\hat{\textbf{n}}} \hat{\textbf{n}} \, .
\end{equation}
Due to the fact that $ \textbf{a} \in \mathcal{T}_p(\Sigma_t) $, the 4-acceleration vector $ \textbf{a} $ can be rewritten in terms of the \emph{lapse function $ N $}:
\begin{equation*}
\begin{aligned}
a_\alpha &= n^\mu \nabla_\mu n_\alpha = n^\mu \nabla_\mu [-N \nabla_\alpha t ] 
 = -n^\mu (\nabla_\mu N )(\nabla_\alpha t)-N n^\mu \nabla_\mu \nabla_\alpha t \\
 &= -n^\mu \nabla_\mu N [-\frac{1}{N} n_\alpha ]- N n^\mu \nabla_\alpha \nabla_\mu t \\ 
&= \frac{1}{N} n_\alpha n^\mu (\nabla_\mu N )- N n^\mu \nabla_\alpha (- \frac{1}{N} n_\mu ) \\
& = \frac{1}{N} n_\alpha n^\mu (\nabla_\mu N ) + N n^\mu \nabla_\alpha ( \frac{1}{N} n_\mu ) \\
&= \frac{1}{N} n_\alpha n^\mu (\nabla_\mu N )- N \frac{1}{N^2} n^\mu n_\mu \nabla_\alpha N + N \frac{1}{N} n^\mu \nabla_\alpha n_\mu \\
& =\frac{1}{N} n_\alpha n^\mu (\nabla_\mu N ) + \frac{1}{N} \nabla_\alpha N \\
&= \frac{1}{N}[ \delta^\mu {_\alpha} \nabla_\mu N + n_\alpha n^\mu \nabla_\mu N] 
= \frac{1}{N} [\delta^\mu {_\alpha} + n_\alpha n^\mu ] \nabla_\mu N \\
& = \frac{1}{N} \gamma^\mu {_\alpha} \nabla_\mu N \Longrightarrow a_\alpha = \frac{1}{N} D_\alpha N=D_\alpha \ln N \, ,
\end{aligned}
\end{equation*}
or in compact form
\begin{equation}
  \tilde{\textbf{a}} = \tilde{\textbf{D}} \ln N \, . 
\label{a}
\end{equation}

\item Gradient of the 1-form $ \tilde{\textbf{n}} $ and $ \tilde{\textbf{m}} $

We are going to deduce very important two relations  of \emph{the gradient of $ \hat{\textbf{n}} $ and of \textbf{m} in terms of the extrinsic curvature and the lapse function}. Firstly, the substitution of the equation (\ref{a}) into the equation (\ref{k}) gives us \emph{\textbf{the gradient of the 1-form $ \tilde{\textbf{n}} $ }}, 
\begin{equation}
 \nabla \tilde{\textbf{n}} = -\textbf{K} - \tilde{D} \ln N \otimes \tilde{\textbf{n}} \Longrightarrow
 \nabla_\beta n_\alpha = -K_{\alpha\beta} - D_\alpha \ln N n_\beta \, .
\label{n}
\end{equation}
Secondly, the equations (\ref{N}) and (\ref{n}) are used to find the gradient of the dual of the \textbf{m}, 
\begin{equation*}
\begin{aligned}
 \nabla \tilde{\textbf{m}} &= \nabla (N \tilde{\textbf{n}} ) \\ 
&= N \nabla ( \tilde{\textbf{n}} ) + \tilde{\textbf{n}} \otimes \nabla N \\
&= N \Big ( \textbf{K} - \tilde{\textbf{D}} \ln N \otimes \tilde{\textbf{n}} \Big ) + \tilde{\textbf{n}} \otimes \nabla N \\ 
&= N \Big ( - \textbf{K} - \frac{1}{N} \tilde{\textbf{D}} N \otimes \tilde{\textbf{n}} \Big ) + \tilde{\textbf{n}} \otimes \nabla N \, ,
\end{aligned}
\end{equation*}
so we get the gradient of the dual of \textbf{m} as 
\begin{equation}
 \nabla \tilde{\textbf{m}} =  -\textbf{K} - \tilde{\textbf{D}} N \otimes \tilde{\textbf{n}} + \tilde{\textbf{n}} \otimes \nabla N \, .
\end{equation}
And its vector form as
\begin{equation}
\nabla \textbf{m} =  - \textbf{K} - \textbf{D} N \otimes \textbf{n} + \textbf{n} \otimes \nabla N \Rightarrow 
\nabla_\beta m_\alpha = -N K_{\alpha\beta} - D_\alpha N n_\beta + n_\alpha \nabla_\beta N \, .
\label{m}
\end{equation}

\item Evolution of the Induced 3-Metric

Under the flow of the hypersurfaces, the induced 3-metric is being carried by the Lie derivative along \textbf{m}. With the help of the equation the equation (\ref{m}), the Lie derivative of $ \gamma $ along \textbf{m} is deduced from
\begin{equation*}
\begin{aligned}
\mathcal{L}_\textbf{m} \gamma_{\alpha\beta} &= m^\mu \nabla_\mu \gamma_{\alpha\beta} + \gamma_{\mu\beta} \nabla_\alpha m^\mu + 
\gamma_{\alpha\mu} \nabla_\beta m^\mu \\
 &= m^\mu [ \nabla_\mu g_{\alpha\beta} + \nabla_\mu (n_\alpha n_\beta ) ] + \gamma_{\mu\beta}[-N K^\mu {_\alpha} -D^\mu N n_\alpha 
+ n^\mu \nabla_\alpha N ] \\
& \quad  + \gamma_{\alpha\mu} [-NK^\mu {_\beta} - D^\mu N n_\beta + n^\mu \nabla_\beta N ] \\
& = m^\mu [(\nabla_\mu n_\alpha ) n_\beta + n_\alpha (\nabla_\mu n_\beta ) ] - N \gamma_{\mu\beta} K^\mu {_\alpha} 
- \gamma_{\mu\beta} D^\mu N n_\alpha \\
& \quad + \gamma_{\mu\beta} n^\mu \nabla_\alpha N  
 - N \gamma_{\alpha\mu} K^\mu {_\beta} - \gamma_{\alpha\mu} D^\mu N n_\beta + \gamma_{\alpha\mu} n^\mu \nabla_\beta N \\
& = N n^\mu (\nabla_\mu n_\alpha ) n_\beta + N n_\alpha n^\mu (\nabla_\mu n_\beta ) - N K_{\beta\alpha} 
- D_\beta N n_\alpha - N K_{\alpha\beta} \\ 
& \quad - D_\alpha N n_\beta \\
&  = N a_\alpha n_\beta + N a_\beta n_\alpha - 2 N K_{\alpha\beta} 
- D_\beta N n_\alpha - D_\alpha N n_\beta \\
& = N D_\alpha \ln N n_\beta + N D_\beta \ln N n_\alpha - 2 N K_{\alpha\beta} - D_\beta N n_\alpha -D_\alpha N n_\beta \\
& = D_\alpha N n_\beta +  D_\beta N n_\alpha - 2 N K_{\alpha\beta} - D_\beta N n_\alpha -D_\alpha N n_\beta \, ,\\
\end{aligned}
\end{equation*}
so we get
\begin{equation}
\mathcal{L}_\textbf{m} \gamma_{\alpha\beta} = -2 N K_{\alpha\beta} \Longrightarrow \mathcal{L}_\textbf{m} \gamma = -2 N \textbf{K} \, .
\label{mm}
\end{equation}
Now,
\begin{equation}
\begin{aligned}
 \mathcal{L}_\textbf{m} \gamma_{\alpha\beta} &= \mathcal{L}_{N\textbf{n}}\gamma_{\alpha\beta} = N n^\mu \nabla_\mu \gamma_{\alpha\beta}
 + \gamma_{\mu\beta} \nabla_\alpha (N n^\mu ) + \gamma_{\alpha\mu} \nabla_\beta (N n^\mu ) \\
& = N n^\mu \nabla_\mu \gamma_{\alpha\beta} + \gamma_{\mu\beta} N \nabla_\alpha n^\mu + \gamma_{\mu\beta} n^\mu \nabla_\alpha N
+ \gamma_{\alpha\mu} N \nabla_\beta n^\mu \\
& \quad + \gamma_{\alpha\mu} n^\mu \nabla_\beta N \\
&= N \{ n^\mu \nabla_\mu \gamma_{\alpha\beta} + \gamma _{\mu\beta} \nabla_\alpha n^\mu
 + \gamma_{\alpha\mu} \nabla_\beta n^\mu \} \\
&= N \mathcal{L}_\textbf{n} \gamma_{\alpha\beta} \, .
\label{nn}
\end{aligned}
\end{equation}
Thus, the equations (\ref{mm}) and (\ref{nn}) lead us to rewrite \emph{the extrinsic curvature in terms of Lie derivative of the induced 3-metric along the unit normal vector}:
\begin{equation}
 \mathcal{L}_\textbf{n} \gamma_{\alpha\beta} = \frac{1}{N} \mathcal{L}_\textbf{m} \gamma_{\alpha\beta}
\Longrightarrow K_{\alpha\beta} = - \frac{1}{2}\mathcal{L}_\textbf{n} \gamma_{\alpha\beta} \, .
\end{equation}

\item Evolution of the Orthogonal Projector

In order to find the evolution of the tensor field \textbf{T} of $ \mathcal{T}(\Sigma) $, we need first to show what happens to the \emph{orthogonal projection operator} under the flow of the hypersurfaces. Again, the evolution of the orthogonal projection operator is done by the Lie derivative along \textbf{m}
\begin{equation*}
\begin{aligned}
 \mathcal{L}_\textbf{m} \gamma^\alpha {_\beta} &= m^\mu \nabla_\mu \gamma^\alpha {_\beta} - \gamma^\mu {_\beta}
\nabla_\mu m^\alpha + \gamma^\alpha {_\mu} \nabla_\beta m^\mu \\
& = N n^\mu \nabla_\mu [\delta^\alpha {_\beta} + n^\alpha n_\beta ] - \gamma^\mu {_\beta} [-N K^\alpha {_\mu} 
- D^\alpha N n_\mu + n^\alpha \nabla_\mu N ]\\ 
& \quad+ \gamma^\alpha {_\mu} [-N K^\mu {_\beta} - D^\mu N n_\beta + n^\mu \nabla_\beta N ] \, .
\end{aligned}
\end{equation*}
Because the projection of the normal vector and its dual onto the hypersurface is zero, we have
\begin{equation*}
\begin{aligned}
 \mathcal{L}_\textbf{m} \gamma^\alpha {_\beta} &= N (n^\mu \nabla_\mu n^\alpha ) n_\beta + N (n^\mu \nabla_\mu n_\beta )n^\alpha + N K^\alpha {_\beta} 
+ (D^\alpha N ) n_\beta \\
& \quad -n^\alpha D_\beta N - N K^\alpha {_\beta} - (D^\alpha N ) n_\beta \\
& = N a^\alpha n_\beta + N a_\beta n^\alpha - n^\alpha (D_\beta N) - (D^\alpha N ) n_\beta \\
& = N \frac{1}{N} D^\alpha N n_\beta + N \frac{1}{N} D_\beta N n^\alpha - n^\alpha (D_\beta N ) - (D^\alpha N ) n_\beta \\
& = 0 \, ,
\end{aligned}
\end{equation*}
 or in compact form 
\begin{equation}
 \mathcal{L}_\textbf{m} \overset{\rightarrow}{\gamma}=0 \, .
\label{proc}
\end{equation}
The equation (\ref{proc}) implies that \emph{if initially a tensor field \textbf{T} is an element of hypersurface, it will remain to be an element of the hypersurface throughout the evolution}. Moreover, let us show this in detail: if a tensor field \textbf{T} is a tensor field belonging to the hypersurface, then, the orthogonal projection operator acts as an identity operator,
\begin{equation}
 \overrightarrow{\gamma}^* \textbf{T} = \textbf{T} \, .
\label{ccc}
\end{equation}
Now, \emph{we are seeking after the Lie dragging whether there is any projected part of the carried tensor field of \textbf{T} along the normal unit vector $ \hat{\textbf{n}} $ or not}. If there is any, then, the Lie dragging will disturb the tensor \textbf{T} which is tangent to $ \Sigma $ to be still tangent to $ \Sigma $ during the flow of the hypersurfaces. Therefore, let us show on a simple sample of \textbf{T} of type $ \binom{1}{1} $. Then, the transformation in terms of the components is
\begin{equation}
 \gamma^\alpha {_\mu} \gamma^\nu {_\beta} T^\mu {_\nu} = T^\alpha {_\beta} \, .
\label{cc}
\end{equation}
Let us carry the transformation along \textbf{m},
\begin{equation}
 \mathcal{L}_\textbf{m} [ \gamma^\alpha {_\mu} \gamma^\nu {_\beta} T^\mu {_\nu} ] = (\mathcal{L}_\textbf{m} \gamma^\alpha {_\mu}) \gamma^\nu {_\beta} T^\mu {_\nu} 
+ \gamma^\alpha {_\mu}( \mathcal{L}_\textbf{m} \gamma^\nu {_\beta}) +  \gamma^\alpha {_\mu} \gamma^\nu {_\beta} \mathcal{L}_\textbf{m} T^\mu {_\nu} \, .
\label{f}
\end{equation}
Hence, the property of (\ref{proc}) allows to write (\ref{f}) as
\begin{equation}
  \gamma^\alpha {_\mu} \gamma^\nu {_\beta} \mathcal{L}_\textbf{m} T^\mu {_\nu} = \mathcal{L}_\textbf{m} T^\alpha {_\beta} \, ,
\end{equation}
or in compact form
\begin{equation}
 \overset{\rightarrow}{\gamma}^* \mathcal{L}_\textbf{m} \textbf{T}=\mathcal{L}_\textbf{m} \textbf{T} \, .
\label{tang}
\end{equation}
\emph{Thus, the Lie derivative along \textbf{m} of any tensor field \textbf{T} tangent to $ \Sigma_t $ is a tensor field tangent to $ \Sigma_t $}.
\end{enumerate}

\subsubsection{The Foliation of Hypersurfaces}

\begin{enumerate}
 \item Foliation of Spacelike Hypersurface $ \Sigma_t $ 

The coordinates adapted to the foliation $ x^\alpha=x^\alpha(y^a) $ is assumed to be the parametrized curves where the parametrization is the induced coordinates $ (y^a) $ of the hypersurface $ \Sigma_t $ \cite{Poisson:2004ue}. The projection operator can be written as 
\begin{equation}
 e^\alpha_a = \frac{\partial x^\alpha}{\partial y^a} \, .
\end{equation}
Since the extended line element of $ \Sigma_t $
\begin{equation}
 d s^2_{\Sigma_t} = g_{\alpha\beta} dx^\alpha dx^\beta 
= g_{\alpha\beta} \Big ( \frac{\partial x^\alpha}{\partial y^a} dy^a \Big ) \Big ( \frac{\partial x^\beta}{\partial y^b} dy^b \Big ) 
= h_{ab} dy^a dy^b \, ,  
\end{equation}
so the induced 3-metric is
\begin{equation}
  h_{ab} = g_{\alpha\beta} e^\alpha_a e^\beta_b \, ,
\end{equation}
where $ h_{ab} $ is the metric component of the spacelike hypersurface $ \Sigma_t $. Therefore, contravariant form of the spacetime metric can be decomposed as
\begin{equation}
 g^{\alpha\beta} = -n^\alpha n^\beta + h^{ab} e^\alpha_a e^\beta_b \, ,
\end{equation}
where $ n_\alpha $ is \emph{timelike normal unit vector} to $ \Sigma_t $. Further inside, the spacelike hypersurface can be decomposed into its spacelike unit normal vector plus its boundary 2-surface $ S_t $: suppose that the induced coordinates $ y^a $  on $ \Sigma_t $ is parametrized curves where the parametrization is the coordinates $ y^a (\theta^A) $ of the  $ S_t $. Then, the corresponding projection operator can be taken as
\begin{equation}
 e^a_A = \frac{\partial y^a}{\partial \theta^A} \, .
\end{equation}
The extended line element of $ S_t $ 
\begin{equation}
 ds^2_{S_t} = h_{ab} dx^a dx^b = h{ab}\Big ( \frac{\partial x^a}{\partial \theta^A} d\theta^A \Big )
\Big ( \frac{\partial x^b}{\partial \theta^B} d\theta^B \Big ) = \sigma_{AB} d\theta^A d\theta^B \, ,
\end{equation}
so the induced 2-metric is
\begin{equation}
 \sigma_{AB} = h_{ab} e^a_A e^b_B \, .
\end{equation}
And the decomposition of the spacelike induced 3-metric  is
\begin{equation}
 h^{ab} = r^a r^b + \sigma^{AB} e^a_A e^b_B \, .
\end{equation}
Therefore, the contravariant form of the spacetime metric can be written in terms of the timelike and spacelike unit normal vectors as
\begin{equation}
\begin{aligned}
g^{\alpha\beta} &= -n^\alpha n^\beta + h^{ab} e^\alpha_a e^\beta_b \\
&= -n^\alpha n\beta + \Big [ r^a r^b + \sigma^{AB} e^a_A e^b_B \Big ] e^\alpha e^\beta_b \\
&= -n^\alpha n\beta + r^\alpha r^\beta + \sigma^{AB} \Big ( e^\alpha_a e^a_A \Big ) \Big(e^\beta_b e^b_B \Big ) \\
&= -n^\alpha n^\beta + r^\alpha r^\beta + \sigma^{AB} e^\alpha_A e^\beta_B \, \, .
\end{aligned}
\end{equation}
Because $ e^\alpha_a $ is tangent to $ \Sigma_t $ and $ n^\alpha $ is normal to $ \Sigma_t $, we have
\begin{equation}
 r_\alpha = r_a e^a_\alpha \Longrightarrow r_\alpha n^\alpha = r_a e^a_\alpha n^\alpha =0 
\Longrightarrow r_\alpha \perp n^\alpha \, ,
\end{equation}
where $ n^\alpha $  is the unit timelike vector which is normal to $ \Sigma_t $ and $ r^\alpha  $ is the unit spacelike vector normal to the boundary of $ \Sigma_t $ (that is, $ S_t $ ).

\item Foliation of the Timelike Hypersurface $ \mathcal{B} $

Let $ z^i $ be the coordinates on the timelike hypersurface $ \mathcal{B} $  \cite{Poisson:2004ue} and the corresponding projection operator to $ \Sigma_t $
\begin{equation}
 e^\alpha_i = \frac{\partial x^\alpha}{\partial z^i} \, .
\end{equation}
The induced 3-metric of $ \mathcal{B} $ is obtained from the projection of the spacetime metric \textbf{g}:
\begin{equation}
 \gamma_{ij} = g_{\alpha\beta} e^\alpha_i e^\beta_j \, .
\end{equation}
And the contravariant form of the spacetime metric can be decomposed as
\begin{equation}
 g^{\alpha\beta}= r^\alpha r^\beta + \gamma^{ij} e^\alpha_i e^\beta_j \, .
\end{equation}
Because $ z^i $ is arbitrary, let us choose them as $ z^i = ( t,\theta^A ) $ 
\begin{equation}
 dx^\alpha = \Big ( \frac{\partial x^\alpha}{\partial t} \Big )_{\theta^A} dt + 
\Big ( \frac{\partial x^\alpha}{\partial \theta^A} \Big )_t d\theta^A
= Nn^\alpha dt + e^\alpha_A d \theta^A \, .
\end{equation}
Here $ \theta^A \,(A=1,2) $ are the adapted coordinates of the boundary of the spacelike hypersurface $ \Sigma_t $. Then, the extended line element of $ \mathcal{B} $ is
\begin{equation}
\begin{aligned}
 d s^2_\mathcal{B} &= g_{\alpha\beta} dx^\alpha dx^\beta \\
&= g_{\alpha\beta} \Big ( Nn^\alpha dt + e^\alpha_A d \theta^A \Big ) \Big ( Nn^\beta dt + e^\beta_B d \theta^B )\\
&= g_{\alpha\beta} \Big \{ N^2 n^\alpha n^\beta dt^2 + N n^\alpha e^\beta_B dt + d\theta^B + Nn^\beta e^\alpha_A dt
+ d\theta^A + e^\alpha_A e^\beta_B d\theta^A d\theta^B \Big \} \\
&=-N^2 dt^2 + \sigma_{AB} d\theta^A d\theta^B \\
&=\gamma_{ij} dz^i dz^j \, .
\end{aligned}
\end{equation}
Thus, we get the 2+1 decomposition of the 3-metric of the timelike hypersurface $ \mathcal{B} $ as
\begin{equation}
 \gamma_{ij} dz^i dz^j = -N^2 dt^2 + \sigma_{AB} d\theta^A d\theta^B \, .
\end{equation}
\end{enumerate}

\subsection{Coordinate Adapted to the Foliation}

Here, it is assumed that there is a coordinate system $( x^i=x^1,x^2,x^3) $ ) on each $ \Sigma_t $ such that it smoothly varies during the flow of the hypersurfaces. Then, we take $ [ x^\alpha = (t,x^i)] $ as a proper coordinate system on $ \mathcal{M} $. 

\begin{figure}[h]
\centering
\includegraphics[width=0.8 \textwidth]{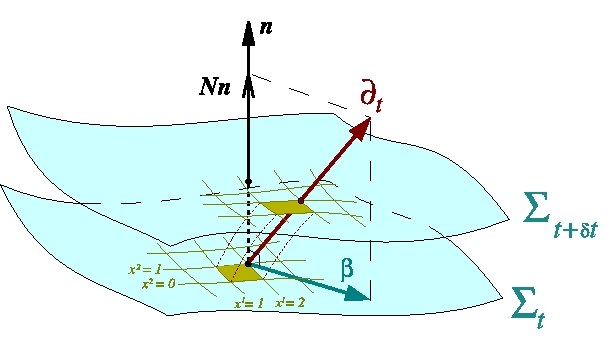}
\caption{The Shift Vector $ \beta $}
\label{evo_distor}
\end{figure}

Naturally, these adapted coordinate systems induce another set of coordinate systems for the $ \mathcal{T}_p(\mathcal{M}) $. Basically, the corresponding partial derivatives are often chosen as the bases of $ \mathcal{T}_p(\mathcal{M}) $ :
\begin{equation}
 \partial_t=\frac{\partial}{\partial t} \,,\,\partial_i = \frac{\partial}{\partial x^i} \, ,
\label{ff}
\end{equation}
where (i=1,2,3). Moreover, because of the shift vector that we are going to define, $ \partial_t $ \emph{does not have to be a timelike vector}.

\subsubsection{The Shift Vector $ \beta $}

In addition to the $ \mathcal{T}_p(\mathcal{M}) $, the coordinates adapted to the foliation also induces a set of basis gradient 1-form $ \textbf{d} x^\alpha $ for $ \mathcal{T}^*_p(\mathcal{M}) $ such that it obeys
\begin{equation}
 < d x^\alpha,\partial_\beta>=\delta^\alpha{_\beta} \, .
\end{equation}
Because of $ <dt,\partial_t>=1 $\,,\,$ \partial_t $ \emph{drags the hypersurfaces as \textbf{m} does, too}. However, in general, they do not have to be collinear. They are collinear only if $ (x^i) $ of $ \Sigma_t $ are orthogonal to each other. Otherwise, as in the figure (\ref{evo_distor}), there is assumed to be a deviation vector (\emph{\textbf{shift vector}} $ \beta $) \cite{Wheeler:1964og} between them
\begin{equation}
 \overset{\rightarrow}{\partial}_t = \textbf{m}+\overset{\rightarrow}{\beta} \, .
\label{shft}
\end{equation}
And it is easy to show that \emph{the shift vector $ \beta $ is an element of the hypersurface} $ \Sigma_t $. As it is said before $ \partial_t $ does not have to be timelike. This property is determined by the square of the $ \beta $: From the equation (\ref{shft}), the norm of $ \overset{\rightarrow}{\partial}_t $ is
\begin{equation}
 \overset{\rightarrow}{\partial}_t.\overset{\rightarrow}{\partial}_t=-N^2 + \overset{\rightarrow}{\beta}.\overset{\rightarrow}{\beta} \, ,
\end{equation}
so the if $  \overset{\rightarrow}{\beta}.\overset{\rightarrow}{\beta}<N^2 $ then $ \partial_t $ is a timelike vector, or if $  \overset{\rightarrow}{\beta}.\overset{\rightarrow}{\beta}>N^2 $ then $ \partial_t $ is a spacelike vector and finally, if  $  \overset{\rightarrow}{\beta}.\overset{\rightarrow}{\beta}=N^2 $ then $ \partial_t $ is a null vector. 

\subsubsection{3+1 Form of the Metric}

The 3+1 decomposition of the spacetime metric and the 2+1 decomposition of the induced 3-metric of the timelike hypersurface $ \mathcal{B} $ play a crucial role during the construction of the Hamiltonian form of the general relativity. Therefore, we need to find the 3+1 form of the spacetime metric \textbf{g}, too: After defining the suitable coordinate systems adapted to $ \mathcal{M} $, let us expand the induced 3-metric $ \gamma $ relative to these coordinates $ (x^i ) $ of $ \Sigma_t $ 
\begin{equation}
 \gamma = \gamma_{ij} dx^i \otimes dx^j \, . 
\end{equation}
Because the shift vector $ \beta $ is tangent to $ \Sigma_t $, we can raise its indices with the help of the components of the induced 3-metric $ \gamma_{ij} $ in this coordinate system,
\begin{equation}
 \beta_i = \gamma_{ij} \beta^j \, .
\end{equation}
Also, the expansion of the spacetime metric \textbf{g} in the corresponding coordinates is
\begin{equation*}
 \textbf{g} = g_{\alpha\beta}\, dx^\alpha \otimes dx^\beta \, .
\end{equation*}
Basically, \textbf{g} can be assumed as a machine which has two-slots for the corresponding vectors. Therefore, the components of it are obtained by
\begin{equation}
 g_{\alpha\beta} = \textbf{g}( \overrightarrow{\partial}_\alpha,\overrightarrow{\partial}_\beta ) \, .
\label{metric}
\end{equation}
Therefore, with the help of the equations (\ref{shft}) and (\ref{metric}), the components $ g_{\alpha\beta} $ in this coordinate systems are 
\begin{equation}
 g_{00} = \textbf{g} (\overrightarrow{\partial_t},\overrightarrow{\partial_t}) = \overrightarrow{\partial}_t . \overrightarrow{\partial}_t
= -N^2 + \beta_i \beta^i \, ,
\end{equation}
\begin{equation}
\begin{aligned}
\quad g_{0i} &= \textbf{g}(\overrightarrow{ \partial_t} ,\overrightarrow{ \partial_i} ) = \overrightarrow{\partial}_t . \overrightarrow{\partial}_i 
= ( \textbf{m} + \overrightarrow{\beta} ). \overrightarrow{\partial}_i \\
&= \textbf{m} . \overrightarrow{\partial}_i + \overrightarrow{\beta} . \overrightarrow{\partial}_i 
= \overrightarrow{\beta} . \overrightarrow{\partial}_i = < \tilde{\beta},\overrightarrow{\partial}_i >  \\
&= <\beta_j dx^j , \overrightarrow{\partial}_i > = \beta_j < dx^j , \overrightarrow{\partial}_i > \\
& = \beta_j \delta^j {_i} \\
& = \beta_i \, ,
\end{aligned}
\end{equation}
\begin{equation}
 g_{ij} = \textbf{g} ( \overrightarrow{\partial}_i,\overrightarrow{\partial}_j ) = \overrightarrow{\partial}_i . \overrightarrow{\partial}_j 
= \gamma_{ij} \, .\qquad \quad 
\end{equation}
Thus, the 3+1 decomposition of the $ g_{\alpha\beta} $ in matrix representation is
\[
 g_{\alpha\beta} = 
\left ( {\begin{array}{cc}
g_{00} & g_{0j} \\
g_{i0} & g_{ij} \\ 
\end{array} } \right) =
\left ( {\begin{array}{cc}
-N^2 + \beta_i \beta^i & \beta_j \\
\beta_i & \gamma_{ij} \\
\end{array} } \right ) \, .
\]
Furthermore, let us evaluate the explicit 3+1 form of spacetime metric:
\begin{equation*}
\begin{aligned}
 g_{\mu\nu} dx^\mu \otimes dx^\nu &= \Big ( -N^2 + \beta_i \beta^i \Big ) dt \otimes dt 
+ \beta_j dt \otimes dx^j + \beta_i dt \otimes dx^i + \gamma_{ij} dx^i \otimes dx^j \\
 &= \Big ( -N^2 + \beta_i \beta^i \Big ) dt \otimes dt + \gamma_{ij} \beta^i dt \otimes dx^j + \gamma_{ij} \beta^i dt \otimes dx^i \\
 & \qquad + \gamma_{ij} dx^i \otimes dx^j \, ,
\end{aligned}
\end{equation*}
 we get
\begin{equation}
 g_{\mu\nu} dx^\mu \otimes dx^\nu = -N^2 dt \otimes dt + \gamma_{ij} \Big [ dx^i + \beta^i dt \Big ] \otimes \Big [dx^j + \beta^j dt \Big ] \, .
\label{dfg}
\end{equation}
The ability of writing \textbf{g} in terms of its parts(\ref{dfg}) is something like the beginning of the 3+1 formalism and the Hamiltonian form of general relativity. Now, let us find the matrix form of the components of the dual spacetime metric $ g^{\alpha\mu} $. Suppose that the dual metric has the form of
\[
 g^{\alpha\mu} =
\left ( {\begin{array}{cc}
g^{00}& g^{0j} \\
g^{i0} & g^{ij} \\
\end{array} } \right ) =
\left ( {\begin{array}{cc}
a & v^k \\
v^j & b^{jk} \\          
\end{array} }\right ) \, .
\]
The matrix multiplication (In general, $ A_{ij} B_{jk} = C_{ik} $) between the covariant and the contravariant forms of the spacetime metric is given by
\[
 \left ( {\begin{array}{cc}
-N^2 + \beta_k \beta^k & \beta_j \\
\beta_i & \gamma_{ij} \\
\end{array} } \right )
\left ( {\begin{array}{cc}
 a & v^k \\
v^j & b^{jk}
\end{array} } \right ) =
\left ( {\begin{array}{cc}
1 & 0 \\
0 & \delta_{ik} \\
\end{array} } \right ) \, .
\] 
The multiplication of the $ 1^{st} $ row of $ g_{\alpha\mu} $ with the $ 1^{st} $ column of  $ g^{\alpha\mu} $  gives
\begin{equation}
 a\beta_i + \gamma_{ij} v^j = 0 \Longrightarrow a \beta_i = - v_i \, ,
\label{o}
\end{equation}
\begin{equation}
(-N^2 + \beta_j \beta^j)a + b_j v^j = 1 \Longrightarrow (-N^2 + \beta_j \beta^j)a -a \beta_j \beta^j= 1 \, .
\label{oo}
\end{equation} 
From the equations (\ref{o}) and (\ref{o}), we get $ a = - \frac{1}{N^2} $ and $  v^j = \frac{\beta^j}{N^2} $. The multiplication between $ 2^{nd} $ row of $ g_{\alpha\mu} $ with the $ 2^{nd} $ column of  $ g^{\alpha\mu} $ will give us the components $ b^{jk} $:
\begin{equation}
 \beta_i v^k + \gamma_{ij} b^{jk} = \delta_{ik} \Longrightarrow \gamma_{ij} b^{jk} = \delta_{ik} - \beta_i v^k
 = \delta_{ik} - \frac{\beta_i \beta^k}{N^2} \, .
\label{ooo}
\end{equation}
Let us multiply (\ref{ooo}) with $ \gamma^{li} $,
\begin{equation*}
\begin{aligned}
\gamma^{li} \gamma_{ij} b^{jk} &= \gamma^{li}\delta_{ik} - \gamma^{li} \frac{\beta_i \beta^k}{N^2} \\
b^{lk} &= \gamma^{lk} - \frac{\beta^l\beta^k}{N^2} \, , \\
 l \rightarrow &i , k \rightarrow j \, \, .
\end{aligned}
\end{equation*}
With the help of the previous change of indices, we get  
\begin{equation}
 b^{ij} = \gamma^{ij}-\frac{\beta^i \beta^j}{N^2} \, .
\end{equation}
Thus, \emph{the decomposition of the components of the dual metric in matrix form} is
\[
 g^{\alpha\beta}=
\left ( {\begin{array}{cc}
g^{00} & g^{0j} \\
g^{i0} & g^{ij} \\
\end{array} } \right )=
\left ( {\begin{array}{cc}
-\frac{1}{N^2} & \frac{\beta^j}{N^2} \\
\frac{\beta^i}{N^2} & \gamma^{ij} - \frac{\beta^i\beta^j}{N^2}
 \end{array} } \right ) \, .
\]
Notice that $ g_{ij} = \gamma_{ij} $ but that in general $ g^{ij} \ne \gamma^{ij} $. Alternatively, the dual spacetime metric $ g^{\alpha\beta} $ in matrix form can be obtained from the Cramer's rule. Now, let us denote the determinant of $ \gamma $ and \textbf{g} as
\begin{equation}
 g=det(g_{\alpha\beta}) \, ,
\end{equation}
\begin{equation}
 \gamma=det(\gamma_{ij}) \, .
\end{equation}
Observe that because of the lapse function N and the shift vector $ \beta $, the related determinants of $ \gamma_{ij} $ and $ g_{\alpha\beta} $ are coordinate dependent. Because of this, they are not tensors rather they are tensor densities. Now, let us evaluate $ g^{00} $ component by using the Cramer's rule, 
\begin{equation}
 g^{00} = \frac{C_{00}} {det(g_{\alpha\beta})} = \frac{M_{00}}{g} = \frac{\gamma}{g} \, .
\label{oooo}
\end{equation}
The (\ref{oooo}) must be equal to the $ g^{00} $ component of the dual spacetime metric that we have obtained from the matrix multiplication,
\begin{equation}
 g^{00} = \frac{\gamma}{g} \Longrightarrow -\frac{1}{N^2} = \frac{\gamma}{g} \, .
\end{equation}
Thus, \emph{the relation between the determinant of the induced 3-metric and of the spacetime metric} is
\begin{equation}
 \sqrt{-g} = N \sqrt{\gamma} \, .
\label{deco_metri}
\end{equation}

\section{THE GAUSS-CODAZZI RELATIONS AND THE 3+1 DECOMPOSITION OF SPACETIME RICCI SCALAR}
\label{sec: The Gauss-Codazzi Relations And The Decomposition of Spacetime Ricci Scalar}

In this chapter we will deduce the fundamental relations that are in the center of the 3+1 formalism of general relativity \cite{Gourgoulhon:2007ue}. From now, unless it is emphasized, the hypersurface that we will work on must be taken as the spacelike hypersurface $ \Sigma_t $ (i.e. whose unit normal vector $ \hat{\textbf{n}} $ is timelike ):

\subsection{Gauss and Codazzi Relations}

\subsubsection{Gauss Relations}

In order to find the first fundamental equation of the 3+1 formalism (i.e. Gauss relation), we start with equation (\ref{intr_ricc_ident}) (or the intrinsic Ricci identity) and use the related transformation (\ref{trans}) between connections of \textbf{D} and $ \nabla $: The Ricci identity on $ \Sigma $ is 
\begin{equation}
D_\alpha D_\beta v^\gamma-D_\beta D_\alpha v^\gamma=R^\gamma {_{\mu\alpha\beta}} v^\mu \, ,
\label{g}
\end{equation}
here $ \mathbf{v} \in \mathcal{T}_p(\mathcal{M}) $. Since the second term in the equation (\ref{g}) is obtained by interchanging the indices of $ \alpha $ and $ \beta $ of the first term, it is better to work just on the $ 1^{st} $ term of the equation (\ref{g}): With the help of the equation (\ref{orth_proj_oper}), the $ 1^{st} $ term of the (\ref{g}) is obtained as follow:
\begin{equation*}
\begin{aligned}
 D_\alpha D_\beta v^\gamma &= D_\alpha ( D_\beta v^\gamma )
= \gamma^\mu {_\alpha} \gamma^\nu {_\beta} \gamma^\gamma {_\rho} \nabla_\mu [D_\nu v^\rho ] 
= \gamma^\mu {_\alpha} \gamma^\nu {_\beta} \gamma^\gamma {_\rho} \nabla_\mu[\gamma^\sigma{_
\nu}\gamma^\rho{_\lambda}\nabla_\sigma v^\lambda ] \\
 &= \gamma^\mu {_\alpha} \gamma^\nu {_\beta} \gamma^\gamma {_\rho}\{(\nabla_\mu \gamma^\sigma
 {_\nu} )\gamma^\rho{_\lambda}\nabla_\sigma v^\lambda
 + \gamma^\sigma {_\nu} (\nabla_\mu \gamma^\rho {_\lambda}) \nabla_\sigma v^\lambda \\
 & \qquad \qquad \qquad + \gamma^\sigma {_\nu} \gamma^\rho {_\lambda} \nabla_\mu \nabla_\sigma v^\lambda \} \\
 &= \gamma^\mu {_\alpha} \gamma^\nu {_\beta} \gamma^\gamma {_\rho} \{ \nabla_\mu [\delta^\sigma
 {_\nu} + n^\sigma n_\nu ] \gamma^\rho {_\lambda} \nabla_\sigma v^\lambda
 +  \gamma^\sigma {_\nu} \nabla_\mu [\delta^\rho {_\lambda} + n^\rho n_\lambda ] \nabla_\sigma
 v^\lambda \\
 & \qquad \qquad \qquad + \gamma^\sigma {_\nu} \gamma^\rho {_\lambda} \nabla_\mu \nabla_\sigma
 v^\lambda \} \\
 &= \gamma^\mu {_\alpha} \gamma^\nu {_\beta} \gamma^\gamma {_\rho}\{ [\nabla_\mu \delta^\sigma {_\nu} 
 + (\nabla_\mu n^\sigma ) n_\nu + n^\sigma (\nabla_\mu n_\nu )] \gamma^\rho {_\lambda} \nabla_\sigma
 v^\lambda \\
& \qquad \qquad \qquad + \gamma^\sigma {_\nu} [\nabla_\mu \delta^\rho {_\lambda} 
  + (\nabla_\mu n^\rho )n_\lambda + n^\rho (\nabla_\mu n_\lambda ) ] 
\nabla_\sigma v^\lambda \\ 
 &\qquad \qquad \qquad + \gamma^\sigma {_\nu} \gamma^\rho {_\lambda} \nabla_\mu \nabla_\sigma v^\lambda \} \\
&=  \gamma^\mu {_\alpha} \gamma^\nu {_\beta} \gamma^\gamma {_\rho}
 \{ (\nabla_\mu n^\sigma ) n_\nu \gamma^\rho {_\lambda} \nabla_\sigma v^\lambda
+ n^\sigma ( \nabla_\mu n_\nu ) \gamma^\rho {_\lambda} \nabla_\sigma v^\lambda \\
& \qquad \qquad \qquad + \gamma^\sigma
 {_\nu} ( \nabla_\mu n^\rho ) n_\lambda \nabla_\sigma v^\lambda 
 + \gamma^\sigma {_\nu} n^\rho (\nabla_\mu n_\lambda ) \nabla_\sigma v^\lambda \\
&\qquad \qquad \qquad  +  \gamma^\sigma {_\nu} \gamma^\rho {_\lambda} \nabla_\mu \nabla_\sigma v^\lambda \} \\
&= \gamma^\mu {_\alpha} \gamma^\nu {_\beta} \gamma^\gamma {_\rho} (\nabla_\mu n^\sigma ) n_\nu
 \gamma^\rho {_\lambda} \nabla_\sigma v^\lambda
+ \gamma^\mu {_\alpha} \gamma^\nu {_\beta} \gamma^\gamma {_\rho} n^\sigma ( \nabla_\mu n_\nu ) 
\gamma^\rho {_\lambda} \nabla_\sigma v^\lambda \\
& + \gamma^\mu {_\alpha} \gamma^\nu {_\beta} \gamma^\gamma {_\rho} \gamma^\sigma
 {_\nu} ( \nabla_\mu n^\rho ) n_\lambda \nabla_\sigma v^\lambda
+ \gamma^\mu {_\alpha} \gamma^\nu {_\beta} \gamma^\gamma {_\rho}  \gamma^\sigma {_\nu} n^\rho (\nabla_\mu
 n_\lambda ) \nabla_\sigma v^\lambda \\
& + \gamma^\mu {_\alpha} \gamma^\nu {_\beta} \gamma^\gamma {_\rho}
\gamma^\sigma {_\nu} \gamma^\rho {_\lambda} \nabla_\mu \nabla_\sigma v^\lambda \, .
\end{aligned}
\end{equation*}
Because the projection of the unit normal vector (and its dual) on the hypersurface is zero, then, the previous equation becomes 
\begin{equation*}
\begin{aligned}
D_\alpha D_\beta v^\gamma &= \gamma^\mu {_\alpha} \gamma^\nu {_\beta} \gamma^\gamma {_\rho} \gamma^\rho 
{_\lambda} n^\sigma ( \nabla_\mu n_\nu ) \nabla_\sigma v^\lambda
+ \gamma^\mu {_\alpha} \gamma^\nu {_\beta} \gamma^\gamma {_\rho} \gamma^\sigma
 {_\nu} ( \nabla_\mu n^\rho ) n_\lambda \nabla_\sigma v^\lambda \\
& \quad + \gamma^\mu {_\alpha} \gamma^\nu {_\beta} \gamma^\gamma {_\rho}
\gamma^\sigma {_\nu} \gamma^\rho {_\lambda} \nabla_\mu \nabla_\sigma v^\lambda \\
&= \gamma^\mu {_\alpha} \gamma^\nu {_\beta} \gamma^\gamma {_\lambda} 
n^\sigma ( \nabla_\mu n_\nu ) \nabla_\sigma v^\lambda
+ \gamma^\mu {_\alpha} \gamma^\sigma {_\beta}\gamma^\gamma {_\rho} ( \nabla_\mu n^\rho ) n_\lambda
 \nabla_\sigma v^\lambda \\
& \quad + \gamma^\mu {_\alpha} \gamma^\sigma {_\beta} \gamma^\gamma {_\lambda}
 \nabla_\mu \nabla_\sigma v^\lambda \\
&=-K_{\alpha\beta} \gamma^\gamma {_\lambda} n^\sigma \nabla_\sigma v^\lambda 
-K^\gamma {_\alpha} \gamma^\sigma {_\beta} n_\lambda \nabla_\sigma v^\lambda +
\gamma^\mu {_\alpha} \gamma^\sigma {_\beta} \gamma^\gamma {_\lambda} \nabla_\mu \nabla_\sigma v^\lambda\\
&=-K_{\alpha\beta} \gamma^\gamma {_\lambda} n^\sigma \nabla_\sigma v^\lambda
+ K^\gamma {_\alpha} \gamma^\sigma {_\beta} v^\lambda \nabla_\sigma n_\lambda +
\gamma^\mu {_\alpha} \gamma^\sigma {_\beta} \gamma^\gamma {_\lambda} \nabla_\mu \nabla_\sigma v^\lambda \\
&=-K_{\alpha\beta} \gamma^\gamma {_\lambda} n^\sigma \nabla_\sigma v^\lambda
+ K^\gamma {_\alpha} \gamma^\sigma {_\beta} \gamma^\lambda {_\xi} v^\xi \nabla_\sigma n_\lambda
+ \gamma^\mu {_\alpha} \gamma^\sigma {_\beta} \gamma^\gamma {_\lambda} \nabla_\mu \nabla_\sigma v^\lambda \, ,
\end{aligned}
\end{equation*}
where we used the equation (\ref{kkkk1}). Let us interchange the dummy index of $ \xi $ with $ \lambda $,
\begin{equation}
D_\alpha D_\beta v^\gamma = -K_{\alpha\beta} \gamma^\gamma {_\lambda} n^\sigma \nabla_\sigma v^\lambda
-K^\gamma {_\alpha} K_{\beta\lambda} v^\lambda
+ \gamma^\mu {_\alpha} \gamma^\sigma {_\beta} \gamma^\gamma {_\lambda} \nabla_\mu \nabla_\sigma v^\lambda \, .
\label{first}
\end{equation}
In order to find the $ 2^{nd} $ of the equation (\ref{g}), it is enough to just interchange the indices $ \alpha $ and $ \beta $ of the equation (\ref{first}):
\begin{equation}
D_\beta D_\alpha v^\gamma = -K_{\beta\alpha} \gamma^\gamma {_\lambda} n^\sigma \nabla_\sigma 
v^\lambda-K^\gamma {_\beta} K_{\alpha\lambda} v^\lambda + 
\gamma^\mu {_\beta} \gamma^\sigma {_\alpha} \gamma^\gamma {_\lambda} \nabla_\mu \nabla_\sigma v^\lambda \, .
\label{2}
\end{equation}
If we interchange the indices $ \mu $ and $ \sigma $ of (\ref{2}) and then subtract it from the equation ( \ref{first}), we will get
\begin{equation*}
 R^\gamma {_{\xi\alpha\beta}} v^\xi = [K^\gamma {_\beta} K_{\alpha\lambda} -
K^\gamma {_\alpha} K_{\beta\lambda} ]v^\lambda + \gamma^\mu {_\alpha} \gamma^\sigma {_\beta}
\gamma^\gamma {_\lambda} [\nabla_\mu \nabla_\sigma -\nabla_\sigma \nabla_\mu ] v^\lambda \, .
\end{equation*}
The last term is nothing but the projection of spacetime curvature tensor. Then, we have
\begin{equation}
 R^\gamma {_{\xi\alpha\beta}} v^\xi = [K^\gamma {_\beta} K_{\alpha\lambda} -
K^\gamma {_\alpha} K_{\beta\lambda} ]v^\lambda + \gamma^\mu {_\alpha} \gamma^\sigma
 {_\beta}
\gamma^\gamma {_\lambda}\,{^4}R^\lambda {_{\rho\mu\sigma}} v^\rho \, .
\label{3}
\end{equation}
We can rewrite $ v^\rho $ as $ v^\rho = \gamma^\rho {_\xi} v^\xi $. Then, the equation (\ref{3}) becomes 
\begin{equation*}
\begin{aligned}
\gamma^\mu {_\alpha} \gamma^\sigma {_\beta}
\gamma^\gamma {_\lambda} \gamma^\rho {_\xi}\,{^4}R^\lambda {_{\rho\mu\sigma}} v^\xi
 = R^\gamma {_{\xi\alpha\beta}} v^\xi + [ K^\gamma {_\alpha} K_{\beta\lambda} -K^\gamma {_\beta}
 K_{\alpha\lambda} ]v^\lambda \, , \\  
 \lambda \rightarrow\rho,\, \rho\rightarrow\sigma, \, \sigma\rightarrow\nu,\,\xi\rightarrow\delta
  \qquad \xi\rightarrow\delta \qquad \quad \quad \lambda \rightarrow \delta \qquad \quad \quad 
\end{aligned}
\end{equation*}
by changing of the dummy indices, the previous equation turns into
\begin{equation}
\gamma^\mu {_\alpha} \gamma^\nu {_\beta}
\gamma^\gamma {_\rho} \gamma^\sigma {_\delta}\,{^4}R^\rho {_{\sigma\mu\nu}} v^\delta 
= R^\gamma {_{\delta\alpha\beta}} v^\delta + [ K^\gamma {_\alpha} K_{\beta\delta} -K^\gamma {_\beta}
 K_{\alpha\delta} ]v^\delta \, ,
\end{equation}
since the vector $ v^\delta $ is arbitrary, then, we can drop it to get \emph{the full projection of the spacetime Riemann curvature tensor $ {^4}{\textbf{R}} $ onto the hypersurface $ \Sigma_t $} as
\begin{equation}
\gamma^\mu {_\alpha} \gamma^\nu {_\beta}
\gamma^\gamma {_\rho} \gamma^\sigma {_\delta}\,{^4}R^\rho {_{\sigma\mu\nu}} 
= R^\gamma {_{\delta\alpha\beta}} + K^\gamma {_\alpha} K_{\beta\delta} -K^\gamma {_\beta}
 K_{\alpha\delta} \, .
\label{gauss}
\end{equation}
The equation (\ref{gauss}) is known as the \emph{\textbf{Gauss relation}}. Let us continue by contracting the indices $ \alpha $ and $ \gamma $ of Gauss relation (\ref{gauss}):
\begin{equation}
 \gamma^\mu {_\rho} \gamma^\nu {_\beta} \gamma^\sigma {_\delta}\,{^4}R^\rho{ _{\sigma\mu\nu}} =
R_{\delta\beta} +K K_{\beta\delta} - K_{\alpha\delta} K^\alpha {_\beta} \, .
\label{4}
\end{equation}
The left hand side of the equation (\ref{4}) can be rewritten in terms of the sum of the full projection of the spacetime Ricci tensor onto $ \Sigma $ and the mixed projection of the spacetime Riemann curvature tensor $ {^4}{\textbf{R}} $:
\begin{equation*}
\begin{aligned}
 \gamma^\mu {_\rho} \gamma^\nu {_\beta} \gamma^\sigma {_\delta}\,{^4}R^\rho{ _{\sigma\mu\nu}} 
& = [ \delta^\mu {_\rho} + n^\mu n_\rho ]  \gamma^\nu {_\beta} \gamma^\sigma {_\delta}\,{^4}R^\rho{ _{\sigma\mu\nu}} \\
&= \gamma^\nu {_\beta} \gamma^\sigma {_\delta}\,{^4}R _{\sigma\nu} +
n^\mu n_\rho \gamma^\nu {_\beta} \gamma^\sigma {_\delta}\,{^4}R^\rho{ _{\sigma\mu\nu}} \\
& = \gamma^\nu {_\beta} \gamma^\sigma {_\delta}\,{^4}R _{\sigma\nu} +
n^\mu n^\rho \gamma^\nu {_\beta} \gamma^\sigma {_\delta}\,{^4}R_{\rho\sigma\mu\nu} \\
& = \gamma^\nu {_\beta} \gamma^\sigma {_\delta}\,{^4}R _{\sigma\nu} + 
n^\mu n^\rho \gamma^\nu {_\beta} \gamma{_{\sigma\delta}}\,{^4}R ^\sigma{_{\rho\nu\mu}} \\
& \qquad \qquad \qquad \qquad \sigma \leftrightarrow \mu,\rho \leftrightarrow \nu 
\end{aligned}
\end{equation*}
\begin{equation}
\qquad \qquad \qquad \qquad \qquad= \gamma^\nu {_\beta} \gamma^\sigma {_\delta}\,{^4}R _{\sigma\nu} +
\gamma_{\mu\delta} \gamma^\rho {_\beta} 
n^\sigma n^\nu \,{^4}R ^\mu { _{\nu\rho\sigma}}  \, .\qquad
\label{5}
\end{equation}
The substitution of the equation (\ref{5}) into the equation (\ref{4}) gives 
\begin{equation*}
\begin{aligned}
& \gamma^\nu {_\beta} \gamma^\sigma {_\delta}\,{^4}R _{\sigma\nu} +\gamma_{\mu\delta} \gamma^\rho {_\beta} 
n^\sigma n^\nu\, {^4}R ^\mu { _{\nu\rho\sigma}} = R_{\delta\beta} +K K_{\beta\delta} - K_{\alpha\delta} K^\alpha {_\beta} \, .\\
& \quad \sigma \rightarrow \mu \qquad \qquad \delta \rightarrow \alpha \quad \quad \qquad \qquad \qquad \qquad \alpha \rightarrow \mu 
\end{aligned}
\end{equation*}
By applying the given change of the dummy indices in the previous equation, we get the well-known relation of \emph{\textbf{contracted Gauss relation}} as:
\begin{equation}
\gamma^\mu {_\alpha} \gamma^\nu {_\beta}\, {^4}R_{\mu\nu} + \gamma_{\alpha\mu} \gamma^\rho {_\beta} n^\sigma n^\nu 
\,{^4}R^\mu {_{\nu\rho\sigma}} =R_{\alpha\beta} + KK_{\alpha\beta} - K_{\alpha\mu}K^\mu {_\beta} \, .
\label{6}
\end{equation}
Now, let us multiply the equation (\ref{6}) by $ \gamma^{\alpha\beta} $ (i.e. taking trace with the dual induced 3-metric $ \gamma^{\alpha\beta} $)  
\begin{equation*}
\begin{aligned} 
\gamma^{\alpha\beta} \gamma^\mu {_\alpha} \gamma^\nu {_\beta} \, {^4}R_{\mu\nu} + \gamma^{\alpha\beta} \gamma_{\alpha\mu} \gamma^\rho {_\beta} n^\sigma n^\nu 
\,{^4}R^\mu {_{\nu\rho\sigma}} &=\gamma^{\alpha\beta}R_{\alpha\beta} + \gamma^{\alpha\beta}KK_{\alpha\beta} \\
& \quad - \gamma^{\alpha\beta} K_{\alpha\mu}K^\mu {_\beta} \, ,
\end{aligned}
\end{equation*}
it becomes
\begin{equation}
\gamma^{\mu\nu} \,{^4}R_{\mu\nu} + \gamma^\rho {_\mu} n^\nu n^\sigma \, {^4}R^\mu {_{\nu\rho\sigma}} = R + K^2 - K_{\alpha\mu}K^{\alpha\mu} \, .
\label{7}
\end{equation}
We need to first modify the $ 1^{st} $ and $ 2^{nd} $ terms of the left hand side of the equation (\ref{7}):
\begin{equation}
\gamma^{\mu\nu} \,{^4}R_{\mu\nu} = [g^{\mu\nu} + n^\mu n^\nu]\, {^4}R_{\mu\nu} = {^4}R + n^\mu n^\nu \, {^4}R_{\mu\nu} \, ,
\label{8}
\end{equation}
and
\begin{equation*}
\begin{aligned}
 \gamma^\rho {_\mu} n^\nu n^\sigma \, {^4}R^\mu {_{\nu\rho\sigma}} &= [\delta^\rho {_\mu} + n^\rho n_\mu ]
 n^\nu n^\sigma \, {^4}R^\mu {_{\nu\rho\sigma}} \\ 
&= \delta^\rho {_\mu} n^\nu n^\sigma \, {^4}R^\mu {_{\nu\rho\sigma}} +
n^\rho n_\mu n^\nu n^\sigma \, {^4}R^\mu {_{\nu\rho\sigma}} \, .
\end{aligned}
\end{equation*}
Because the Riemann tensor is antisymmetric in its first two and second two indices, the last term in the previous equation is zero. Therefore, it turns into
\begin{equation}
\gamma^\rho {_\mu} n^\nu n^\sigma \, {^4}R^\mu {_{\nu\rho\sigma}} = n^\nu n^\sigma \,{^4}R_{\nu\sigma} \, .
\label{9}
\end{equation}
By substituting the related results of (\ref{8}) and (\ref{9}) into the main equation of (\ref{7}), we get the another well-known relation of \emph{\textbf{scalar Gauss relation}} as
\begin{equation}
 {^4}R + 2 {^4}R_{\nu\sigma} n^\nu n^\sigma \ = R + K^2 - K_{ij}K^{ij} \, .
\label{dddd}
\end{equation}

\subsubsection{Codazzi Relation}

In order to find the second fundamental relation of 3+1 formalism (i.e. \emph{Codazzi relation}), we start with the Ricci identity of the four-dimensional spacetime for the unit normal vector  $ \hat{\textbf{n}} $. And, then, we will \emph{project it 3-times onto the hypersurface $ \Sigma_t $} : Now, the Ricci identity in four dimension is
\begin{equation}
(\nabla_\alpha \nabla_\beta - \nabla_\beta \nabla_\alpha ) \,n^\gamma ={^4}R^\gamma {_{\sigma\alpha\beta}} n^\sigma \, .
\label{coda}
\end{equation}
Let us start by projecting the equation (\ref{coda}) \emph{three-times onto the hypersurface $ \Sigma_t $}
\begin{equation}
 \gamma^\mu {_\alpha} \gamma^\nu {_\beta} \gamma^\gamma {_\rho}(\nabla_\mu \nabla_\nu - \nabla_\nu \nabla_\mu ) n^\rho 
= \gamma^\mu {_\alpha} \gamma^\nu {_\beta} \gamma^\gamma {_\rho} {^4}R^\rho {_{\sigma\mu\nu}} n^\sigma \, .
\label{codaz}
\end{equation}
Again, as we did in the part of Gauss relation, let us just work on \emph{the $ 1^{st} $ term of the equation (\ref{codaz})} because the $ 2^{nd} $ term is obtained simply by interchanging the indices $ \alpha $ and $ \beta $ of the $ 1^{st} $ term: Therefore,
\begin{equation*}
\begin{aligned}
 \gamma^\mu {_\alpha} \gamma^\nu {_\beta} \gamma^\gamma {_\rho} \nabla_\mu \nabla_\nu n^\rho
 &= \gamma^\mu {_\alpha} \gamma^\nu {_\beta} \gamma^\gamma {_\rho} \nabla_\mu[-K^\rho {_\nu} - a^\rho n_\nu ] \\
&= \gamma^\mu {_\alpha} \gamma^\nu {_\beta} \gamma^\gamma {_\rho} \{-\nabla_\mu K^\rho {_\nu} -( \nabla_\mu a^\rho ) n_\nu
- a^\rho (\nabla_\mu n_\nu )\} \\
& = - \gamma^\mu {_\alpha} \gamma^\nu {_\beta} \gamma^\gamma {_\rho} \nabla_\mu K^\rho {_\nu} 
- \gamma^\mu {_\alpha} \gamma^\nu {_\beta} \gamma^\gamma {_\rho} ( \nabla_\mu a^\rho ) n_\nu \\
 & \quad - \gamma^\mu {_\alpha} \gamma^\nu {_\beta} \gamma^\gamma {_\rho} a^\rho (\nabla_\mu n_\nu ) \, .
\end{aligned}
\end{equation*}
Since the projection of the dual of the unit vector is zero, the $ 2^{nd} $ term on the last part of the previous equation vanishes. Then, it becomes
\begin{equation}
\begin{aligned}
 \gamma^\mu {_\alpha} \gamma^\nu {_\beta} \gamma^\gamma {_\rho} \nabla_\mu \nabla_\nu n^\rho 
&= - \gamma^\mu {_\alpha} \gamma^\nu {_\beta} \gamma^\gamma {_\rho} \nabla_\mu K^\rho {_\nu} - 
\gamma^\mu {_\alpha} \gamma^\nu {_\beta} \gamma^\gamma {_\rho} a^\rho (\nabla_\mu n_\nu ) \\
& =-D_\alpha K^\gamma {_\beta} - \gamma^\mu {_\alpha} \gamma^\nu {_\beta} \gamma^\gamma {_\rho} a^\rho (\nabla_\mu n_\nu ) \, ,
\label{cc}
\end{aligned}
\end{equation}
where we used the general transformation relation (\ref{trans}) between connections. Because of the equation (\ref{kkkk1}), the equation (\ref{cc}) turns into
\begin{equation}
 \gamma^\mu {_\alpha} \gamma^\nu {_\beta} \gamma^\gamma {_\rho} \nabla_\mu \nabla_\nu n^\rho
= - D_\alpha K^\gamma {_\beta} + K_{\alpha\beta} a^\gamma \, .
\label{ccc}
\end{equation}
As we said before the interchange of the indices $ \alpha $ and $ \beta $ of the projected equation (\ref{ccc}) of the $ 1^{st} $ term of (\ref{coda}) gives \emph{the projected version of the $ 2^{nd} $ term of the equation (\ref{coda})},
\begin{equation}
 \gamma^\mu {_\alpha} \gamma^\nu {_\beta} \gamma^\gamma {_\rho} \nabla_\nu \nabla_\mu n^\rho 
= -  D_\beta K^\gamma {_\alpha} + K_{\beta\alpha} a^\gamma \, .
\label{cccc}
\end{equation}
Finally, the subtraction of (\ref{cccc}) from (\ref{ccc}) provides us the famous \emph{ \textbf{Codazzi-Mainardi relation}}:
\begin{equation}
 \gamma^\mu {_\alpha} \gamma^\nu {_\beta} \gamma^\gamma {_\rho} \,{^4}R^\rho {_{\sigma\mu\nu}} n^\sigma
 = D_\beta K^\gamma {_\alpha} - D_\alpha K^\gamma {_\beta} \, .
\label{codazi}
\end{equation}
Now, Let us contract the Codazzi-Mainardi relation (\ref{codazi}) on the indices $ \alpha $ and $ \gamma $
\begin{equation*}
\begin{aligned}
 \gamma^\mu {_\rho} \gamma^\nu {_\beta} \,{^4}R^\rho {_{\sigma\mu\nu}} n^\sigma
 &= D_\beta K - D_\alpha K^\alpha {_\beta}  \, ,\\
  [ \delta^\mu {_\rho} + n^\mu n_\rho ]
\gamma^\nu {_\beta} \,{^4}R^\rho {_{\sigma\mu\nu}} n^\sigma & = D_\beta K - D_\alpha K^\alpha {_\beta} \, ,\\
 \gamma^\nu {_\beta} \delta^\mu {_\rho} \,{^4}R^\rho {_{\sigma\mu\nu}} n^\sigma 
+ n^\mu n_\rho n^\sigma \gamma^\nu {_\beta} \,{^4}R^\rho {_{\sigma\mu\nu}} & = D_\beta K - D_\alpha K^\alpha {_\beta} \, .
\end{aligned}
\end{equation*}
Because the multiplication between \emph{symmetric} and \emph{antisymmetric} tensors are zero, the second term on the left hand side of the previous equation vanishes and we get
\begin{equation}
\begin{aligned}
\gamma^\nu {_\beta} \delta^\mu {_\rho} \,{^4}R^\rho {_{\sigma\mu\nu}} n^\sigma  = D_\beta K - D_\alpha K^\alpha {_\beta}
 \Rightarrow & \gamma^\nu {_\beta} \,{^4}R_{\sigma\nu} n^\sigma  = D_\beta K - D_\alpha K^\alpha {_\beta} \\
& \sigma \rightarrow \nu,\nu\rightarrow \mu \qquad \qquad \alpha\rightarrow\mu \, ,
\label{cccs}
\end{aligned}
\end{equation}
and with the given change of the dummy indices in the equation (\ref{cccs}), we reach 
\begin{equation}
\gamma^\mu {_\beta} n^\nu \,{^4}R_{\mu\nu} = D_\beta K- D_\mu K^\mu {_\beta} \, ,
\label{cont_coda}
\end{equation}
which is known as \emph{\textbf{the contracted Codazzi relation}}.

As we deduced above, the Gauss and Codazzi (or Codazzi-Mainardi) relations \emph{which are obtained from the various numbers of the projection of the spacetime Riemann curvature tensor  $ {^4}{\textbf{R}} $ onto or normal to a \textbf{single hypersurface $ \Sigma_t $}}. That's, \emph{\underline{the full-projection}} of $ {^4}{\textbf{R}} $ onto the hypersurface $ \Sigma_t $ gives \underline{the Gauss relation}. And \emph{\underline{the three-times projections} of $ {^4}{\textbf{R}} $ onto the $ \Sigma_t $ with \underline{the one-times projection} of it along the unit normal vector $ \hat{\textbf{n}} $ gives \underline{the Codazzi-Mainardi relation}}.

\subsection{Last Fundamental Relations of the 3+1 Decomposition}

\subsubsection{The Ricci Equation}

Contrary to the Gauss-Codazzi relations, the contraction of the \emph{Ricci equation that we will deduce in this section results in the third fundamental relation of the 3+1 decomposition of the spacetime Ricci scalar and is essentially based on \textbf{the flow of the hypersurfaces}}. Moreover, we will see that the two-times projection of $ {^4}{\textbf{R}} $ onto $ \Sigma_t $ with the two-times projection of it along $ \hat{\textbf{n}} $ will lead us to the Ricci equation. As we did in the previous section, the starting point is the Ricci identity in four dimension of $ \hat{\textbf{n}} $ but, here, we will project it two-times onto $ \Sigma_t $ and one-times along $ \hat{\textbf{n}} $: The related four-dimensional Ricci identity for $ \hat{\textbf{n}} $ is 
\begin{equation}
(\nabla_\nu \nabla_\sigma - \nabla_\sigma \nabla_\nu ) n^\mu = {^4}R^\mu {_{\rho\nu\sigma}} n^\rho \, .
\label{ricc_iden}
\end{equation}
Let us project the equation (\ref{ricc_iden}) \emph{two-times onto  $ \Sigma_t $ and one-times along $ \hat{\textbf{n}} $},
\begin{equation}
\begin{aligned}
\gamma_{\alpha\mu} n^\sigma \gamma^\nu {_\beta} \,{^4}R^\mu {_{\rho\nu\sigma}} n^\rho 
&=  \gamma_{\alpha\mu} n^\sigma \gamma^\nu {_\beta} (\nabla_\nu \nabla_\sigma - \nabla_\sigma \nabla_\nu ) n^\mu  \\
&= \gamma_{\alpha\mu} n^\sigma \gamma^\nu {_\beta} \{ \nabla_\nu [ -K^\mu {_\sigma} - D^\mu \ln N n_\sigma ] 
- \nabla_\sigma [ -K^\mu {_\nu} - D^\mu \ln N n_\nu ] \} \\
& = \gamma_{\alpha\mu} n^\sigma \gamma^\nu {_\beta} \{ -\nabla_\nu K^\mu {_\sigma} - \nabla_\nu [ D^\mu \ln N n_\sigma ] 
+ \nabla_\sigma K^\mu {_\nu} \\
&\qquad \qquad \qquad  + \nabla_\sigma [D^\mu \ln N n_\nu ]\} \\
&=  \gamma_{\alpha\mu} n^\sigma \gamma^\nu {_\beta} \{ -\nabla_\nu K^\mu {_\sigma} -(\nabla_\nu D^\mu \ln N ) n_\sigma 
- D^\mu \ln N (\nabla_\nu n_\sigma) \\
& \qquad \qquad \qquad+ \nabla_\sigma K^\mu {_\nu} + (\nabla_\sigma D^\mu \ln N ) n_\nu 
+ D^\mu \ln N (\nabla_\sigma n_\nu) \} \\ 
&= \gamma_{\alpha\mu} \gamma^\nu {_\beta} \{ -n^\sigma \nabla_\nu K^\mu {_\sigma} +\nabla_\nu ( D^\mu \ln N) 
+ n^\sigma \nabla_\sigma K^\mu {_\nu} \\
&\qquad \qquad \quad + D^\mu \ln N n^\sigma \nabla_\sigma n_\nu \} \, ,
\label{r}
\end{aligned}
\end{equation}
where we have used the equation (\ref{n}) of the gradient of $ \hat{\textbf{n}} $. Because of the relation
\begin{equation*}
 n^\sigma K^\mu {_\sigma} =0 \Longrightarrow n^\sigma \nabla_\nu K^\mu {_\sigma} =-K^\mu {_\sigma} \nabla_\nu n^\sigma \, ,
\end{equation*}
the equation (\ref{r}) becomes
\begin{equation}
\begin{aligned}
\gamma_{\alpha\mu} n^\sigma \gamma^\nu {_\beta}\,{^4}R^\mu {_{\rho\nu\sigma}} &= \gamma_{\alpha\mu} \gamma^\nu {_\beta}
\{ K^\mu {_\sigma} \nabla_\nu n^\sigma + \nabla_\nu ( D^\mu \ln N) + n^\sigma \nabla_\sigma K^\mu {_\nu} \\
& \qquad \qquad \quad+ (D^\mu \ln N)(D_\nu \ln N) \} \\
&= \gamma_{\alpha\mu} \gamma^\nu {_\beta} \{ K^\mu {_\sigma} [- K^\sigma {_\nu} - D^\sigma \ln N n_\nu ] 
+ \nabla_\nu (D^\mu \ln N) \\
& \qquad \qquad \quad + n^\sigma \nabla_\sigma K^\mu {_\nu} + (D^\mu \ln N )(D_\nu \ln N)\} \, .
\label{rr}
\end{aligned}
\end{equation}
Since the projection of dual of the normal vector is zero, the equation (\ref{rr}) reduces to
\begin{equation}
\begin{aligned}
\gamma_{\alpha\mu} n^\sigma \gamma^\nu {_\beta}\,{^4}R^\mu {_{\rho\nu\sigma}} n^\rho &= \gamma_{\alpha\mu} \gamma^\nu {_\beta} \{- K^\mu {_\sigma} K^\sigma {_\nu} + \nabla_\nu (D^\mu \ln N) 
 + n^\sigma \nabla_\sigma K^\mu {_\nu} \\
& \qquad \qquad \quad + (D^\mu \ln N )(D_\nu \ln N)\} \\
&= -\gamma_{\alpha\mu} \gamma^\nu {_\beta} K^\mu {_\sigma} K^\sigma {_\nu} + \gamma_{\alpha\mu} \gamma^\nu {_\beta} \nabla_\nu (D^\mu \ln N) \\
 & \quad + \gamma_{\alpha\mu} \gamma^\nu {_\beta} n^\sigma \nabla_\sigma K^\mu {_\nu} 
 + \gamma_{\alpha\mu} \gamma^\nu {_\beta} (D^\mu \ln N )(D_\nu \ln N) \\
&= -K_{\alpha\sigma}K^\sigma {_\beta} + D_\beta (D_\alpha \ln N) 
 + \gamma^\mu {_\alpha} \gamma^\nu {_\beta} n^\sigma \nabla_\sigma K_{\mu\nu} \\
 & \quad + (D_\alpha \ln N )(D_\beta \ln N) \\
&= -K_{\alpha\sigma}K^\sigma {_\beta} - \frac{1}{N^2} (D_\beta N) (D_\alpha N) + \frac{1}{N}D_\beta D_\alpha N \\ 
 & \quad + \gamma^\mu {_\alpha} \gamma^\nu {_\beta} n^\sigma \nabla_\sigma K_{\mu\nu} + \frac{1}{N^2} (D_\alpha N )(D_\beta N) \\
&= -K_{\alpha\sigma}K^\sigma {_\beta} + \frac{1}{N}D_\beta D_\alpha N 
 + \gamma^\mu {_\alpha} \gamma^\nu {_\beta} n^\sigma \nabla_\sigma K_{\mu\nu} \, .
\label{rrr}
\end{aligned}
\end{equation}
The aim is to find such a projection of $ {^4}{\textbf{R}} $ which is totally composed of the intrinsic quantities of $ \Sigma_t $. Therefore, we need to get rid off the last term of the equation (\ref{rrr}): 
\begin{equation}
\begin{aligned}
\mathcal{L}_\textbf{m} K_{\alpha\beta} &= m^\mu \nabla_\mu K_{\alpha\beta} + K_{\mu\beta} \nabla_\alpha m^\mu + K_{\alpha\mu} \nabla_\beta m^\mu \\ 
&= N n^\mu \nabla_\mu K_{\alpha\beta} + K_{\mu\beta} [-N K^\mu {_\alpha} - D^\mu N n_\alpha + n^\mu \nabla_\alpha N] \\
& \quad + K_{\alpha\mu} [-N K^\mu {_\beta} - D^\mu N n_\beta + n^\mu \nabla_\beta N ] \\
&= N n^\mu \nabla_\mu K_{\alpha\beta} -N K_{\mu\beta} K^\mu {_\alpha} -K_{\mu\beta} D^\mu N n_\alpha + K_{\mu\beta} n^\mu \nabla_\alpha N \\
&- N K_{\alpha\mu} K^\mu {_\beta} - K_{\alpha\mu} D^\mu N n_\beta + K_{\alpha\mu} n^\mu \nabla_\beta N \\
&= N n^\mu \nabla_\mu K_{\alpha\beta} - 2N K_{\alpha\mu} K^\mu {_\beta} -K_{\mu\beta} D^\mu N n_\alpha 
- K_{\alpha\mu} D^\mu N n_\beta \, ,
\label{j}
\end{aligned}
\end{equation}
where we have used the fact that the projection of the extrinsic curvature along the unit normal vector is zero. Due to the property of (\ref{tang}), the full projection of the equation (\ref{j}) onto $ \Sigma_t $ is
\begin{equation}
\begin{aligned}
\overrightarrow{\gamma}^* \mathcal{L}_\textbf{\textbf{m}} K_{\alpha\beta} &= \mathcal{L}_{\textbf{m}} K_{\alpha\beta} \\ 
&= N \gamma^\mu {_\alpha} \gamma^\nu {_\beta} n^\sigma \nabla_\sigma K_{\mu\nu} - 2 N  \gamma^\mu {_\alpha} \gamma^\nu {_\beta}
K_{\mu\sigma} K^\sigma {_\nu} \\
& \quad -  \gamma^\mu {_\alpha} \gamma^\nu {_\beta} K_{\sigma\nu} D^\sigma N n_\mu
- \gamma^\mu {_\alpha} \gamma^\nu {_\beta} K_{\mu\sigma} D^\sigma N n_\nu \\
&= N \gamma^\mu {_\alpha} \gamma^\nu {_\beta} n^\sigma \nabla_\sigma K_{\mu\nu} - 2 N  \gamma^\mu {_\alpha} \gamma^\nu {_\beta}
K_{\mu\sigma} K^\sigma {_\nu} \, ,
\end{aligned} 
\end{equation}
hence, we obtain
\begin{equation}
\gamma^\mu {_\alpha} \gamma^\nu {_\beta} n^\sigma \nabla_\sigma K_{\mu\nu} = \frac{1}{N} \mathcal{L}_{\textbf{m}} K_{\alpha\beta}
+ 2K_{\alpha\sigma} K^\sigma {_\beta} \, .
\label{jj}
\end{equation}
By substituting the equation (\ref{jj}) into the main equation (\ref{rrr}), we obtain the crucial \emph{\textbf{Ricci equation}} as
\begin{equation}
\gamma_{\alpha\mu} n^\rho \gamma^\nu {_\beta} n^\sigma \,{^4}R^\mu {_{\rho\nu\sigma}} = \frac{1}{N} \mathcal{L}_{\textbf{m}} K_{\alpha\beta}
+ \frac{1}{N} D_\alpha D_\beta N + K_{\alpha\sigma} K^\sigma {_\beta} \, ,
\label{jjj}
\end{equation}
where we benefited from the fact that the intrinsic connection \emph{D} is torsion-free (That's. $ D_\alpha D_\beta f = D_\beta D_\alpha f $ ; here, f is a scalar field). Furthermore, \emph{we can replace the first term of the equation (\ref{jjj}) that is a projection of the spacetime Riemann tensor  $ {^4}R^\mu {_{\rho\nu\sigma}} $ with the full projection of spacetime Ricci tensor $ {^4}R_{\mu\nu} $ onto $ \Sigma_t $ by using the contracted Gauss relation (\ref{6})}: Now, the contracted Gauss relation that we have found is
\begin{equation*}
\gamma^\mu {_\alpha} \gamma^\nu {_\beta}\, {^4}R_{\mu\nu} + \gamma_{\alpha\mu} \gamma^\rho {_\beta} n^\sigma n^\nu 
\,{^4}R^\mu {_{\nu\rho\sigma}} =R_{\alpha\beta} + KK_{\alpha\beta} - K_{\alpha\mu}K^\mu {_\beta} \, ,
\end{equation*}
then
\begin{equation}
 \gamma_{\alpha\mu} \gamma^\rho {_\beta} n^\sigma n^\nu \,{^4}R^\mu {_{\nu\rho\sigma}} 
= R_{\alpha\beta}-\gamma^\mu {_\alpha} \gamma^\nu {_\beta}\, {^4}R_{\mu\nu} + KK_{\alpha\beta} - K_{\alpha\mu}K^\mu {_\beta} \, ,
\label{jjjj}
\end{equation}
by interchanging the indices $ \nu $ and $ \rho $ of the previous equation (\ref{jjjj}) and then substituting it into the Ricci Equation (\ref{jjj}), we get
\begin{equation}
 \gamma^\mu {_\alpha} \gamma^\rho {_\beta} \,{^4}R_{\mu\rho} = - \frac{1}{N} \mathcal{L}_\textbf{m} K_{\alpha\beta}
-\frac{1}{N} D_\alpha D_\beta N + R_{\alpha\beta} +K K_{\alpha\beta} -2 K_{\alpha\mu} K^\mu {_\beta} \, .
\label{jjjjj}
\end{equation}
For convention, let us do the operation of $ \rho \rightarrow \nu $ in the first term of the equation (\ref{jjjjj}). Then, it becomes
\begin{equation}
 \gamma^\mu {_\alpha} \gamma^\nu {_\beta} \,{^4}R_{\mu\nu} = - \frac{1}{N} \mathcal{L}_\textbf{m} K_{\alpha\beta}
-\frac{1}{N} D_\alpha D_\beta N + R_{\alpha\beta} +K K_{\alpha\beta} -2 K_{\alpha\mu} K^\mu {_\beta} \, .
\label{jjjjjjj}
\end{equation}
This is \emph{\textbf{the equation of the full projection of the Ricci tensor onto the hypersurface $ \Sigma_t $}}. And \emph{the compact form of the equation (\ref{jjjjjjj})} is given by
\begin{equation}
 \overrightarrow{\gamma}^*\, {^4}\textbf{R} = -\frac{1}{N} \mathcal{L}_\textbf{m} \textbf{K}-\frac{1}{N} \textbf{D} \textbf{D} N
+ \textbf{R}+K \tilde{\textbf{K}} - 2 \tilde{\textbf{K}} . \overrightarrow{ \textbf{K}} \, .
\label{ricc_proj}
\end{equation}

\subsubsection{3+1 Expression of the Spacetime Scalar Curvature}

The Einstein equation contains the spacetime scalar curvature tensor $ {^4}R $ so we are inevitably forced to find its 3+1 decomposition. Otherwise, the 3+1 decompositions of the Einstein equation can not to be constructed. Therefore, let us take the trace of the equation (\ref{jjjjjjj}) with the dual induced 3-metric   $ \gamma^{\alpha\beta} $:
\begin{equation}
\gamma^{\alpha\beta}\gamma^\mu {_\alpha} \gamma^\nu {_\beta} \,{^4}R_{\mu\nu} = - \frac{1}{N} \gamma^{ij} \mathcal{L}_\textbf{m} K_{ij}
-\frac{1}{N} D_{i} D^{i} N + R +K^2 -2 K_{ij} K^{ij} \, ,
\label{dd}
\end{equation}
it reduces to
\begin{equation}
 \gamma^{\mu\nu}\,{^4}R_{\mu\nu} = - \frac{1}{N} \gamma^{ij} \mathcal{L}_\textbf{m} K_{ij}
-\frac{1}{N} D_{i} D^{i} N + R +K^2 -2 K_{ij} K^{ij} \, .
\label{l}
\end{equation}
The $ 1^{st} $ term of the equation (\ref{l}) contains $ \gamma^{\mu\nu} $ and $ {^4}R_{\mu\nu} $. In order to take the trace of $ {^4}R_{\mu\nu} $, we need to rewrite $ \gamma^{\mu\nu} $  in terms of $ g^{\mu\nu} $ which is given in the equation (\ref{indu}):
\begin{equation}
 \gamma^{\mu\nu}\,{^4}R_{\mu\nu} = [g^{\mu\nu} + n^\mu n^\nu ]\,{^4}R_{\mu\nu} = {^4}R + {^4}R_{\mu\nu} n^\mu n^\nu \, .
\label{d}
\end{equation}
Observe that (\ref{l}) is a scalar equation. Because of this, we need to find the explicit form of the $ 2^{nd} $ term of (\ref{l}): Now,
\begin{equation}
 \gamma^{ij} \mathcal{L}_\textbf{m} K_{ij} = \mathcal {L}_\textbf{m} (\gamma^{ij} K_{ij} ) - K_{ij} \mathcal{L}_\textbf{m} \gamma^{ij}
= \mathcal {L}_\textbf{m} K - K_{ij} \mathcal{L}_\textbf{m} \gamma^{ij} \, .
\label{ll}
\end{equation}
The last term of the equation (\ref{ll}) contains \emph{the Lie derivative of the dual induced 3-metric, \\ $\mathcal{L}_\textbf{m} \gamma^{ij} $, along the normal evolution vector \textbf{m}} but we do not know what $\mathcal{L}_\textbf{m} \gamma^{ij} $, directly. Fortunately, we know the Lie derivative of the induced 3-metric along \textbf{m} [equation (\ref{mm})]. Therefore, to find $\mathcal{L}_\textbf{m} \gamma^{ij} $, we will use $\mathcal{L}_\textbf{m} \gamma_{ij} $ by starting from the relation $ \gamma_{ik} \gamma^{kj} = \delta^{j} {_{i}} $ : 
\begin{equation}
\begin{aligned}
 \gamma_{ik} \gamma^{kj} &= \delta^{j} {_{i}} \\
 (\mathcal{L}_\textbf{{m}}\gamma_{ik} ) \gamma^{kj} + \gamma_{ik} (\mathcal{L}_\textbf{{m}}\gamma^{kj} )&= 0 \\ 
\gamma_{ik} (\mathcal{L}_\textbf{{m}}\gamma^{kj} ) &=-\gamma^{kj} (\mathcal{L}_\textbf{{m}}\gamma_{ik} )  \, .
\label{lll}
\end{aligned}
\end{equation}
Let us multiply the equation (\ref{lll}) by $ \gamma^{il} $
\begin{equation*}
\begin{aligned}
 \gamma^{il} / \gamma_{ik} (\mathcal{L}_\textbf{{m}}\gamma^{kj} )=-\gamma^{kj} (\mathcal{L}_\textbf{{m}}\gamma_{ik} )  
\Longrightarrow \mathcal{L}_\textbf{m} &\gamma^{lj} = -\gamma^{il} \gamma^{kj} (\mathcal{L}_{\textbf{m}}\gamma_{ik} ) \, , \\
 & l \rightarrow i, i \rightarrow k,k \rightarrow l
\end{aligned}
\end{equation*}
\begin{equation}
\begin{aligned}
\qquad \qquad \qquad \qquad \qquad \qquad \qquad \quad  \mathcal{L}_\textbf{m} \gamma^{ij}
 &= - \gamma^{ik} \gamma^{jl} \mathcal{L}_\textbf{m} \gamma_{kl} \\
& = 2 N \gamma^{ik} \gamma^{jl} K_{kl} \\
& =2 N K^{ij} \, .
\end{aligned}
\end{equation}
With this result, the equation (\ref{ll}) turns into the form that we want
\begin{equation}
 \gamma^{ij} \mathcal{L}_\textbf{m} K_{ij} = \mathcal{L}_\textbf{m} K - 2 N K_{ij} K^{ij} \, .
\label{llll}
\end{equation}
Finally, by substituting the results of (\ref{d}) and (\ref{llll}) into the main equation (\ref{l}), we get
\begin{equation}
 {^4}R + {^4}R_{\mu\nu} n^\mu n^\nu = - \frac{1}{N} \mathcal{L}_\textbf{m} K - \frac{1}{N} D_{i} D^{i} N + R + K^2 \, .
\label{ddd}
\end{equation}
To find the 3+1 decomposition of the spacetime scalar curvature $ {^4}R $, we need to rewrite the term in the equation (\ref{ddd}) which contains the spacetime Ricci tensor. This is done with the help of the scalar Gauss relation (\ref{dddd}):
\begin{equation} 
\begin{aligned} 
&{^4}R + 2 \,{^4}R_{\mu\nu} n^\mu n^\nu =R + K^2 -K_{ij} K^{ij} \, ,\\ 
& {^4}R_{\mu\nu} n^\mu n^\nu = - \frac{1}{2} {^4}R + \frac{1}{2}R + \frac{1}{2} K^2 - \frac{1}{2} K_{ij} K^{ij} \, .
\label{ddddd}
\end{aligned}
\end{equation}
Thus, by substituting the results of (\ref{ddddd}) into the equation (\ref{ddd}), we will get the \emph{\textbf{3+1 decomposition of spacetime Scalar curvature}} as
\begin{equation}
{ ^4}R = R + K^2 + K_{ij}K^{ij} -\frac{2}{N} \mathcal{L}_\textbf{m} K -\frac{2}{N} D_i D^i N \, .
\label{mama}
\end{equation}

\section{THE 3+1 DECOMPOSITION OF EINSTEIN EQUATION}
\label{sec:The 3+1 Decomposition of Einstein Equation}

In this chapter, we will suppose that the spacetime that we are going to deal with is a solution of the Einstein equation with zero cosmological constant. Now, the Einstein equation: 
\begin{equation}
 {^4}\textbf{R}-\frac{1}{2}\, {^4}R \,\textbf{g} =8 \pi \, \textbf{T} \, .
\label{e1}
\end{equation}
It can be assumed as an equation of machines where each machine has two slots for inputs. Alternatively, it can be written in terms of matter parts
\begin{equation}
 {^4}\textbf{R} = 8 \pi \Big [\textbf{T} - \frac{1}{2} T \,\textbf{g} \Big ] \, .
\label{e2}
\end{equation}
where \textbf{T}  stands for the second rank stress-energy tensor and $ T $ is its trace. Then, by substituting the corresponding inputs into the slots, one can reach the \emph{3+1 decomposition of the Einstein equation}. In order to find it, as we see from the equations (\ref{e1}) and  (\ref{e2}), we need first to find what the 3+1 decomposition of \emph{the stress-energy tensor \textbf{T}}:

\subsection{The 3+1 Decomposition of the Stress-Energy Tensor}

Now, the full projection of the \textbf{T} along the unit normal vector $ \hat{} $ is nothing but the \emph{\textbf{energy density}} $ E $ evaluated by the observer,
\begin{equation}
 E=\textbf{T}(\hat{\textbf{n}} \,,\, \hat{\textbf{n}}) \, .
\label{s1}
\end{equation}
And the full projection of  \textbf{T} onto $ \Sigma_t $ is \emph{\textbf{bilinear form stress tensor}} \textbf{S}: $ \forall (\textbf{u},\textbf{v}) \in \mathcal{T}_p(\Sigma_t) $,
\begin{equation}
 \textbf{S}=\overset{\rightarrow}{\gamma}^* \textbf{T}=\textbf{T}(\textbf{u}\,,\,\textbf{v}) \, ,
\label{strs}
\end{equation}
or in components form
\begin{equation}
 S_{\alpha\beta}=\gamma^\mu{_\alpha} \gamma^\nu {_\beta} T_{\mu\nu} \, .
\label{s2}
\end{equation}
What the equation (\ref{s2}) says is that the action of the component of the force along $ e_\alpha $ onto the infinitesimal surface of the normal along $ e_\beta $ is given by $ S_{\alpha\beta} $. Here, $ e_\alpha $ and $ e_\alpha $ are the spacelike vectors with respect to the frame which does the measurements.

Finally, the mixed projection of \textbf{T} is the $ p_\alpha $ \emph{\textbf{component of the 1-form momentum density}} \textbf{p}: That's, $ \forall\, \textbf{v} \in \mathcal{T}_p(\Sigma) $,
\begin{equation}
 \textbf{p}=-\textbf{T}(\hat{\textbf{n}},\textbf{v}) \, ,
\label{mome}
\end{equation}
or in terms of the components
\begin{equation}
 p_\alpha=-n^{\mu}\gamma^\nu {_\alpha} T_{\mu\nu} \, .
\end{equation}
Now, let us think reversely. We naturally expect that the stress-energy tensor must be the union of the corresponding projections of it: That's,
\begin{equation}
 \textbf{T}=E\,\tilde{\textbf{n}} \otimes \tilde{\textbf{n}} + \textbf{S}+\textbf{p}\otimes \tilde{\textbf{n}}+ \tilde{\textbf{n}} \otimes \textbf{p} \, .
\label{ste}
\end{equation}
From the equation (\ref{ste}), it is easy to show that \emph{the important relation between the traces of} \textbf{T} (i.e. $ T $)\,, \textbf{S} (i.e. $ S $) [\emph{with respect to} \textbf{g}] and $ E $ is given by
\begin{equation}
 T=S-E \, .
\label{gg}
\end{equation}

\subsection{The Projection of the Einstein Equation}

With the help of the fundamental relations of the 3+1 formalism that we have found in the previous chapter and the 3+1 decomposition of the stress-energy tensor, we will find \emph{the 3+1 decomposition of the Einstein  equation which are known as \, \textbf{the dynamical Einstein equation, the Hamiltonian and the Momentum constraint equations}}. 

\subsubsection{Full Projection onto $ \Sigma_t $ }

First of all, let us start by fully projecting the Einstein equation of (\ref{e2}) onto the hypersurface $ \Sigma_t $:
\begin{equation}
 \overrightarrow{\gamma}^* {^4}\textbf{R} = 8 \pi \Big( \overrightarrow{\gamma}^* \textbf{T} - \frac{1}{2} T \, \overrightarrow{\gamma}^* \textbf{g} \Big) \, .
\end{equation}
Now, the knowledge of $ T $ is equal to the difference between $ S $ and $ E $ (see the equation \ref{gg}), the full projection of the \textbf{T} onto $ \Sigma_t $ (the equation \ref{strs}) is \textbf{S} and the full projection of the \textbf{g} is the induced 3-metric $ \gamma $ with the equation (\ref{ricc_proj}) of
\begin{equation*}
 \overrightarrow{\gamma}^* {^4}\textbf{R} = -\frac{1}{N} \mathcal{L}_\textbf{m} \textbf{K}-\frac{1}{N} \textbf{D} \textbf{D} N
+ \textbf{R} + K\, \textbf{K} - 2 \textbf{K}.\overrightarrow{\textbf{K}} 
\end{equation*}
leads us
\begin{equation}
\begin{aligned}
& -\frac{1}{N} \mathcal{L}_\textbf{m} \textbf{K} - \frac{1}{N} \textbf{D} \textbf{D} N + \textbf{R} + K\, \textbf{K}
- 2 \textbf{K} \overset{\rightarrow}{\textbf{K}} = 8 \pi \Big ( \textbf{S} - \frac{1}{2} [S-E] \mathbf{\gamma} \Big ) \, , \\
&\quad \frac{1}{N} \mathcal{L}_\textbf{m} \textbf{K} = -\frac{1}{N} \textbf{D} \textbf{D} N + \textbf{R} +K \textbf{K}
- 2 \textbf{K}.\overrightarrow{\textbf{K}} + 8 \pi \Big ( \frac{1}{2} [S-E] \mathbf{\gamma} -\textbf{S} \Big) \, ,\\
& \quad \mathcal{L}_\textbf{m} \textbf{K} = -\textbf{D} \textbf{D} N + N \textbf{R} +K \textbf{K}
- 2 \textbf{K}.\overrightarrow{\textbf{K}} + 4 \pi \Big( [S-E] \mathbf{\gamma} - 2 \textbf{S} \Big) \, , \\
\end{aligned}
\end{equation}
or in the components form
\begin{equation}
\begin{aligned}
\mathcal{L}_\textbf{m} K_{\alpha\beta}  = -D_{\alpha\beta} D_{\alpha\beta} N + N & \Big \{R_{\alpha\beta} +K K_{\alpha\beta}
- 2 K_{\alpha\mu} K^\mu {_\beta} \\ 
&+ 4 \pi ( [S-E] \gamma_{\alpha\beta} - 2S_{\alpha\beta} ) \Big \}  \, .
\label{full}
\end{aligned}
\end{equation}
Since the equation (\ref{full}) is totally composed of the quantities of $ \Sigma_t $, we can rewrite it in terms of the spatial indices,
\begin{equation}
\begin{aligned}
\mathcal{L}_\textbf{m} K_{ij}  = -D_i D_j N + N & \Big \{R_{ij} +K K_{ij}
- 2 K_{ik} K^k {_j} \\ 
&+ 4 \pi ( [S-E] \gamma_{ij} - 2S_{ij} ) \Big \}  \, .
\label{dynm1}
\end{aligned}
\end{equation}
The equation (\ref{dynm1}) is known as \emph{\textbf{ the dynamical part of the Einstein equation} and it has 6 independent components}.

\subsubsection{Full Projection Perpendicular to $ \Sigma_t $ }

Secondly, let us fully project the Einstein equation (\ref{e1}) along the normal unit vector $ \hat{\textbf{n}} $ 
\begin{equation}
{^4} \textbf{R} (\hat{\textbf{n}},\hat{\textbf{n}}) - \frac{1}{2}{^4}R \, \textbf{g} 
(\hat{\textbf{n}},\hat{\textbf{n}}) = 8 \pi \textbf{T} (\hat{\textbf{n}},\hat{\textbf{n}}) \, .
\label{fg}
\end{equation}
From the equation (\ref{s1}), the component form of the equation (\ref{fg}) becomes
\begin{equation}
 {^4}R_{\mu\nu} n^\mu n^\nu + \frac{1}{2}{^4}R  = 8 \pi E \, .
\end{equation}
By using the scalar Gauss relation (\ref{dddd}), we get \emph{\textbf{the Hamiltonian constraint part of the Einstein equation}} as
 \begin{equation}
 R + K^2 -K_{ij} K^{ij} = 16 \pi E \, .
\label{gh}
\end{equation}
Since \emph{the Hamiltonian constraint equation (\ref{gh}) is a scalar equation, it has only 1 independent component}.

\subsubsection{Mixed Projection}

Let us project the Einstein equation (\ref{e1}) either one-times onto the hypersurface or one-times orthogonal to it 
\begin{equation*}
 {^4}\textbf{R} \Big ( \hat{\textbf{n}},\overrightarrow{\gamma}(.)\Big )
 - \frac{1}{2} {^4}R \,\textbf{g} \Big( \hat{\textbf{n}},\overrightarrow{\gamma}(.) \Big) = 8 \pi \textbf{T} \Big(\hat{\textbf{n}},\overrightarrow{\gamma}(.) \Big)
\Rightarrow  {^4}\textbf{R} \Big( \hat{\textbf{n}},\overrightarrow{\gamma}(.)\Big)
 =  8 \pi \textbf{T} \Big(\hat{\textbf{n}},\overrightarrow{\gamma}(.)\Big) \, .
\end{equation*}
From the equation (\ref{mome}) we know that the mixed projection of the \textbf{T} is \textbf{p}. By using this knowledge, the component form of the previous equation becomes
\begin{equation}
 \gamma^\mu {_\alpha} n^\nu \, {^4}R_{\mu\nu} = -8 \pi p_\alpha \, .
\label{momen}
\end{equation}
Now, by substituting \emph{the Contracted Codazzi Relation} (\ref{cont_coda}) into the equation (\ref{momen}), we get \emph{\textbf{the momentum constraint part of the Einstein equation}} (or simply \emph{\textbf{Momentum constraint equation}}) as
\begin{equation}
 D_\mu K^\mu {_\alpha} - D_\alpha K = 8 \pi p_\alpha \, ,
\label{mome}
\end{equation}
where (\ref{mome}) has 3 independent components. Moreover, it can also be rewritten in terms of the spatial indices
\begin{equation}
 D_j K^j {_i} - D_i K = 8 \pi p_i \, .
\label{momen_cost}
\end{equation}

\subsection{The 3+1 Dimensional Einstein Equation as a PDEs System}

Now, the coordinate systems adapted to the flow of the hypersurfaces (\ref{ff}) allow us to transform the 3+1 Einstein equation into a set of partial differential equations (PDEs).

\subsubsection{Lie Derivatives Along ''\textbf{m}'' as Partial Derivatives}

\begin{enumerate}
 \item For $ \mathcal{L}_\textbf{m} $ $ \textbf{K} $:

The equation (\ref{shft}) provides us to decompose the Lie derivative of the extrinsic curvature along \textbf{m} in terms of the Lie derivatives of \textbf{K} along $ \overset{\rightarrow}{\partial}_t $ and along the shift vector $ \overset{\rightarrow}{\beta} $ 
\begin{equation}
 \mathcal{L}_\textbf{m} \textbf{K} = \mathcal{L}_{\overset{\rightarrow}{\partial}_t} \textbf{K} - \mathcal{L}_{\overset{\rightarrow}{\beta}} \textbf{K} \, .
\label{i}
\end{equation}
The equation (\ref{i}) says that $ \mathcal{L}_{\overset{\rightarrow}{\partial}_t} \textbf{K} $ is also a tensor field of the hypersurface. We wish to write the tensor components relative to the well-defined coordinates. Then, the $ \mathcal{L}_{\overset{\rightarrow}{\partial}_t} \textbf{K} $ turns into the partial derivate relative to the time coordinate t
\begin{equation}
\mathcal{L}_{\overset{\rightarrow}{\partial}_t} K_{ij} = \frac{\partial K_{ij}}{\partial t} \, . 
\end{equation}
Similarly, the Lie derivative of \textbf{K} along $ \beta $ turns into the partial derivatives relative to the spatial coordinates
\begin{equation}
 \mathcal{L}_{\overset{\rightarrow}{\beta}} K_{ij} = \beta^k \frac{\partial K_{ij}}{\partial x^i} + K_{kj} \frac{\partial \beta^k}{\partial x^i}
+ K_{ik} \frac{\partial \beta^k}{\partial x^j} \, .
\end{equation}
\item For $ \mathcal{L}_\textbf{m} \gamma $ :

Identically, the equation (\ref{shft}) provides us to decompose the Lie derivative $ \mathcal{L}_\textbf{m} \gamma $ as
\begin{equation}
\mathcal{L}_\textbf{m} \gamma_{ij} = \mathcal{L}_{\overset{\rightarrow}{\partial}_t} \gamma_{ij} - \mathcal{L}_{\overset{\rightarrow}{\beta}} \gamma_{ij} = -2 N K_{ij} \, .
\label{jk}
\end{equation}
Again, because of the same reason, the Lie derivative of the induced 3-metric becomes the partial derivative,
\begin{equation}
 \mathcal{L}_{\overset{\rightarrow}{\partial}_t} \gamma_{ij} = \frac{\partial \gamma_{ij}}{\partial t} \, ,
\end{equation}
and also
\begin{equation}
\begin{aligned}
 \mathcal{L}_{\overset{\rightarrow}{\beta}} \gamma_{ij} &= \beta^k D_k \gamma_{ij} + \gamma_{kj} D_i \beta^k + \gamma_{ik} D_j \beta^k \\
& =\gamma_{kj} D_i \beta^k + \gamma_{ik} D_j \beta^k  \\
& = D_i \beta_j + D_j \beta_i \, .
\end{aligned}
\end{equation}
\end{enumerate}
With the help of the form of the corresponding Lie derivatives (\ref{i}) and (\ref{jk}) in the foliation adapted coordinates, we get 
\emph{the 3+1 decomposition of Einstein equation within these coordinates}   
\begin{equation}
\bigg (\frac{\partial}{\partial t} - \mathcal{L}_{\overset{\rightarrow}{\beta}} \bigg ) \gamma_{ij} = -2 N K_{ij} \, ,
\label{tt}
\end{equation}
\begin{equation}
 \bigg (\frac{\partial}{\partial t} - \mathcal{L}_{\overset{\rightarrow}{\beta}} \bigg ) K_{ij} = -D_i D_j N + N  \bigg \{ R_{ij} 
+ K K_{ij}- 2K_{ik} K^k {_j} + 4 \pi [(S-E)\gamma_{ij} - 2 S_{ij}] \bigg \} \, , 
\end{equation}
\begin{equation}
R + K^2 - K_{ij} K^{ij} = 16 \pi E  \, ,
\end{equation}
\begin{equation}
D_j K^j {_i} -D_i K = 8 \pi p_i \, .
\end{equation}
This set of equations is known as the \emph{\textbf{3+1 dimensional Einstein system}}.

\section{CONFORMAL DECOMPOSITION OF THE 3+1 EINSTEIN EQUATION}
\label{sec:Conformal Decomposition of The 3+1 Einstein Equation}

In the Ricci and Cotton flows \cite{Hamilton:1982}, \cite{Kisisel:2008jx}, the Riemannian metric is being mapped by particular diffeomorphisms. Furthermore, if the mapped metric is a scale times the initial metric then it is called the gradient Ricci and the gradient Cotton solitons. As in the Ricci and Cotton flows, we will examine the evolution of the hypersurfaces as if there is a well-defined Riemannian conformal background metric $ \tilde{\gamma} $ and it is smoothly mapped into the 3-metric of $ \Sigma_t $ by a positive scalar field $ \Psi $ during the flow. This type of the flow of the metric is known as the \emph{conformal transformation (or flow) of the induced metric of} $ \Sigma_t $ ( see Lichnerowicz \cite{Lichnerowicz:1939vy})     
\begin{equation}
 \gamma = \Psi^4 \tilde{\gamma} \, .
\label{conf}
\end{equation}
Now, let us take a look at what Eric Gourgoulhon \cite{Gourgoulhon:2007ue} says about the York's publications \cite{York:1979cf}, \cite{York:1981bg}  which give the importance of the conformal transformation in gravity : '' In $ 1971-72 $, York has shown that conformal decompositions are also important for the \emph{time evolution} problem, by demonstrating that two degrees of freedom of the gravitational field are carried by the conformal equivalence classes of the 3-metrics. \emph{A conformal equivalence} is defined as the set of all metrics that can be related to a given metric $  \gamma_{ij} $ by a conformal transformation like the relation of (\ref{conf})". Now, we know that the Weyl tensor gives whether a given spacetime is conformally flat or not [that's, \emph{the background metric in  the equation (\ref{conf}}) is flat] and it is valid for the spacetimes whose dimension is greater than 3. Here, in the lower dimensional cases the conformally invariant Cotton tensor, $ C_{ijk} $, \cite{Cotton:1899hj} takes the role of the Weyl tensor:
 \begin{equation}
 C_{ijk}=D_k \Big(R_{ij}-\frac{1}{4}R \gamma_{ij} \Big)-D_j \Big(R_{ik}-\frac{1}{4}R\gamma_{ik} \Big ) \, .
\label{cotton}
\end{equation}
Moreover, the York \cite{York:1979cf}, \cite{York:1981bg} showed that the $ C^{ij}_* $ is \emph{another conformally invariant tensor},
\begin{equation}
 C^{ij}_*=\gamma^{5/6} C^{ij} \, ,
\label{conf_inva}
\end{equation}
where the $ C^{ij} $ is the well-known Cotton-York tensor \cite{York:1979cf}, \cite{York:1981bg}, \cite{Cotton:1899hj} which is constructed from the Cotton tensor (\ref{cotton}) \cite{Cotton:1899hj} as
\begin{equation}
 C^{ij}=-\frac{1}{2} \epsilon^{ikl} C_{mkl} \gamma^{mj}= \epsilon^{ikl} D_k \Big ( R^j{_l}-\frac{1}{4} R \,\delta^j{_l} \Big ) \, ,
\end{equation}
which satisfies the following properties
\begin{enumerate}
 \item  \emph{Symmetric}:
\begin{equation}
\begin{aligned}
 \epsilon_{ijm} C^{ij}&= \epsilon_{ijm} \epsilon^{ikl} D_k \Big ( R^j{_l}-\frac{1}{4} R \,\delta^j{_l} \Big ) \\
&= \Big (\delta^k{_j} \delta^l{_m} - \delta^k{_m} \delta^l{_j} \Big ) D_k \Big ( R^j{_l}-\frac{1}{4} R \,\delta^j{_l} \Big ) \\
&=D_j \Big ( R^j{_m}-\frac{1}{4} R \,\delta^j{_m} \Big ) - D_m \Big ( R-\frac{1}{4} R \,\delta^j{_j} \Big )  \\
&=D_j G^j{_m} \\
&=0 \, .
\end{aligned}
\end{equation}
\item \emph{The traceless}: The Cotton-York tensor is traceless, then, let us first rewrite this in terms of the sum of the Einstein tensor:
\begin{equation}
\begin{aligned}
C^{ij} &=\frac{1}{2} \epsilon^{ikl} D_k \Big ( R^j{_l}-\frac{1}{4} R \,\delta^j{_l} \Big ) + \frac{1}{2} \epsilon^{jkl} D_k \Big ( R^i{_l}-\frac{1}{4} R \,\delta^i{_l} \Big ) \\
&= \frac{1}{2} \epsilon^{ikl} D_k \Big ( G^j{_l}+\frac{1}{4} R \,\delta^j{_l} \Big )+ \frac{1}{2} \epsilon^{jkl} D_k \Big ( G^i{_l}+\frac{1}{4} R \,\delta^i{_l} \Big ) \\
&=\frac{1}{2} \epsilon^{ikl} D_k G^j{_l}+ \frac{1}{8} \epsilon^{ikj} D_k R + \frac{1}{2} \epsilon^{jkl} D_k G^i{_l}+ \frac{1}{8} \epsilon^{jki} D_k R  \\
&=\frac{1}{2} \epsilon^{ikl} D_k G^j{_l} + \frac{1}{2} \epsilon^{jkl} D_k G^i{_l} \, .
\label{trc}
\end{aligned}
\end{equation}
By taking the trace of (\ref{trc}) with respect to $ \gamma_{ij} $:
\begin{equation}
 \gamma_{ij} C^{ij} =\frac{1}{2} \epsilon^{ikl} D_k G_{il} + \frac{1}{2} \epsilon^{jkl} D_k G_{jl} \, .
\label{trce}
\end{equation}
 Because of the symmetric and antisymmetric relations in the equation (\ref{trce}), we get 
\begin{equation}
 \gamma_{ij} C^{ij} = 0  \, .
\end{equation}
\item \emph{Divergence-free, (i.e. transverse)}: $ D_j C^{ij}=0  \, . $
\end{enumerate}
\newpage 
Now, let us prove that \emph{the tensor defined in the equation (\ref{conf_inva}) is really conformally invariant under the transformation (\ref{conf}) by using the following particular conformal transformations}: 
\begin{enumerate}
\item $ \gamma_{ij} \rightarrow \Psi^4 \tilde{\gamma}_{ij} $ , $ \gamma^{ij} \rightarrow \Psi^{-4} \tilde{\gamma}^{ij} $ . 
\item $ det(\gamma_{ij}) \rightarrow (\Psi^4 )^3 det(\tilde{\gamma}_{ij})  \Rightarrow \gamma= \Psi^{12} \tilde{\gamma} $ .
\item $ \epsilon \rightarrow \frac{1}{\sqrt{\gamma}} \Rightarrow \epsilon^{ikl}=\Psi^{-6} \tilde{\epsilon}^{ikl} $ where $ \gamma=det(\gamma_{ij}) $ .
\item $ C_{mkl} = \tilde{C}_{mkl} $ .
\end{enumerate}
Then
\begin{equation}
\begin{aligned}
C^{ij}_* &= \gamma^{5/6} C^{ij} \\
&=-\frac{1}{2}\gamma^{5/6} \epsilon^{ikl} C_{mkl} \gamma^{mj} \\
&=-\frac{1}{2} \Big [ \Psi^{12}\tilde{\gamma} \Big]^{5/6} \Psi^{-6} \tilde{\epsilon}^{ikl} \tilde{C}_{mkl} \Psi^{-4} \tilde{\gamma}^{mj} \\
&=\tilde{\gamma}^{5/6} \bigg \{-\frac{1}{2} \tilde{\epsilon}^{ikl} \tilde{C}_{mkl} \Psi^{-4} \tilde{\gamma}^{mj} \bigg \} \\
&=\tilde{\gamma}^{5/6} \tilde{C}^{ij} \\
&= \tilde{C}^{ij}_* \, .
\end{aligned}
\end{equation}

From here, we are going to find what happens to the 3+1 form of the Einstein equation under the conformal transformation. Therefore, in order to find the explicit form of the conformal 3+1 expression of the Einstein equation, we first need to find the conformal form of the fundamental quantities:

\subsection{The Conformal Form of the Intrinsic Quantities}

Because the determinant of $ \gamma $ depends upon the choice of coordinates so the conformal factor is not a scalar field. Thanks to the flat background metric, we achieve to make the conformal factor a scalar field \cite{Gourgoulhon:2007ue}.

\subsubsection{Conformal Connection }

Now, suppose that there is a conformal background metric $ \tilde{\gamma} $ on the hypersurface whose intrinsic connection is metric-compatible
\begin{equation*}
 \tilde{D} \tilde{\gamma} = 0 \, .
\end{equation*}
And associated Christoffel symbols are defined as
\begin{equation}
 \tilde{\Gamma}^k_{ij} = \frac{1}{2} \tilde{\gamma}^{kl} \Big \{ \frac{\partial \tilde{\gamma}_{jl}}{\partial x^i}
+ \frac{\partial \tilde{\gamma}_{il}}{\partial x^j} - \frac{\partial \tilde{\gamma}_{ij}}{\partial x^l} \Big \} \, .
\end{equation}
Here notice that the partial derivatives are taken with respect to the coordinates $ (x^i) $. Since the intrinsic covariant derivative of any tensor field \textbf{T} can be taken relative to two connections  $ \tilde{D} $ and $ D $, then, it is normal to expect that there must be a transformation between them. And this transformation is given by
\begin{equation}
\begin{aligned}
 D_k T^{i_1...i_p} {_{j_1 ...j_q}} &=\tilde{D}_k T^{i_1 ...i_p} {_{j_1 ...j_q}} 
+ \sum_{r=1}^{p} C^{i_r} {_kl} T^{i_1 ...l...i_p} {_{j_1 ...j_q}} \\
& \qquad \qquad  \qquad \qquad  -\sum_{r=1}^{q} C^l {_{k{j_r}}} T^{i_1 ...i_p} {_{j_1 ...l...j_q}} \, ,
\label{tran1}
\end{aligned}
\end{equation}
Here $ C^k {_{ij}} = \Gamma^k_{ij} - \tilde{\Gamma}^k_{ij} $. Also, it a \emph{tensor field} because it is defined as the difference between the Christoffel symbols. And its explicit form is
\begin{equation}
\begin{aligned}
 C^k {_{ij}} &= \Gamma^k_{ij} - \tilde{\Gamma}^k_{ij} \\
&= \Gamma^k_{ij} -\delta^k {_m} \tilde{\Gamma}^m_{ij} \\
&=\Gamma^k_{ij} + \frac{1}{2} (-2) \gamma^{kl} \tilde{\Gamma}^m_{ij} \gamma_{ml} \\
& = \frac{1}{2} \gamma^{kl} \bigg \{ \frac{ \partial \gamma_{lj}}{ \partial x^i} + \frac{\partial \gamma_{li}}{ \partial x^j}
- \frac{\partial \gamma_{ij}}{ \partial x^l} \bigg \} + \frac{1}{2} (-2) \gamma^{kl} \tilde{\Gamma}^m_{ij} \gamma_{ml} \\
& = \frac{1}{2} \gamma^{kl} \bigg \{ \Big [ \tilde{D}_i \gamma_{lj} + \tilde{\Gamma}^m_{il} \gamma_{mj} + \tilde{\Gamma}^m_{ij} \gamma_{ml} \Big ]
 + \Big [ \tilde{D}_j \gamma_{il}+ \tilde{\Gamma}^m_{ji} \gamma_{ml} + \tilde{\Gamma}^m_{jl} \gamma_{im} \Big ] \\
& \qquad \quad- \Big [-\tilde{D}_l \gamma_{ij} - \tilde{\Gamma}^m_{li} \gamma_{mj} - \tilde{\Gamma}^m_{lj} \gamma_{mi} \Big ]
- 2 \tilde{\Gamma}^m_{ij} \gamma_{ml}  \bigg \} \\
& = \frac{1}{2} \gamma^{kl} \bigg \{ \tilde{D}_i \gamma_{lj} + \tilde{D}_j \gamma_{il} -\tilde{D}_l \gamma_{ij} \bigg \} \, . \\
\label{rstg}
\end{aligned}
\end{equation}
Because of the conformal transformation of induced 3-metric and its dual,
\begin{equation}
 \gamma_{ij} =  \Psi^4 \tilde{\gamma}_{ij} \, ,
\label{conf3}
\end{equation}
\begin{equation}
 \gamma^{ij} = \Psi^{-4} \tilde{\gamma}^{ij} \, .
\end{equation}
We can rewrite the equation (\ref{rstg}) fully in terms of the conformal quantities as  
\begin{equation}
\begin{aligned}
 C^k {_{ij}}& = \frac{1}{2} \gamma^{kl} \bigg \{ \tilde{D}_i \gamma_{lj} + \tilde{D}_j \gamma_{il} -\tilde{D}_l \gamma_{ij} \bigg \}   \\
& =\frac{1}{2} \Psi^{-4} \tilde{\gamma}^{kl} \bigg \{ \tilde{D}_i (\Psi^4 \tilde{\gamma}_{lj}) + \tilde{D}_j (\Psi^4 \tilde{\gamma}_{il} ) -\tilde{D}_l(\Psi^4 \tilde{\gamma}_{ij} \bigg \}   \\
& =\frac{1}{2} \Psi^{-4} \tilde{\gamma}^{kl} \bigg \{\tilde{\gamma}_{lj} \tilde{D}_i (\Psi^4 ) + \tilde{\gamma}_{il} \tilde{D}_j (\Psi^4 ) - \tilde{\gamma}_{ij}\tilde{D}_l(\Psi^4 ) \bigg \}   \\
& =\frac{1}{2} \Psi^{-4} \bigg \{\delta^k {_j} \tilde{D}_i (\Psi^4 ) + \delta^k {_i} \tilde{D}_j (\Psi^4 ) - \tilde{\gamma}_{ij}\tilde{D}^k(\Psi^4 ) \bigg \}   \\
& =\frac{1}{2} \bigg \{\delta^k {_j} \tilde{D}_i (\ln \Psi^4 ) + \delta^k {_i} \tilde{D}_j (\ln \Psi^4 ) - \tilde{\gamma}_{ij}\tilde{D}^k(\ln \Psi^4 ) \bigg \}   \\
& =2 \bigg \{\delta^k {_j} \tilde{D}_i (\ln \Psi) + \delta^k {_i} \tilde{D}_j (\ln \Psi ) - \tilde{D}^k(\ln \Psi) \tilde{\gamma}_{ij} \bigg \} \, .  
\label{kpn}
\end{aligned}
\end{equation}
The equation (\ref{kpn}) is playing an important role in the conformal transformation. As a sample, let us find the relation between the $ \textbf{v} \in T(\Sigma_t) $
\begin{equation}
\begin{aligned}
 D_j v^i &= \tilde{D}_j v^i + C^i {_{jk}} v^k \\
& =  \tilde{D}_j v^i + 2 \bigg \{v^k \delta^i {_j} \tilde{D}_k (\ln \Psi) + v^i \tilde{D}_j (\ln \Psi ) - v^k \tilde{\gamma}_{jk}\tilde{D}^i(\ln \Psi) \bigg \} \, . \\
\label{cdc}
\end{aligned}
\end{equation}
Let us do the change of $ j \rightarrow i $ in the equation (\ref{cdc}),
\begin{equation*}
\begin{aligned}
 D_i v^i &= \tilde{D}_i v^i + 2 \bigg \{v^k \delta^i {_i} \tilde{D}_k (\ln \Psi) + v^i \tilde{D}_i (\ln \Psi ) - v^k \tilde{\gamma}_{ik}\tilde{D}^i(\ln \Psi) \bigg \}   \\
&=\tilde{D}_i v^i + 2 \bigg \{ 3 v^k \tilde{D}_k (\ln \Psi) + v^i \tilde{D}_i (\ln \Psi ) - v^k \tilde{D}_k (\ln \Psi) \bigg \} \, .  \\
& \qquad \qquad \qquad \quad k \rightarrow i \qquad  \qquad \qquad \qquad k \rightarrow i
\end{aligned}
\end{equation*}
Thus, we get \emph{\textbf{the conformal form of the divergence of the vector tangent to $ \Sigma $}} as 
\begin{equation}
\begin{aligned}
 D_i v^i &= \tilde{D}_i v^i + 6 v^i \tilde{D}_i \ln \Psi \\
&=\Psi^{-6} \tilde{D}_i (\Psi^6 v^i ) \, .
\label{diver}
\end{aligned}
\end{equation}

\subsubsection{Conformal Transformation of the Intrinsic Ricci Tensor}

\begin{enumerate}
 \item Conformal Relation of the Ricci Tensors in terms of the Tensor Field \textbf{C}

The corresponding Ricci identity is
\begin{equation}
 (D_iD_j - D_jD_i)v^k = R^k {_{lij}} v^l \, .
\label{rica1}
\end{equation}
By doing the suitable operation of contraction and change of dummy indices, the equation (\ref{rica1}) becomes
\begin{equation}
 R_{ij} v^j = D_j D_i v^j -D_i D_j v^j \, .
\label{tran2}
\end{equation}
Now, with the help of the general transformation relation (\ref{tran1}), the equation (\ref{tran2}) can be written in terms of the conformal tensors of $ \tilde{\textbf{R}} $ , \textbf{C} and the conformal covariant derivative of \textbf{C}, 
\begin{equation}
\begin{aligned}
 R_{ij} v^j &= \tilde{D}_j (D_i v^j) - C^k {_{ji}} D_k v^j + C^j {_{jk}} D_i v^k - \tilde{D}_i ( D_j v^j ) \\
&= \tilde{D}_j \Big \{\tilde{D}_i v^j + C^j {_{ik}} v^k \Big \} - C^k {_{ji}} \Big \{\tilde{D}_k v^j + C^j {_{kl}} v^l \Big \} 
+ C^j {_{jk}} \Big \{\tilde{D}_i v^k  + C^k {_{il}} v^l \Big \} \\ 
& -\tilde{D}_i \Big \{\tilde{D}_j v^j + C^j {_{jk}} v^k \Big \} \\
&= \tilde{D}_j \tilde{D}_i v^j - \tilde{D}_i \tilde{D}_j v^j + \tilde{D}_j C^j {_{ik}} v^k - C^k {_{ji}} C^j {_{kl}} v^l
+C^j {_{jk}} C^k {_{il}} v^l- \tilde{D}_i C^j {_{jk}} v^k \\
&= \tilde{R}_{ij} v^j +  \tilde{D}_j C^j {_{ik}} v^k - C^k {_{ji}} C^j {_{kl}} v^l
+C^j {_{jk}} C^k {_{il}} v^l- \tilde{D}_i C^j {_{jk}} v^k \\
& \qquad \qquad \quad k \rightarrow j \qquad  l \rightarrow j \quad \qquad l \rightarrow j  \quad \qquad k \rightarrow j \\
&= \tilde{R}_{ij} v^j +  \tilde{D}_k C^k {_{ij}} v^j - C^k {_{li}} C^l {_{kj}} v^j
+C^l {_{lk}} C^k {_{ij}} v^j- \tilde{D}_i C^k {_{kj}} v^j \, .
\end{aligned}
\end{equation}
Because $ v^j $ is an arbitrary vector field, we get
\begin{equation}
 R_{ij} = \tilde{R}_{ij} +  \tilde{D}_k C^k {_{ij}} - \tilde{D}_i C^k {_{kj}} +C^k {_{ij}} C^l {_{lk}} - C^k {_{il}} C^l {_{kj}} \, . 
\label{tran3}
\end{equation}
\item The Conformal Transformation of the Intrinsic Ricci Tensors in terms of the Conformal Factor

The conformal equation (\ref{tran3}) can be rewritten in terms of the Conformal factor $ \Psi $  by using the equation of (\ref{kpn}). For simplicity, let us work on the terms of the equation (\ref{tran3}) which contain the conformal covariant derivative of tensor field \textbf{C} :
\begin{enumerate}
 \item For $ \tilde{D}_i C^k {_{kj}} $ :
\begin{equation}
 C^k {_{ij}}= 2 \bigg \{\delta^k {_i} \tilde{D}_j (\ln \Psi) + \delta^k {_j} \tilde{D}_i (\ln \Psi ) - \tilde{D}^k(\ln \Psi) \tilde{\gamma}_{ij} \bigg \} \, . 
\label{tran4}
\end{equation}
Let us do the change $ i \rightarrow k $ in the equation (\ref{tran4}), then, it becomes
\begin{equation}
\begin{aligned}
C^k {_{kj}} &= 2 \bigg \{\delta^k {_k} \tilde{D}_j (\ln \Psi) + \delta^k {_j} \tilde{D}_k (\ln \Psi ) - \tilde{D}^k(\ln \Psi) \tilde{\gamma}_{kj} \bigg \}  \\
&=2 \bigg \{3 \tilde{D}_j (\ln \Psi) + \tilde{D}_j (\ln \Psi ) - \tilde{D}_j(\ln \Psi) \bigg \} \\
&=6 \tilde{D}_j (\ln \Psi) \, ,  
\end{aligned}
\end{equation}
so we get
\begin{equation}
 \tilde{D}_i C^k {_{kj}} = 6 \tilde{D}_i \tilde{D}_j (\ln \Psi) \, .
\label{tran5}
\end{equation}

\newpage

\item For $ \tilde{D}_k C^k {_{ij}} $ :
\begin{equation}
\begin{aligned}
 \tilde{D}_k C^k {_{ij}} &= \tilde{D}_k \bigg \{ 2 \Big \{ {\delta^k {_i} \tilde{D}_j (\ln \Psi) 
+ \delta^k {_j} \tilde{D}_i (\ln \Psi ) - \tilde{D}^k(\ln \Psi) \tilde{\gamma}_{ij}} \Big \} \bigg \} \\  
&=2 \Big \{ \tilde{D}_i \tilde{D}_j (\ln \Psi) 
+ \tilde{D}_j \tilde{D}_i (\ln \Psi ) - \tilde{D}_k \tilde{D}^k (\ln \Psi) \tilde{\gamma}_{ij} \Big \} \\  
& = 4 \tilde{D}_i \tilde{D}_j (\ln \Psi) - 2 \tilde{D}_k \tilde{D}^k (\ln \Psi) \tilde{\gamma}_{ij} \, .
\label{tran6}
\end{aligned}
\end{equation}
\end{enumerate}
Let us substitute the results of (\ref{tran5}), (\ref{tran6}) and the explicit formula (\ref{kpn}) for the components of \textbf{C} into the main equation (\ref{tran3})
\begin{equation}
 \begin{aligned}
 R_{ij} &= \tilde{R}_{ij} +  \tilde{D}_k C^k {_{ij}} - \tilde{D}_i C^k {_{kj}} +C^k {_{ij}} C^l {_{lk}} - C^k {_{il}} C^l {_{kj}}  \\
&= \tilde{R}_{ij} + 4 \tilde{D}_i \tilde{D}_j (\ln \Psi) - 2 \tilde{D}_k \tilde{D}^k (\ln \Psi) \tilde{\gamma}_{ij}
-6 \tilde{D}_i\tilde{D}_j (\ln \Psi) \\
& \quad + 2 \bigg \{\delta^k {_i} \tilde{D}_j (\ln \Psi) + \delta^k {_j} \tilde{D}_i (\ln \Psi ) - \tilde{D}^k(\ln \Psi) \tilde{\gamma}_{ij} \bigg \}
 \times 6 \tilde{D}_k \ln \Psi \\
& \quad-4 \bigg \{\delta^k {_i} \tilde{D}_l (\ln \Psi) + \delta^k {_l} \tilde{D}_i (\ln \Psi ) - \tilde{D}^k(\ln \Psi) \tilde{\gamma}_{il} \bigg \} \times \\
& \qquad \quad \bigg \{\delta^l {_k} \tilde{D}_j (\ln \Psi) + \delta^l {_j} \tilde{D}_k (\ln \Psi ) - \tilde{D}^l(\ln \Psi) \tilde{\gamma}_{kj} \bigg \} \\
&= \tilde{R}_{ij} -2 \tilde{D}_i \tilde{D}_j (\ln \Psi) - 2 \tilde{D}_k \tilde{D}^k (\ln \Psi) \tilde{\gamma}_{ij} \\
& \quad + 12 \delta^k {_i} (\tilde{D}_j (\ln \Psi)(\tilde{D}_k (\ln \Psi) + 12 \delta^k {_j} (\tilde{D}_i \ln \Psi)(\tilde{D}_k \ln \Psi) \\
&\quad-12 (\tilde{D}^k \ln \Psi)(\tilde{D}_k \ln \Psi)\tilde{\gamma}_{ij} - 4  \delta^k {_i}  \delta^l {_k} (\tilde{D}_l \ln \Psi)
(\tilde{D}_j \ln \Psi) \\
& \quad- 4 \delta^k {_i}  \delta^l {_j} (\tilde{D}_l \ln \Psi) (\tilde{D}_k \ln \Psi)
+ 4 \delta^k {_i} (\tilde{D}_l \ln \Psi)(\tilde{D}^l \ln \Psi ) \tilde{\gamma}_{kj} \\
& \quad - 4 \delta^k {_l}  \delta^l {_k} (\tilde{D}_i \ln \Psi) (\tilde{D}_j \ln \Psi) - 4 \delta^k {_l}  \delta^l {_j} (\tilde{D}_i \ln \Psi) (\tilde{D}_k \ln \Psi)  \\
&\quad+ 4 \delta^k {_l} (\tilde{D}_i \ln \Psi)(\tilde{D}^l \ln \Psi ) \tilde{\gamma}_{kj}
+ 4 \delta^l {_k} (\tilde{D}^k \ln \Psi)(\tilde{D}_j \ln \Psi ) \tilde{\gamma}_{il} \\
&\quad + 4 \delta^l {_j} (\tilde{D}^k \ln \Psi)(\tilde{D}_k \ln \Psi ) \tilde{\gamma}_{il}
-4 (\tilde{D}^k \ln \Psi)(\tilde{D}^l \ln \Psi )\tilde{\gamma}_{il} \tilde{\gamma}_{kj} \, ,
\label{tran7}
\end{aligned}
\end{equation}
by collecting the identical terms of the equation (\ref{tran7}) in the each corresponding clusters, we get \emph{\textbf{conformal transformation of the Ricci tensor}} as
\begin{equation}
\begin{aligned}
 R_{ij} &= \tilde{R}_{ij}-2 \tilde{D}_i \tilde{D}_j (\ln \Psi) - 2 \tilde{D}_k \tilde{D}^k (\ln \Psi) \tilde{\gamma}_{ij}
+ (\tilde{D}_i \ln \Psi) (\tilde{D}_j \ln \Psi) \\ 
& \quad - 4 (\tilde{D}_k \ln \Psi )(\tilde{D}^k \ln \Psi )\tilde{\gamma}_{ij} \, . 
\label{tran8}
\end{aligned}
\end{equation}
\end{enumerate}

\subsubsection{Conformal Transformation of the Scalar Intrinsic Curvature}

Let us first take the trace of the equation (\ref{tran8}) with respect to the dual induced 3-metric $ \gamma^{ij} $:
\begin{equation}
\begin{aligned}
 R &=\gamma^{ij} R_{ij} \\
&= \Psi^{-4} \tilde{\gamma}^{ij} R_{ij} \\ 
&= \Psi^{-4} \bigg \{ \tilde{\gamma}^{ij} \tilde{R}_{ij}-2 \tilde{\gamma}^{ij} \tilde{D}_i \tilde{D}_j (\ln \Psi) - 2 \tilde{\gamma}^{ij}  \tilde{\gamma}_{ij} \tilde{D}_k \tilde{D}^k (\ln \Psi) \\
& \qquad \qquad + 4 \tilde{\gamma}^{ij} (\tilde{D}_i \ln \Psi) (\tilde{D}_j \ln \Psi)  
 -4 \tilde{\gamma}^{ij} \tilde{\gamma}_{ij} (\tilde{D}_k \ln \Psi )(\tilde{D}^k \ln \Psi ) \bigg \} \\ 
& =\Psi^{-4} \bigg \{ \tilde{R} - 8 \Big [ \tilde{D}_i \tilde{D}^i \ln \Psi + 
(\tilde{D}_i \ln \Psi )(\tilde{D}^i \ln \Psi ) \Big ] \bigg \} \, .
\label{tran9}
\end{aligned}
\end{equation}
Here we need to modify the term $ \tilde{D}_i \tilde{D}^i \ln \Psi $ of the equation (\ref{tran9}) :
\begin{equation}
\begin{aligned}
\tilde{D}_i \tilde{D}^i \ln \Psi = \tilde{D}_i \Big [ \frac{\tilde{D}^i \Psi}{\Psi} \Big ]
&= \Psi^{-1} \tilde{D}_i \tilde{D}^i \Psi - \Psi^{-2} \tilde{D}_i \Psi \tilde{D}^i \Psi \\
&= \Psi^{-1} \tilde{D}_i \tilde{D}^i \Psi -( \tilde{D}_i \ln \Psi )(\tilde{D}^i \ln \Psi ) \, .
\label{tran10}
\end{aligned}
\end{equation}
Thus, the substitution of the equation (\ref{tran10})  into the equation (\ref{tran9}) results in \emph{\textbf{the conformal transformation of the intrinsic scalar curvature}} of
\begin{equation}
 R = \Psi^{-4} \tilde{R} - 8 \Psi^{-5} \tilde{D}_i \tilde{D}^i \Psi \, .
\label{conf_scall}
\end{equation}

\subsubsection{Conformal Transformation of the Extrinsic Curvature}

\emph{Since the trace and traceless parts of the 3+1 Dynamical Einstein equation transform differently under conformal transformation, we need first to decompose the extrinsic curvature into the trace part and traceless part}.
\begin{enumerate}
\item The Extrinsic Curvature in terms of Trace and Traceless Parts

Now, the traceless part of the extrinsic curvature is defined as
\begin{equation}
\textbf{A} =\textbf{K}-\frac{1}{3} K \gamma \, ,
\end{equation}
such that  $ tr_{\gamma} \textbf{A} = \gamma^{ij} A_{ij} = 0 $. Therefore, the covariant and conravariant components of the extrinsic curvature can be rewritten in terms of the trace and traceless parts,
\begin{equation}
 K_{ij} = A_{ij} + \frac{1}{3} K \gamma_{ij} \quad  and \quad  K^{ij} = A^{ij} + \frac{1}{3} K \gamma^{ij} \, .
\label{yuh}
\end{equation}
\item Conformal Transformation of the Traceless Part

As in the conformal transformation of the induced 3-metric $ \gamma_{ij} $, the conformal transformation of the traceless part of the extrinsic curvature must be something like 
\begin{equation}
 A^{ij} = \Psi^\alpha \tilde{A}^{ij} \, .
\end{equation}
We will see that the choice $ \alpha = -4 $ gives us \emph{the evolution equations for the conformal factor $ \Psi $, the conformal 3-metric $ \tilde{\gamma}_{ij} $ and its dual. On the other hand, the choice $ \alpha = -10 $ gives the conformal form of the moment constraint equation.}
\begin{enumerate}
  \item For the $ 1^{st} $ choice of $ \alpha = -4 $ 

The Lie derivative of the induced 3-metric is given by the equation (\ref{mm}). Let us find what happens to it under the conformal transformation by using the equations (\ref{conf3}) and (\ref{yuh}):
\begin{equation}
\begin{aligned}
 \mathcal{L}_\textbf{m} \gamma_{ij} &= -2 N K_{ij} \\
\mathcal{L}_\textbf{m} \Big ( \Psi^4 \tilde{\gamma}_{ij} \Big )
&= -2 N \Big \{ A_{ij} + \frac{1}{3} K \gamma_{ij} \Big \} \\ 
\Psi^4 \mathcal{L}_\textbf{m} \tilde{\gamma}_{ij}  + \Big ( \mathcal{L}_\textbf{m} \Psi^4 \Big ) \tilde{\gamma}_{ij}
&= -2 N A_{ij} - \frac{2}{3} N K \gamma_{ij} \, ,
\end{aligned}
\end{equation}
then
\begin{equation}
\begin{aligned}
\Psi^4 \mathcal{L}_\textbf{m} \tilde{\gamma}_{ij} &= -2 N A_{ij} - \frac{2}{3} N K \gamma_{ij} 
- \Big ( \mathcal{L}_\textbf{m} \Psi^4 \Big ) \tilde{\gamma}_{ij} \\   
&= -2 N A_{ij} - \frac{2}{3} N K \Psi^4 \tilde{\gamma}_{ij} 
- \Big ( \mathcal{L}_\textbf{m} \Psi^4 \Big ) \tilde{\gamma}_{ij} \\
\mathcal{L}_\textbf{m} \tilde{\gamma}_{ij} &= -2 N \Psi^{-4} A_{ij} - \frac{2}{3} N K \tilde{\gamma}_{ij} 
- \mathcal{L}_\textbf{m} \Big (\ln \Psi^4 \Big ) \tilde{\gamma}_{ij} \\
&=-2 N \Psi^{-4} A_{ij} - \frac{2}{3} N K \tilde{\gamma}_{ij} 
- 4 \mathcal{L}_\textbf{m} \Big (\ln \Psi \Big ) \tilde{\gamma}_{ij} \, . 
\label{yuh2}
\end{aligned}
\end{equation}
Therefore, the equation (\ref{yuh2}) becomes
\begin{equation}
\mathcal{L}_\textbf{m} \tilde{\gamma}_{ij} = -2 N \Psi^{-4} A_{ij} - \frac{2}{3} \bigg \{ N K + 6 \mathcal{L}_\textbf{m} \Big (\ln \Psi \Big ) \bigg \} \tilde{\gamma}_{ij} \, . 
\label{yuh3}
\end{equation}
Since  the $ A_{ij} $ is traceless, let us multiply the equation (\ref{yuh3}) by the conformal dual 3-metric $ \tilde{\gamma}^{ij} $: 
\begin{equation}
\begin{aligned}
 \tilde{\gamma}^{ij} \mathcal{L}_\textbf{m} \tilde{\gamma}_{ij} &= -2 N \Psi^{-4}  \tilde{\gamma}^{ij} A_{ij}
 - \frac{2}{3} \bigg \{ N K + 6 \mathcal{L}_\textbf{m} \Big (\ln \Psi \Big ) \bigg \}  \tilde{\gamma}^{ij} \tilde{\gamma}_{ij} \\ 
&= - \frac{2}{3} \bigg \{ N K + 6 \mathcal{L}_\textbf{m} \Big (\ln \Psi \Big ) \bigg \} \times 3 \, , 
\end{aligned}
\end{equation}
so we get
\begin{equation}
 \tilde{\gamma}^{ij} \mathcal{L}_\textbf{m} \tilde{\gamma}_{ij}
 = -2 \bigg \{ N K + 6 \mathcal{L}_\textbf{m} \Big (\ln \Psi \Big ) \bigg \} \, .
\label{yuh4}
\end{equation}
Now, the variation of the determinant of an invertible matrix $ \mathcal{A} $ is given by
\begin{equation}
 \delta \Big ( \ln det \mathcal{A} \Big ) = tr \Big ( \mathcal{A}^{-1} \times \delta \mathcal{A} \Big ) \, .
\label{yuh5}
\end{equation}
Let us do the changes of $ \mathcal{A} \rightarrow \tilde{\gamma}_{ij} $ and $ \delta \rightarrow \mathcal{L}_\textbf{m} $ in (\ref{yuh5}) so it turns into the left hand side of the equation  (\ref{yuh4}). Then, the left hand side of the equation (\ref{yuh4}) becomes the Lie derivative of a scalar field along \textbf{m} which allows us to decompose the Lie derivative along \textbf{m} into the time derivative and the Lie derivative along the shift vector $ \beta $
\begin{equation}
 \tilde{\gamma}^{ij} \mathcal{L}_\textbf{m} \tilde{\gamma}_{ij} 
= \mathcal{L}_\textbf{m} \ln det (\tilde{\gamma}_{ij})
= \Big ( \frac{\partial}{\partial t} - \mathcal{L}_\beta \Big ) \ln det ( \tilde{\gamma}_{ij} ) \, .  
\label{yuh6}
\end{equation}
Because the time derivative of the scalar field $ \ln det ( \tilde{\gamma}_{ij} ) $ vanishes, the equation (\ref{yuh6}) reduces to 
\begin{equation}
\begin{aligned}
 \mathcal{L}_\textbf{m} \ln det (\tilde{\gamma}_{ij})  
&= - \mathcal{L}_\beta \ln det ( \tilde{\gamma}_{ij} ) 
= - \tilde{\gamma}^{ij} \mathcal{L}_\beta \tilde{\gamma}_{ij} \\
&=- \tilde{\gamma}^{ij} \Big \{ \beta^k \tilde{D}_k \tilde{\gamma}_{ij} + \tilde{\gamma}_{kj} \tilde{D}_i \beta^k
+ \tilde{\gamma}_{ik} \tilde{D}_j \beta^k \Big \} \\
& =- \tilde{\gamma}^{ij} \Big \{ \tilde{\gamma}_{kj} \tilde{D}_i \beta^k+ \tilde{\gamma}_{ik} \tilde{D}_j \beta^k \Big \} \\
&= -\delta^i {_k} \tilde{D}_i \beta^k - \delta^j {_k} \tilde{D}_j \beta^k \\
& = -2 \tilde{D}_i \beta^i \, .
\label{yuh7}
\end{aligned}
\end{equation}
Furthermore, by substituting the result of (\ref{yuh7}) into the equation (\ref{yuh4}), we get \emph{\textbf{the evolution equation for $ \Psi $ under conformal transformation}} as
\begin{equation}
 \Big ( \frac{\partial}{\partial t} - \mathcal{L}_\beta \Big ) \ln \Psi 
= \frac{1}{6} \Big ( \tilde{D}_i \beta^i - N K \Big ) \, .
\label{yuh8}
\end{equation}
By inserting the equation (\ref{yuh8}) into the equation (\ref{yuh3}), it becomes
\begin{equation}
 \mathcal{L}_\textbf{m} \tilde{\gamma}_{ij}
 = -2 N \Psi^{-4} A_{ij} - \frac{2}{3} \tilde{D}_k \beta^k \tilde{\gamma}_{ij} \, .
\label{yuh9}
\end{equation}
For consistency in the equation (\ref{yuh9}), we must have such a \emph{\textbf{conformal transformation of the traceless part of $ K_{ij} $}} as
\begin{equation}
 \tilde{A}_{ij} = \Psi^{-4} A_{ij} \, ,
\label{alphaa}
\end{equation}
which says that the corresponding \emph{\textbf{$ \alpha $ must be -4}}. Therefore, we find \emph{\textbf{the evolution equation for conformal metric during the conformal transformation}} as 
\begin{equation}
 \Big ( \frac{\partial}{\partial t} - \mathcal{L}_\beta \Big ) \tilde{\gamma}_{ij}
= -2 N \tilde{A}_{ij} - \frac{2}{3} \tilde{D}_k \beta^k \tilde{\gamma}_{ij} \, .
\label{yuh10}
\end{equation}
Now, the conformal transformation for the contravariant component of the traceless part of $ K^{ij} $, $ A^{ij} $, is obtained by
\begin{equation}
\begin{aligned}
 \tilde{A}^{ij} &= \tilde{\gamma}^{ik} \tilde{\gamma}^{jl} \tilde{A}_{ij} \\ 
&= \Psi^4 \gamma^{ik} \Psi^4 \gamma^{jl} \Psi^{-4} A_{kl} \\
& = \Psi^4 \gamma^{ik} \gamma^{jl} A_{kl} \\    
&= \Psi^4 A^{ij} \, .
\label{alphaaa}
\end{aligned}
\end{equation}
Finally, let us see \emph{\textbf{how the dual conformal 3-metric $ \tilde{\gamma}^{ij} $ evolves under the conformal transformation}} by starting from the equation (\ref{yuh10}) : 
\begin{equation}
\begin{aligned}
\tilde{\gamma}^{ik} \tilde{\gamma}^{jl} \mathcal{L}_\textbf{m} \tilde{\gamma}_{kl}
&= -2 N \tilde{A}^{ij} - \frac{2}{3} \tilde{D}_k \beta^k \tilde{\gamma}^{ij} \\
\tilde{\gamma}^{ik} \bigg \{ \mathcal{L}_\textbf{m} \Big ( \tilde{\gamma}^{jl} \tilde{\gamma}_{kl} \Big ) 
- \tilde{\gamma}_{kl} \mathcal{L}_\textbf{m} \tilde{\gamma}^{jl} \bigg \} 
&= -2 N \tilde{A}^{ij} - \frac{2}{3} \tilde{D}_k \beta^k \tilde{\gamma}^{ij} \\
\tilde{\gamma}^{ik} \bigg \{ \mathcal{L}_\textbf{m} \delta^j {_k} 
- \tilde{\gamma}_{kl} \mathcal{L}_\textbf{m} \tilde{\gamma}^{jl} \bigg \} 
&= -2 N \tilde{A}^{ij} - \frac{2}{3} \tilde{D}_k \beta^k \tilde{\gamma}^{ij} \\ 
 - \tilde{\gamma}^{ik} \tilde{\gamma}_{kl} \mathcal{L}_\textbf{m} \tilde{\gamma}^{jl}  
&= -2 N \tilde{A}^{ij} - \frac{2}{3} \tilde{D}_k \beta^k \tilde{\gamma}^{ij} \\
 \delta^i {_l} \mathcal{L}_\textbf{m} \tilde{\gamma}^{jl}  
&= 2 N \tilde{A}^{ij} + \frac{2}{3} \tilde{D}_k \beta^k \tilde{\gamma}^{ij} \, .
\end{aligned}
\end{equation}
Thus, we get 
\begin{equation}
 \Big ( \frac{\partial}{\partial t} - \mathcal{L}_\beta \Big ) \tilde{\gamma}^{ij}
 = 2 N \tilde{A}^{ij} + \frac{2}{3} \tilde{D}_k \beta^k \tilde{\gamma}^{ij} \, .
\end{equation}
\item  For the $ 2^{nd} $ Choice of  $ \alpha = -10 $

In this case, we will start with the decomposed form of the conravariant extrinsic curvature $ K^{ij}  $ [see the equation (\ref{yuh})] and take the divergence of it. That's,
\begin{equation}
 K^{ij} = A^{ij} + \frac{1}{3} K \gamma^{ij} \Longrightarrow D_j K^{ij} = D_j A^{ij}+ \frac{1}{3} D^i K \, .
\label{yuh11}
\end{equation}
Now, the equations of (\ref{tran1}), (\ref{kpn}) and (\ref{tran5}) provide to rewrite the term $ D_j A^{ij} $ of the equation (\ref{yuh11}) in terms of the conformal quantities,
\begin{equation}
\begin{aligned}
 D_j A^{ij} &= \tilde{D}_j A^{ij} + C^i {_{jk}} A^{kj} + C^j {_{jk}} A^{ik} \\
&= \tilde{D}_j A^{ij} + 2 \Big \{ \delta^i {_j} \tilde{D}_k \ln \Psi + \delta^i {_k} \tilde{D}_j \ln \Psi
- \tilde{D}^i \ln \Psi \tilde{\gamma}_{jk} \Big \} A^{kj} \\
& \quad + 6 \tilde{D}_k \ln \Psi A^{ik} \\ 
&= \tilde{D}_j A^{ij} + 2 \delta^i {_j} A^{kj} \tilde{D}_k \ln \Psi + 2 \delta^i {_k} A^{kj} \tilde{D}_j \ln \Psi
- 2 \tilde{\gamma}_{jk} A^{kj} \tilde{D}^i \ln \Psi \\
& \qquad + 6 \tilde{D}_k \ln \Psi A^{ik} \\
& = \tilde{D}_j A^{ij} + 2 A^{ik} \tilde{D}_k \ln \Psi + 2 A^{ij} \tilde{D}_j \ln \Psi
- 2 \Psi^{-4} \gamma_{jk} A^{kj} \tilde{D}^i \ln \Psi \\
& \qquad + 6 \tilde{D}_k \ln \Psi A^{ik} \, .
\label{yuh12}
\end{aligned}
\end{equation}
Because $ A^{kj} $ is traceless, the related term vanishes. Then, (\ref{yuh12}) turns into
\begin{equation}
\begin{aligned}
 D_j A^{ij} &= \tilde{D}_j A^{ij} + 2 A^{ik} \tilde{D}_k \ln \Psi + 2 A^{ij} \tilde{D}_j \ln \Psi
 + 6 \tilde{D}_k \ln \Psi A^{ik} \, , \\
& \qquad \quad \qquad \quad k \rightarrow j \qquad \qquad \qquad  \qquad  k \rightarrow j
\end {aligned}
\end{equation}
with the given change of indices, we get
\begin{equation}
\begin{aligned}
 D_j A^{ij} &= \tilde{D}_j A^{ij} + 10 A^{ij} \tilde{D}_j \ln \Psi \\
&= \Psi^{-10} \tilde{D}_j \Big ( \Psi^{10} A^{ij} \Big ) \, .
\label{yuh13}
\end{aligned}
\end{equation}
Notice that for consistency of   
\begin{equation}
 \hat{A}^{ij} = \Psi^{10} A^{ij} \, ,
\end{equation}
the corresponding \emph{\textbf{$ \alpha $ must be -10}}.

Finally, let us insert the equation (\ref{11}) into \emph{the momentum constraint equation (\ref{momen_cost})}:
\begin{equation}
D_j A^{ij}+ \frac{1}{3} D^i K -D^i K = 8 \pi p^i \Longrightarrow  D_j A^{ij}- \frac{2}{3} D^i K = 8 \pi p^i \, .
\label{yuh14}
\end{equation}
And let us insert the conformal form of the $  D_j A^{ij} $ (\ref{yuh13}) into the equation (\ref{yuh14}):
\begin{equation}
  \Psi^{-10} \tilde{D}_j \Big ( \Psi^{10} A^{ij} \Big ) - \frac{2}{3} D^i K = 8 \pi p^i \, .
\label{yuh15}
\end{equation}
Now, the modification of 
\begin{equation}
\begin{aligned}
 D_j K = \tilde{D}_j K  \Longrightarrow 
\gamma^{ij} D_j K = D^i K &= \gamma^{ij} \tilde{D}_j K \\
&= \Psi^{-4} \tilde{\gamma}^{ij} \tilde{D}_j K \\
&= \Psi^{-4} \tilde{D}^i K 
\end{aligned}
\end{equation}
provides us to rewrite the equation (\ref{yuh15}) as
\begin{equation}
 \Psi^{-10} \tilde{D}_j \Big ( \Psi^{10} A^{ij} \Big ) - \frac{2}{3} \Psi^{-4} \tilde{D}^i K = 8 \pi p^i \, .
\end{equation}
Thus, the \emph{\textbf{the conformal transformation of the momentum constraint}} is
\begin{equation}
\tilde{D}_j \Big ( \hat{A}^{ij} \Big ) - \frac{2}{3} \Psi^6 \tilde{D}^i K = 8 \pi \Psi^{10} p^i \, .
\label{moment_cnb}
\end{equation}
Finally, let us find the conformal transformation of $ \hat{A}_{ij} $
\begin{equation}
\begin{aligned}
 \hat{A}_{ij} &= \tilde{\gamma}^{ik} \tilde{\gamma}^{jl} \hat{A}^{kl} \\
&= \Big ( \Psi^{-4} \gamma^{ik} \Big ) \Big ( \Psi^{-4} \gamma^{jl} \Big ) \Big ( \Psi^{10} A^{kl} \Big ) \\
&= \Psi^2 \gamma^{ik} \gamma^{jl} A^{kl} \, , \\
&=\Psi^2 A_{ij}
\end{aligned}
\end{equation}
then, we get
\begin{equation}
 \hat{A}_{ij} = \Psi^2 A_{ij} \, .
\end{equation}
\end{enumerate}
\end{enumerate}

\subsection{The Conformal Form of the 3+1 Dimensional Einstein System}

As we said before, the trace and traceless part of the 3+1 dynamical Einstein equation (\ref{dynm1}) transform separately under conformal transformation. Therefore, we need to first deduce the related decomposition of it. And then, we will be ready to construct their conformal forms.

\subsubsection{Trace and Traceless Parts of the 3+1 Dynamical Einstein Equation}

\begin{enumerate}
 \item  Trace Part of the 3+1 Dynamical Einstein Equation

Let us take the trace of the 3+1 dynamical Einstein equation (\ref{dynm1}) with respect to $ \gamma^{ij} $ 
\begin{equation}
\gamma^{ij} \mathcal{L}_\textbf{m} K_{ij} = -D_i D^i N + N  \bigg \{ R 
+ K^2 - 2K_{ij} K^{ij} + 4 \pi \Big ( S-3 E \Big ) \bigg \} \, . 
\label{trace_part}
\end{equation}
From the equations (\ref{llll}) and (\ref{trace_part}), we have 
\begin{equation}
\begin{aligned}
\mathcal{L}_\textbf{m} K &= \gamma^{ij} \mathcal{L}_\textbf{m} K_{ij} + 2 N K_{ij} K^{ij} \\
&= -D_i D^i N + N  \bigg \{ R + K^2 - 2K_{ij} K^{ij} + 4 \pi \Big ( S-3 E \Big ) \bigg \} + 2 N K_{ij} K^{ij} \\
&=  -D_i D^i N + N  \bigg \{ R + K^2 + 4 \pi \Big ( S-3 E \Big ) \bigg \} \, .
\label{tracee}
\end{aligned}
\end{equation}
Now, the Hamiltonian constraint equation (\ref{gh})
\begin{equation}
R + K^2 - K_{ij} K^{ij} = 16 \pi E \Longrightarrow  R + K^2 = 16 \pi E + K_{ij} K^{ij} \, ,
\end{equation}
by substituting the previous modification into (\ref{tracee}), we obtain \emph{\textbf{the trace part of the dynamical 3+1 Einstein equation}} as
\begin{equation}
\mathcal{L}_\textbf{m} K = -D_i D^i N + N  \bigg \{ K_{ij} K^{ij} + 4 \pi \Big ( S + E \Big ) \bigg \} \, .
\label{31}
\end{equation}
\item Traceless Part of the 3+1  Dynamical Einstein Equation

Now, let us now decompose the left hand side of the 3+1 dynamical Einstein equation (\ref{dynm1}) by using the equation 
(\ref{yuh})
\begin{equation}
\begin{aligned}
 \mathcal{L}_\textbf{m} K_{ij} &= \mathcal{L}_\textbf{m} \Big ( A_{ij} + \frac{1}{3} K \gamma_{ij} \Big ) \\
&= \mathcal{L}_\textbf{m}  A_{ij} +  \frac{1}{3} \Big (\mathcal{L}_\textbf{m} K \Big ) \gamma_{ij} 
+ \frac{1}{3} K \Big ( \mathcal{L}_\textbf{m} \gamma_{ij} \Big ) \\
&= \mathcal{L}_\textbf{m}  A_{ij} +  \frac{1}{3} \Big (\mathcal{L}_\textbf{m} K \Big ) \gamma_{ij} 
- \frac{2N}{3} K K_{ij} \, ,
\end{aligned}
\end{equation}
where we used (\ref{mm}). Therefore, the Lie derivative of the traceless part $ A_{ij} $ along \textbf{m} is
\begin{equation}
 \mathcal{L}_\textbf{m}  A_{ij} = \mathcal{L}_\textbf{m} K_{ij}-\frac{1}{3} \Big (\mathcal{L}_\textbf{m} K \Big ) \gamma_{ij}
+ \frac{2N}{3} K K_{ij} \, .
\label{traceless1}
\end{equation}
Notice that the first term on the right hand side of (\ref{traceless1}) is the 3+1 dynamical Einstein equation (\ref{dynm1}) and the second term is the trace part of it (\ref{tracee}). Then, the explicit form of (\ref{traceless1}) is
\begin{equation}
\begin{aligned}
 \mathcal{L}_\textbf{m}  A_{ij} &=  -D_i D_j N \\
& \quad + N  \bigg \{ R_{ij} 
+ K K_{ij}- 2K_{ik} K^k {_j} + 4 \pi \Big [(S-E)\gamma_{ij} - 2 S_{ij}\Big ] \bigg \}  \\
& \quad -\frac{1}{3} \bigg \{-D_k D^k N + N  \bigg \{ R + K^2 + 4 \pi \Big ( S-3 E \Big ) \bigg \} \bigg \} \gamma_{ij} \\
& \quad + \frac{2N}{3} K K_{ij} \\
&=  -D_i D_j N \\
& \quad + N  \bigg \{ R_{ij}+\frac{5}{3} K K_{ij}- 2K_{ik} K^k {_j}
- \frac{1}{3} K^2 \gamma_{ij} - 8 \pi \Big ( S_{ij}-\frac{1}{3} S \gamma_{ij} \Big ) \bigg \} \\
 & \quad + \frac{1}{3} \bigg \{ D_k D^k N - N R \bigg \} \gamma_{ij} \, .
\label{wow11}
\end{aligned}
\end{equation}
Since we wish a totally traceless equation, we need to get rid of the terms that contain the \textbf{K} :
\begin{equation}
\begin{aligned}
 \frac{5}{3} K K_{ij}- 2K_{ik} K^k {_j}- \frac{1}{3} K^2 \gamma_{ij} 
&= \frac{5}{3} K \Big ( A_{ij} +\frac{1}{3} K  \gamma_{ij} \Big ) \\
& \quad - 2 \Big ( A_{ik} +\frac{1}{3} K  \gamma_{ik} \Big ) \Big ( A^k {_j} +\frac{1}{3} K  \delta^k {_j} \Big ) \\
& \quad - \frac{1}{3} K^2 \gamma_{ij} \\
& = \frac{1}{3} K A_{ij} - 2 A_{ik} A^k {_j} \, .
\label{wow21}
\end{aligned}
\end{equation}
Thus, by substituting (\ref{wow21}) into (\ref{wow11}), we obtain \emph{\textbf{the traceless part of the dynamical 3+1 Einstein equation}} as
\begin{equation}
\begin{aligned}
\mathcal{L}_\textbf{m}  A_{ij} &=  -D_i D_j N + N  \bigg \{ R_{ij} + \frac{1}{3} K A_{ij} - 2 A_{ik} A^k {_j}
  - 8 \pi \Big ( S_{ij}-\frac{1}{3} S \gamma_{ij} \Big ) \bigg \} \\
 &\qquad \qquad \qquad + \frac{1}{3} \bigg \{ D_k D^k N - N R \bigg \} \gamma_{ij} \, .
\label{trclss}
\end{aligned}
\end{equation}
\end{enumerate}
After constructing the conformal transformation of the fundamental tools, it is time to construct the conformal transformation of the 3+1 Einstein equation :

\subsubsection{Conformal Decomposition of the Trace and Traceless Parts of the Dynamical 3+1 Einstein Equation}

We are trying to construct the corresponding time evolution equations. Therefore, we need use the $ \alpha = -4 $ case.
\begin{enumerate}
\item  Conformal Form of the Trace Part of the 3+1 Dynamical Einstein Equation 

The trace part of the 3+1 dynamical Einstein equation (\ref{31}) is
\begin{equation}
\Big ( \frac{\partial}{\partial t} - \mathcal{L}_\beta \Big ) K = -D_i D^i N + N  \bigg \{ K_{ij} K^{ij} + 4 \pi \Big ( S + E \Big ) \bigg \} \, .
\label{12345}
\end{equation}
For simplicity, let us find only the conformal form of the terms $ D_i D^i N $ and  $  K_{ij} K^{ij} $ separately:
\begin{enumerate}
 \item \emph{The conformal form of the term $ D_i D^i N $}

We have found that the conformal transformation of the divergence of a vector \textbf{v} (\ref{diver}) is given by
\begin{equation}
 D_i v^i = \Psi^{-6} \tilde{D}_i \Big ( \Psi^6 v^i \Big ) \, .
\label{diver1}
\end{equation}
Since the gradient of a scalar field is a vector field, we will take $ v^i = D^i N $,
\begin{equation}
 v^i = D^i N = \gamma^{ij} D_j N = \Psi^{-4} \tilde{\gamma}^{ij} \tilde{D}_j N = \Psi^{-4} \tilde{D}^i N \, .
\label{diver2}
\end{equation}
Let us substitute (\ref{diver2})  into  (\ref{diver1}),
\begin{equation}
\begin{aligned}
 D_i D^i N &= \Psi^{-6} \tilde{D}_i \Big ( \Psi^6 D^i N \Big ) \\
&= \Psi^{-6} \tilde{D}_i \Big ( \Psi^6 \Big [ \Psi^{-4} \tilde{D}^i N \Big ] \Big ) \\
&= \Psi^{-6} \tilde{D}_i \Big ( \Psi^2 \tilde{D}^i N  \Big ) \\ 
&=\Psi^{-6} \tilde{D}_i \Big ( \Psi^2 \tilde{D}_i  \tilde{D}^i N + 2 \Psi \tilde{D}_i \Psi\tilde{D}^i N  \Big ) \\
&=\Psi^{-4} \Big ( \tilde{D}_i  \tilde{D}^i N + 2 \tilde{D}_i \ln \Psi\tilde{D}^i N  \Big ) \, .
\label{321}
\end{aligned}
\end{equation}
\item \emph{The explicit decomposition of the term $ K_{ij} K^{ij} $}

Let us use decomposed form of the extrinsic curvature (\ref{yuh}):
\begin{equation}
\begin{aligned}
 K_{ij} K^{ij} &= \Big ( A_{ij} + \frac{1}{3} K \gamma_{ij} \Big )\Big ( A^{ij} + \frac{1}{3} K \gamma^{ij} \Big ) \\  
&= A_{ij} A^{ij} + \frac{1}{3} K \gamma_{ij} A^{ij} + \frac{1}{3} K \gamma_{ij} A^{ij} + \frac{1}{9} K^2 \gamma_{ij} \gamma^{ij} \, .
\label{second}
\end{aligned}
\end{equation}
Because the trace of $ A_{ij} $ is zero, (\ref{second}) reduces to
\begin{equation}
\begin{aligned}
 K_{ij} K^{ij} &= A_{ij} A^{ij} + \frac{1}{3} K^2 \\ 
&=\Big (\Psi^4 \tilde{A}_{ij} \Big ) \Big ( \Psi^{-4} \tilde{A}^{ij} \Big ) + \frac{1}{3} K^2 \\
&= \tilde{A}_{ij} \tilde{A}^{ij}+ \frac{1}{3} K^2 \, ,
\label{123}
\end{aligned}
\end{equation}
where we used the conformal transformations (\ref{alphaa}) and (\ref{alphaaa}).
\end{enumerate}
Finally, by substituting (\ref{321}) and (\ref{123}) into the fundamental equation (\ref{12345}), we will get the \emph{\textbf{conformal form of the trace part of the 3+1 dynamical Einstein equation}} as
\begin{equation}
\begin{aligned}
\Big ( \frac{\partial}{\partial t} - \mathcal{L}_\beta \Big ) K 
&= -\Psi^{-4} \Big ( \tilde{D}_i  \tilde{D}^i N + 2 \tilde{D}_i \ln \Psi\tilde{D}^i N  \Big ) \\
& \quad + N \Big \{ \tilde{A}_{ij} \tilde{A}^{ij}+ \frac{1}{3} K^2 + 4 \pi \Big ( E + S \Big ) \Big \} \, .
\end{aligned}
\end{equation}
\item Conformal Form of the Traceless Part of the 3+1 Dynamical Einstein Equation

The traceless dynamical 3+1 Einstein equation (\ref{trclss}) is 
\begin{equation}
\begin{aligned}
\mathcal{L}_\textbf{m}  A_{ij} &=  -D_i D_j N + N  \bigg \{ R_{ij} + \frac{1}{3} K A_{ij} - 2 A_{ik} A^k {_j}
  - 8 \pi \Big ( S_{ij}-\frac{1}{3} S \gamma_{ij} \Big ) \bigg \} \\
 &\qquad \qquad \qquad + \frac{1}{3} \bigg \{ D_k D^k N - N R \bigg \} \gamma_{ij} \, .
\label{conf_trcls}
\end{aligned}
\end{equation}
First, let us find the conformal form of the terms $ \mathcal{L}_\textbf{m}  A_{ij} $ and $ D_i D_j N $ 
\begin{enumerate}
 \item \emph{The conformal form of $ \mathcal{L}_\textbf{m}  A_{ij} $ }

\begin{equation}
\begin{aligned}
\mathcal{L}_\textbf{m}  A_{ij} &= \mathcal{L}_\textbf{m} \Big (\Psi^4 \tilde{A}_{ij} \Big ) \\
&=  \Psi^4 \mathcal{L}_\textbf{m} \tilde{A}_{ij} + 4 \Psi^3 \Big ( \mathcal{L}_\textbf{m} \Psi \Big ) \tilde{A}_{ij}  \\
&=  \Psi^4 \mathcal{L}_\textbf{m} \tilde{A}_{ij} + 4 \Psi^4 \Big ( \mathcal{L}_\textbf{m} \ln \Psi \Big ) \tilde{A}_{ij} \\
&= \Psi^4 \mathcal{L}_\textbf{m} \tilde{A}_{ij} + 4 \Psi^4  \Big ( \frac{1}{6} \Big [ \tilde{D}_k \beta^k
-N K \Big ] \Big )\tilde{A}_{ij} \\
&= \Psi^4 \bigg \{ \mathcal{L}_\textbf{m} \tilde{A}_{ij} +  \frac{2}{3} \Big ( \tilde{D}_k \beta^k
-N K \Big ) \tilde{A}_{ij} \bigg \} \, ,
\end{aligned}
\end{equation}
where we used \emph{the conformal evolution equation of $ \Psi $ (\ref{yuh8}) and the conformal transformations (\ref{alphaa}) and (\ref{alphaaa})}. Finally, let us use the equation (\ref{trclss}) of $ \mathcal{L}_\textbf{m}  A_{ij} $ in the following equation 
\begin{equation}
\begin{aligned} 
\mathcal{L}_\textbf{m} \tilde{A}_{ij} &= \Psi^{-4} \mathcal{L}_\textbf{m}  A_{ij} - \frac{2}{3} \Big ( \tilde{D}_k \beta^k
-N K \Big ) \tilde{A}_{ij} \\
&= \Psi^{-4} \bigg \{ -D_i D_j N \\
& \quad + N  \bigg ( R_{ij} + \frac{1}{3} K A_{ij} - 2 A_{ik} A^k {_j}
- 8 \pi \Big [ S_{ij}-\frac{1}{3} S \gamma_{ij} \Big ] \bigg ) \\
& \quad + \frac{1}{3} \bigg ( D_k D^k N - N R \bigg ) \gamma_{ij} \bigg \} 
- \frac{2}{3} \Big ( \tilde{D}_k \beta^k
-N K \Big ) \tilde{A}_{ij} \, .
\label{miyaw1}
\end{aligned}
\end{equation}
\item \emph{The conformal form of $  D_i D_j N $ }

\begin{equation}
\begin{aligned}
 D_i D_j N &= D_i \tilde{D}_j N \\ 
&= \tilde{D}_i \tilde{D}_j N - C^k {_{ij}} \tilde{D}_k N \\
&= \tilde{D}_i \tilde{D}_j N - 2 \Big \{ \delta^k {_i} \tilde{D}_j \ln \Psi + \delta^k {_j} \tilde{D}_i \ln \Psi
- \tilde{D}^k \ln \Psi \tilde{\gamma}_{ij} \Big \} \times \tilde{D}_k N \\
&= \tilde{D}_i \tilde{D}_j N - 2 \Big \{ \tilde{D}_i N  \tilde{D}_j \ln \Psi + \tilde{D}_j N \tilde{D}_i \ln \Psi
-\tilde{D}_k N  \tilde{D}^k \ln \Psi \tilde{\gamma}_{ij} \Big \} \, . 
\label{miyaw2}
\end{aligned}
\end{equation}
\end{enumerate}
Finally, let us substitute the equations (\ref{tran8}), (\ref{321}), (\ref{miyaw1}) and (\ref{miyaw2}) into the equation of (\ref{conf_trcls})
\begin{equation}
 \begin{aligned}
\mathcal{L}_\textbf{m} \tilde{A}_{ij} &= \Psi^{-4} \bigg \{ - \bigg ( \tilde{D}_i \tilde{D}_j N - 
2 \Big \{ \tilde{D}_i N  \tilde{D}_j \ln \Psi + \tilde{D}_j N \tilde{D}_i \ln \Psi
-\tilde{D}_k N  \tilde{D}^k \ln \Psi \tilde{\gamma}_{ij} \Big \}  \bigg ) \\
& \qquad \qquad + N \bigg \{ \bigg (  \tilde{R}_{ij}-2 \tilde{D}_i \tilde{D}_j (\ln \Psi) - 2 \tilde{D}_k \tilde{D}^k (\ln \Psi) \tilde{\gamma}_{ij} \\
& \qquad \qquad \qquad \qquad + 4 (\tilde{D}_i \ln \Psi) (\tilde{D}_j \ln \Psi)  
- 4 (\tilde{D}_k \ln \Psi )(\tilde{D}^k \ln \Psi )\tilde{\gamma}_{ij} \bigg ) \\
& \qquad \qquad \quad + \frac{1}{3} K \Big [ \Psi^4 \tilde{A}_{ij} \Big ] -2 \Psi^4 \tilde{\gamma}^{kl} \tilde{A}_{ik} \tilde{A}_{jl} 
- 8 \pi \Big [ S_{ij} -\frac{1}{3} S \Psi^4 \tilde{\gamma}_{ij} \Big ] \bigg \} \\
& \qquad \qquad + \frac{1}{3} \bigg \{ \Psi^{-4} \Big ( \tilde{D}_k  \tilde{D}^k N + 2 \tilde{D}_k \ln \Psi\tilde{D}^k N  \Big ) \\
& \qquad \qquad \qquad - N \Psi^{-4} \Big ( \tilde{R} - 8 \Big [ \tilde{D}_k \tilde{D}^k \ln \Psi  
 + (\tilde{D}_k \ln \Psi ) 
(\tilde{D}^k \ln \Psi ) \Big ] \Big ) \bigg \} \Psi^4 \tilde{\gamma}^{ij} \bigg \} \\ 
& \qquad - \frac{2}{3} \Big [ \tilde{D}_k \beta^k-N K \Big ] \tilde{A}_{ij} \, .
\end{aligned}
\end{equation}
After some algebra, we reach \emph{\textbf{the conformal transformation of the traceless part of the 3+1 dynamical Einstein equation}} as
\begin{equation}
 \begin{aligned}
\mathcal{L}_\textbf{m} \tilde{A}_{ij} &= -\frac{2}{3} \tilde{D}_k \beta^k \tilde{A}_{ij} 
+ N \bigg \{ K \tilde{A}_{ij} - 2 \tilde{\gamma}^{kl} \tilde{A}_{ik} \tilde{A}_{jl} 
-8 \pi \Big [ \Psi^{-4} S_{ij} -\frac{1}{3} S \tilde{\gamma}_{ij} \Big ] \bigg \} \\
& \quad+ \Psi^{-4} \bigg \{ - \tilde{D}_i \tilde{D}_j N + 2 \tilde{D}_i N  \tilde{D}_j \ln \Psi + 2 \tilde{D}_j N  \tilde{D}_i \ln \Psi \\       
&\quad \qquad \qquad+ \frac{1}{3} \Big [\tilde{D}_k \tilde{D}^k N -4 \tilde{D}_k \ln \Psi \tilde{D}^k N \Big ] \tilde{\gamma}_{ij} \\
&\qquad \qquad \quad + N \Big [\tilde{R}_{ij} - \frac{1}{3} \tilde{R} \tilde{\gamma}_{ij}-2 \tilde{D}_i \tilde{D}_j \ln \Psi
+ 4 \tilde{D}_i \ln \Psi \tilde{D}_j \ln \Psi \\
& \qquad \qquad \qquad \qquad+\frac{2}{3} \Big ( \tilde{D}_k  \tilde{D}^k \ln \Psi - 2 \tilde{D}_k \ln \Psi\tilde{D}^k \ln \Psi \Big ) \tilde{\gamma}_{ij} \Big ] \bigg \} \, . 
\end{aligned}
\end{equation}
\end{enumerate}

\subsubsection{The Conformal Transformation of the Hamiltonian Constraint}

The Hamiltonian Constraint equation (\ref{gh}) is
\begin{equation}
 R + K^2 -K_{ij} K^{ij} = 16 \pi E \, .
\label{rto}
\end{equation}
Now, let us substitute the conformal transformation of $ R $ (\ref{conf_scall}) and the decomposed form of $ K_{ij} K^{ij} $ (\ref{123}) into the Hamiltonian Constraint equation (\ref{rto}),
\begin{equation}
\Psi^{-4} \tilde{R} -8 \Psi^{-5} \tilde{D}_i \tilde{D}^i \Psi + K^2 -\tilde{A}_{ij} \tilde{A}^{ij} -\frac{1}{3} K^2 = 16 \pi E \, .
\label{hamilt}
\end{equation}
From the equation (\ref{hamilt}), we get \emph{\textbf{the conformal transformation of the Hamiltonian constraint}} as
\begin{equation}
 \tilde{D}_i \tilde{D}^i \Psi -\frac{1}{8}\tilde{R} \Psi + \Big \{\frac{1}{8}\tilde{A}_{ij}\tilde{A}^{ij}-\frac{1}{12} K^2 + 2 \pi E \Big \} \Psi^5 =0  \, .             
\label{fdrs}
\end{equation}
Due to the relation of $ \tilde{A}_{ij}\tilde{A}^{ij}= \Psi^{-12} \hat{A}_{ij} \hat{A}^{ij} $, the equation (\ref{fdrs}) becomes
\begin{equation}
 \tilde{D}_i \tilde{D}^i \Psi -\frac{1}{8}\tilde{R} \Psi + \frac{1}{8}\hat{A}_{ij}\hat{A}^{ij} \Psi^{-7}
+ \Big \{2 \pi E -\frac{1}{12} K^2 \Big \} \Psi^5 =0  \, ,             
\end{equation}
which is known as \emph{\textbf{Lichnerowicz Equation}} \cite{Lichnerowicz:1939vy} \cite{Lichnerowicz:1982}.

Thus, \emph{\textbf{the conformal transformation of the 3+1 dimensional Einstein system}} \cite{Gourgoulhon:2007ue} can be summarized as,
\begin{equation}
 \Big ( \frac{\partial}{\partial t} - \mathcal{L}_\beta \Big ) \ln \Psi 
= \frac{1}{6} \Big ( \tilde{D}_i \beta^i - N K \Big ) \, ,
\end{equation}
\begin{equation}
 \Big ( \frac{\partial}{\partial t} - \mathcal{L}_\beta \Big ) \tilde{\gamma}_{ij}
= -2 N \tilde{A}_{ij} - \frac{2}{3} \tilde{D}_k \beta^k \tilde{\gamma}_{ij} \, ,
\end{equation}
\begin{equation}
\begin{aligned}
\Big ( \frac{\partial}{\partial t} - \mathcal{L}_\beta \Big ) K 
&= -\Psi^{-4} \Big ( \tilde{D}_i  \tilde{D}^i N + 2 \tilde{D}_i \ln \Psi\tilde{D}^i N  \Big ) \\
& \quad + N \Big \{ \tilde{A}_{ij} \tilde{A}^{ij}+ \frac{1}{3} K^2 + 4 \pi \Big ( E + S \Big ) \Big \} \, ,
\end{aligned}
\end{equation}

\begin{equation}
 \begin{aligned}
\mathcal{L}_\textbf{m} \tilde{A}_{ij} &= -\frac{2}{3} \tilde{D}_k \beta^k \tilde{A}_{ij} 
+ N \bigg \{ K \tilde{A}_{ij} - 2 \tilde{\gamma}^{kl} \tilde{A}_{ik} \tilde{A}_{jl} 
-8 \pi \Big [ \Psi^{-4} S_{ij} -\frac{1}{3} S \tilde{\gamma}_{ij} \Big ] \bigg \} \\
& \quad+ \Psi^{-4} \bigg \{ - \tilde{D}_i \tilde{D}_j N + 2 \tilde{D}_i N  \tilde{D}_j \ln \Psi + 2 \tilde{D}_j N  \tilde{D}_i \ln \Psi \\       
&\quad \qquad \qquad+ \frac{1}{3} \Big [\tilde{D}_k \tilde{D}^k N -4 \tilde{D}_k \ln \Psi \tilde{D}^k N \Big ] \tilde{\gamma}_{ij} \\
&\qquad \qquad \quad + N \Big [\tilde{R}_{ij} - \frac{1}{3} \tilde{R} \tilde{\gamma}_{ij}-2 \tilde{D}_i \tilde{D}_j \ln \Psi
+ 4 \tilde{D}_i \ln \Psi \tilde{D}_j \ln \Psi \\
& \qquad \qquad \qquad \qquad+\frac{2}{3} \Big ( \tilde{D}_k  \tilde{D}^k \ln \Psi - 2 \tilde{D}_k \ln \Psi\tilde{D}^k \ln \Psi \Big ) \tilde{\gamma}_{ij} \Big ] \bigg \} \, , 
 \end{aligned}
\end{equation}

\begin{equation}
 \tilde{D}_i \tilde{D}^i \Psi -\frac{1}{8}\tilde{R} \Psi + \frac{1}{8}\tilde{A}_{ij}\tilde{A}^{ij} \Psi^{-7}
+ \Big \{2 \pi E -\frac{1}{12} K^2 \Big \}=0  \, ,             
\end{equation}
\begin{equation}
\tilde{D}_j \Big ( \hat{A}^{ij} \Big ) - \frac{2}{3} \Psi^6 \tilde{D}^i K = 8 \pi \Psi^{10} p^i \, .
\end{equation}

\subsection{The Isenberg-Wilson-Mathews Approach to General Relativity(IWM)}

In IWM model \cite{Isenberg:1978}, \cite{Wilson:1989}, the spacetime is assumed to be foliated by a continuous set of $ (\Sigma_t)_{t \in \mathcal{R}} $ such that the foliation is maximally sliced ($ K=0 $). Here, the induced 3-metric is conformally flat which means that its conformal background metric is flat,
\begin{equation}
 \tilde{\gamma}_{ij}=f_{ij} \, .
\label{cotton}
\end{equation}
(\ref{cotton}) is implies that the Cotton-York tensor \cite{York:1979cf},\cite{York:1981bg}, \cite{Cotton:1899hj} vanishes. Furthermore, the conformal Ricci tensor is zero. Thus, the conformal 3+1 Einstein equation turns into,
\begin{equation}
 \Big ( \frac{\partial}{\partial t} - \mathcal{L}_\beta \Big ) \ln \Psi 
= \frac{1}{6} D_i \beta^i ,
\end{equation}
\begin{equation}
 \Big ( \frac{\partial}{\partial t} - \mathcal{L}_\beta \Big ) f_{ij}
= -2 N \tilde{A}_{ij} - \frac{2}{3} D_k \beta^k f_{ij} \, ,
\label{iwm}
\end{equation}
\begin{equation}
\begin{aligned}
0 = -\Psi^{-4} \Big ( D_i  D^i N + 2 D_i \ln \Psi D^i N  \Big ) 
+ N \Big \{ \tilde{A}_{ij} \tilde{A}^{ij} + 4 \pi \Big ( E + S \Big ) \Big \} \, ,
\end{aligned}
\end{equation}

\begin{equation}
 \begin{aligned}
\Big ( \frac{\partial}{\partial t} - \mathcal{L}_\beta \Big ) \tilde{A}_{ij} &= -\frac{2}{3} D_k \beta^k \tilde{A}_{ij} 
+ N \bigg \{- 2 f^{kl} \tilde{A}_{ik} \tilde{A}_{jl} 
-8 \pi \Big [ \Psi^{-4} S_{ij} -\frac{1}{3} S f_{ij} \Big ] \bigg \} \\
& \quad+ \Psi^{-4} \bigg \{ -D_i D_j N + 2 D_i N \, D_j \ln \Psi + 2 D_j N \, D_i \ln \Psi \\       
&\quad \qquad \qquad+ \frac{1}{3} \Big [D_k D^k N -4 D_k \ln \Psi D^k N \Big ] f_{ij} \\
&\qquad \qquad \quad + N \Big [-2 D_i D_j \ln \Psi
+ 4 D_i \ln \Psi D_j \ln \Psi \\
& \qquad \qquad \qquad \qquad+\frac{2}{3} \Big ( D_k D^k \ln \Psi - 2 D_k \ln \Psi D^k \ln \Psi \Big ) f_{ij} \Big ] \bigg \} \, ,  
\end{aligned}
\end{equation}

\begin{equation}
 D_i D^i \Psi + \Big \{ \frac{1}{8}\tilde{A}_{ij}\tilde{A}^{ij} + 2 \pi E \Big \} \Psi^{5}  =0 \, ,              
\end{equation}
\begin{equation}
D_j \tilde{A}^{ij} + 6 \tilde{A}^{ij} D_j \ln \Psi = 8 \pi \Psi^{4} p^i \, .
\label{kjlb}
\end{equation}
Here the equation (\ref{kjlb}) is obtained by using the relation $ \tilde{A}_{ij}\tilde{A}^{ij} = \Psi^{-12} \hat{A}_{ij} \hat{A}^{ij} $ in \emph{the momentum constraint equation} (\ref{moment_cnb}). In order to find the IWM conformal system, let us work on the equation (\ref{iwm}) : Because of the metric-compatibility, we have
\begin{equation}
\begin{aligned}
 \mathcal{L}_\beta f_{ij} &= \beta^k D_k f_{ij} +  f_{kj} D_i \beta^k + f_{ik} D_j \beta^k \\
&=f_{kj} D_i \beta^k + f_{ik} D_j \beta^k \, .
\end{aligned}
\end{equation}
Now, since time derivative of the $ f_{ij} $ is zero, the equation (\ref{iwm}) turns into,
\begin{equation}
 2 N \tilde{A}_{ij} =f_{kj} D_i \beta^k + f_{ik} D_j \beta^k - \frac{2}{3} D_k \beta^k f_{ij} \, .
\label{rlks}
\end{equation}
Let us multiply (\ref{rlks}) by $ f^{im} f^{jn} $ 
\begin{equation}
 2 N  \tilde{A}^{mn}= D^m \beta^n + D^n \beta^m-\frac{2}{3} D_k \beta^k f^{mn}
\label{mamimo}
\end{equation}
By change of the indices $ m \rightarrow i $, $ n \rightarrow j $, we can rewrite the equation  (\ref{mamimo}) as
\begin{equation}
 \tilde{A}^{ij}= \frac{1}{2 N} \Big (\tilde{\mathcal{L}} \beta \Big )^{ij} \, ,
\label{trtst}
\end{equation}
where
\begin{equation}
  \Big (\tilde{\mathcal{L}} \beta \Big )^{ij}= D^i \beta^j + D^j \beta^i-\frac{2}{3} D_k \beta^k f^{ij}
\end{equation}
is known as the \emph{\textbf{the conformal Killing derivative operator}}. Moreover, with the help of (\ref{trtst}) the corresponding momentum constraint equation (\ref{kjlb}) can be rewritten as
\begin{equation}
 \bigtriangleup \beta^i + \frac{1}{3} D^i D_j \beta^j + 2 \tilde{A}^{ij} \Big (6 N D_j \ln \Psi - D_j N \Big )=16 \pi N \Psi^4 p^i \, .
\end{equation}

Thus, we get \emph{\textbf{the conformal IWM system}} as the set of
\begin{equation}
 \bigtriangleup N + 2 D_i \ln \Psi D^i N = N \Big ( 4 \pi ( E + S) + \tilde{A}_{ij} \tilde{A}^{ij} \Big ) \Psi^4 \, ,
\end{equation}
\begin{equation}
\bigtriangleup \Psi + \Big \{ \frac{1}{8}\tilde{A}_{ij}\tilde{A}^{ij} + 2 \pi E \Big \} \Psi^{5}  =0 \, ,               
\end{equation}
\begin{equation}
 \bigtriangleup \beta^i + \frac{1}{3} D^i D_j \beta^j + 2 \tilde{A}^{ij} \Big (6 N D_j \ln \Psi - D_j N \Big )=16 \pi N \Psi^4 p^i \, .
\end{equation}

\section{ASYMPTOTIC FLATNESS AND THE ADM FORMALISM FOR GENERAL RELATIVITY}
\label{sec:Asymptotic Flatness And The ADM Formalism For General Relativity}

\subsection{The Asymptotic Flatness}

In this chapter, we will deduce the conserved quantities of the ADM mass, linear momentum and angular momentum of a given hypersurface $ \Sigma_t $. Since these quantities can be only in the globally-hyperbolic asymptotically flat spacetimes (i.e. the spacetimes which approaches asymptotically to the well-defined spacetimes such as the Minkowski, AdS). Therefore, let us first see review what the asymptotic flatness is: The asymptotic flat spacetime is such a particular spacetime for the massive objects in which it is assumed that there is nothing in the universe except these objects. Now, \emph{a globally-hyperbolic spacetime is called asymptotically flat if each of its Cauchy surface has a background metric \emph{\textbf{f}} with signature (+,+,+)} such that \textbf{f} is flat, can be diagonalized in a particular coordinate system on the $ \Sigma_t $ \cite{Gourgoulhon:2007ue}, \cite{York:1979cf}. Moreover, in the case of spatial infinity, $ r \rightarrow \infty $, the decay of $ \gamma_{ij} $ and their spatial partial derivatives must be something like
\begin{equation}
 \gamma_{ij} = f_{ij} + O[r^{-1}] \, ,
\end{equation}
\begin{equation}
 \frac{\gamma_{ij}}{\partial x^k}=O[r^{-2}] \, .
\end{equation}
And also as $ r \rightarrow \infty $, the decay of $ K_{ij} $ and their spatial partial derivatives must obey
\begin{equation}
 K_{ij}=O[r^{-2}] \, ,
\end{equation}
\begin{equation}
  \frac{\partial K_{ij}}{\partial x^k}=O[r^{-3}] \, .
\end{equation}

\subsection{The Hamiltonian Formalism for the General Relativity}

The Hamiltonian model approaches a physical state at a \emph{certain time} and gives the evolution of the state as t varies. This model is being transformed into the gravitational theory as a state on \emph{a particular spacelike hypersurface}. Now, the gravitational theory is a covariant theory and locally has Lorentz symmetry. The first attempts tried to start with the spacelike hypersurface that is free of choosing coordinates to avoid breaking of the crucial properties of the gravitational theory \cite{Dirac:1958sc}, \cite{Dirac:1958sq}. However, it is then hard to define initial state of practical problems. In order to write the Einstein equations into the Hamiltonian form, people started to give up the main properties of the gravitational theory by choosing a family of particular coordinate systems such that `` $ x^0 = $ constant``  corresponds a spacelike hypersurface. Contrary to the unknowns ($ \gamma_{ij}\,,\, K_{ij}\,,\,N\,,\, \beta^i $ ) in the PDEs form of 3+1 Einstein system, Arnowitt, Deser and Misner have proposed the ADM formalism of the General Relativity \cite{Arnowitt:1959ah} in which \emph{conjugate momentum of the induced three-metric $ \gamma_{ij} $,} $ \pi^{ij}=\sqrt{\gamma}(K\gamma^{ij}-K^{ij}) $, is used instead of $ K_{ij} $. Moreover, in the ADM formalism, $ \pi^{ij} $ and $ \gamma_{ij} $ are the dynamical variables and the Lapse function N and the shift vector $ \beta $ are taken as Lagrange multipliers \cite{Arnowitt:1959ah}, \cite{Gourgoulhon:2007ue}.

In this section, we will first deduce the corresponding Hamiltonian form  of the vacuum field equation by mean of the 3+1 decomposition of the spacetime metric that we have found in the $ 1^{st} $ chapter (the equation \ref{deco_metri}) and the knowledge that the boundary term is zero. Secondly, we will deal with the general case. That's, we will deduce the corresponding Hamiltonian form of the Einstein equation when the boundary term does not vanish by using the 2+1 decomposition of the timelike $ \mathcal{B} $ hypersurface that we have found in the $ 1^{st} $ chapter. This of the general case will lead us to get the explicit form of the famous ADM formulas for conserved quantities of $ \Sigma_t $ \cite{Poisson:2004ue}.

\subsubsection{3+1 Decomposition of the Einstein-Hilbert Action and the corresponding \\ Hamiltonian Form of the Vacuum Field Equation}

\begin{enumerate}
\item 3+1 Decomposition of The Einstein-Hilbert Action

The action for the four-dimensional vacuum field equation is of the Einstein-Hilbert action \cite{Deruelle:2006gh}, \cite{Gourgoulhon:2007ue}
\begin{equation}
 S=\int_\mathcal{V} {^4}R \sqrt{-g} d^4 x \, ,
\label{hilb}
\end{equation}
where the infinitesimal volume element $ \mathcal{V} $ is composed of the union of the neighboring hypersurfaces $ \Sigma_{t_1} $ and $ \Sigma_{t_1} $. Symbolically,
\begin{equation}
 \mathcal{V} = \overset{t_2}{\underset{t=t_1}{\bigcup}} \,\Sigma_t \, .
\end{equation}
Let us substitute the equations (\ref{mama}) of the 3+1 form of the spacetime Ricci scalar and (\ref{deco_metri}) of the 3+1 decomposition of \textbf{g} into the action (\ref{hilb}),
\begin{equation}
\begin{aligned}
 S &= \int_\mathcal{V}  \bigg \{ R + K^2 + K_{ij}K^{ij} - \frac{2}{N} \mathcal{L}_\textbf{m} K - \frac{2}{N} D_i D^i N \bigg \} N \sqrt{\gamma} d^4 x \\
& = \int_\mathcal{V} \bigg \{ N[R + K^2 + K_{ij}K^{ij}]- 2 \mathcal{L}_\textbf{m} K - 2 D_i D^i N \bigg \} \sqrt{\gamma} d^4 x \, .
\label{hilbe}
\end{aligned}
\end{equation}
Let us convert the term $ \mathcal{L}_\textbf{m} K $ into of the boundary and substitute it into the action (\ref{hilbe})
\begin{equation}
\begin{aligned}
 \mathcal{L}_\textbf{m} K = m^\mu \nabla_\mu K = N n^\mu \nabla_\mu K &= N [ \nabla_\mu (K n^\mu ) - K \nabla_\mu n^\mu ] \\
& = N [ \nabla_\mu (K n^\mu ) + K^2 ] \, .
\end{aligned}
\end{equation}
Then, the action (\ref{hilbe}) becomes
\begin{equation}
\begin{aligned}
&\qquad \qquad  S= \int_\mathcal{V} \bigg \lbrace N[R + K^2 + K_{ij}K^{ij}]-2 N \nabla_\mu (K n^\mu ) -2 N K^2  - 2 D_i D^i N \bigg \rbrace \sqrt{\gamma} d^4 x \\
&\qquad \qquad \quad= \int_\mathcal{V} \bigg \{ N[R + K_{ij}K^{ij} - K^2 ]- 2 D_i D^i N \bigg \} \sqrt{\gamma} d^4 x \\
&\qquad \qquad  \qquad - 2 \int_\mathcal{V} N \nabla_\mu ( K n^\mu ) \sqrt{\gamma} d^4 x \, .
\label{hilber}
\end{aligned}
\end{equation}
Here we need to show that because of the boundary condition, the last integral of the action (\ref{hilber}) vanishes:
\begin{equation}
\begin{aligned}
 \int_\mathcal{V} N \nabla_\mu ( K n^\mu ) \sqrt{\gamma} d^4 x &=\int_\mathcal{V} \nabla_\mu ( K n^\mu ) \sqrt{-g} d^4 x \\
&= \int_\mathcal{V} \frac{\partial}{\partial x^\mu} (\sqrt{-g}K n^\mu) d^4 x \\
&=0 \, ,
\end{aligned}
\end{equation}
so the action (\ref{hilber}) reduces to
\begin{equation}
 \qquad \qquad \quad S= \int_\nu \bigg \{ N[R + K_{ij}K^{ij} - K^2 ]- 2 D_i D^i N \bigg \} \sqrt{\gamma} d^4 x \, .
\label{hilbert}
\end{equation}
Observe that the action (\ref{hilbert}) is fully composed of the intrinsic quantities of $ \Sigma_t $. This provides us to decompose the four-dimensional integral into of the spatial one and of the time coordinate
\begin{equation}
 S= \int_{t_1}^{t_2} \bigg \{ \int_{\Sigma_t} \bigg ( N [ R + K_{ij}K^{ij} - K^2 ]- 2 D_i D^i N \bigg ) \sqrt{\gamma} d^3 x \bigg \} dt \, .
\label{mkhkn}
\end{equation}
Again, the boundary term vanishes and we get the \emph{\textbf{3+1 decomposition of the Einstein-Hilbert action}} as
\begin{equation}
 S= \int_{t_1}^{t_2} \bigg \{ \int_{\Sigma_t} N \bigg ( R + K_{ij}K^{ij} - K^2 \bigg ) \sqrt{\gamma} d^3 x \bigg \} dt \, .
\label{hilbert_e}
\end{equation}
\item The Corresponding Hamiltonian Form of the Vacuum Field Equation

\emph{The variables of the action in the configuration}  are $ q= (\gamma_{ij},N,\beta^i ) $ and $ \overset{.}{q} = (\overset{.}{\gamma}_{ij} , \overset{.}{N} , \overset{.}{\beta^i} ) $ (\ref{hilbert_e}) \cite{Gourgoulhon:2007ue}. That's,
\begin{equation*}
 S=S[q,\, \overset{.}{q}] \, .
\end{equation*}
The Lagrangian density contains the extrinsic curvature $ K_{ij} $ [see the 3+1 decomposition of the action (\ref{hilbert_e})]. However, in the Hamiltonian approach, it is replaced with the configuration variables. Now, from the $ 1^{st} $ equation (\ref{tt}) of the 3+1 dimensional Einstein system, we have 
\begin{equation}
\bigg ( \frac{\partial}{\partial t} - \mathcal{L}_\beta \bigg ) \gamma_{ij} = -2 N K_{ij} \Longrightarrow 
K_ {ij}=\frac{1}{2N} \bigg [\mathcal{L}_\beta \gamma_{ij} - \overset{.}{\gamma}_{ij} \bigg ] \, .
\label{t}
\end{equation}
We know  from the (\ref{jk}) that $ \mathcal{L}_\beta \gamma_{ij} = D_i \beta_j + D_j \beta_i $. Then, the equation (\ref{t}) becomes 
\begin{equation}
 K_ {ij}=\frac{1}{2N} \bigg [ D_i \beta_j + D_j \beta_i - \overset{.}{\gamma}_{ij} \bigg ]
= \frac{1}{2N} \bigg [ \gamma_{jk} D_i \beta^k + \gamma_{ik} D_j \beta^k - \overset{.}{\gamma}_{ij} \bigg ] \, .
\label{is}
\end{equation}
And the Lagrangian density of the gravitational field (\ref{hilbert_e}) turns into
\begin{equation}
\begin{aligned}
 L(q,\overset{.}{q} ) &= N \sqrt{\gamma} \bigg ( R + K_{ij}K^{ij} - K^2 \bigg ) \\
&= N \sqrt{\gamma} 
\bigg ( R +\Big [ \gamma^{ik} \gamma^{jl} - \gamma^{ij} \gamma^{kl} \Big ] K_{ij} K_{kl} \bigg ) \, .
\label{se}
\end{aligned}
\end{equation}
\emph{As we see from (\ref{se}), the Lagrangian density does not depend on the time derivative of N and $ \beta^i $ so they are not \textbf{\textbf{dynamical variables}}. They are just the Lagrange multipliers. On the other hand, \textbf{the remaining variable $ \gamma_{ij} $ is just the dynamical variable in phase space. And the corresponding conjugate momentum of it}} is 
\begin{equation}
\begin{aligned}
 \pi^{ij} = \frac{\partial L}{\partial \overset{.}{\gamma}_{ij}} &=\frac{\partial K_{ab}}{\partial \overset{.}{\gamma}_{ij}}
\frac{\partial L}{\partial K_{ab}} \\
&=-\frac{1}{2N} \delta_{ia} \delta_{jb} N \sqrt{\gamma} \Big ( \gamma^{ik} \gamma^{jl} - \gamma^{ij} \gamma^{kl} \Big)
\Big ( \delta_{ia} \delta_{jb} K_{kl} + \delta_{ka} \delta_{lb} K_{ij} \Big ) \\
& = -\frac{1}{2} \sqrt{\gamma} \delta_{ia} \delta_{jb} \bigg \{ \,\gamma^{ka} \gamma^{lb} K_{kl} + \gamma ^{ia} \gamma^{jb} K_{ij}
-\gamma^{ab} \gamma^{kl} K_{kl} -\gamma^{ij} \gamma^{ab} K_{ij} \, \bigg \} \\
& \quad \qquad \qquad \qquad  k \rightarrow i , l \rightarrow j \qquad \qquad \qquad  k \rightarrow i , l \rightarrow j \\
& =- \sqrt{\gamma} \delta_{ia} \delta_{jb} \bigg \{  \gamma ^{ia} \gamma^{jb} K_{ij} - \gamma^{ij} \gamma^{ab} K_{ij} \, \bigg \} \\
& = -\sqrt{\gamma} \bigg \{ K ^{ij} -\gamma^{ij} K \bigg \} \, .
\end{aligned}
\end{equation}
Thus, we get
\begin{equation}
 \pi^{ij} = \sqrt{\gamma} \bigg \{ \gamma^{ij} K - K ^{ij} \bigg \} \, .
\label{conj}
\end{equation}
Finally, let us find the corresponding Hamiltonian: The Legendre transformation is defined as
\begin{equation}
 \mathcal{H} = \pi^{ij} \overset{.}{\gamma}_{ij} -L \, .
\label{asa}
\end{equation}
Let us substitute the explicit form of $ \pi^{ij} $ (\ref{conj}) and $ \overset{.}{\gamma}_{ij} $ (\ref{is}) into the Hamiltonian density (\ref{asa}):
\begin{equation*}
\begin{aligned}
 \mathcal{H} &= \sqrt{\gamma} \bigg ( \gamma^{ij} K - K ^{ij} \bigg ) \bigg (-2 N K_{ij} + D_j \beta_i + D_i \beta_j \bigg ) \\
& \quad -N \sqrt{\gamma} \bigg (R + K_{ij} K^{ij} - K^2 \bigg ) \\
& = \sqrt{\gamma} \bigg \{-2 N K \gamma^{ij} K_{ij} + K \gamma^{ij} D_j \beta_i + K \gamma^{ij} D_i \beta_j
2 N K_{ij} K^{ij} - K^{ij} D_j \beta_i \\
& \qquad \qquad - K^{ij} D_i \beta_j -N R - N K_{ij} K^{ij} + N K^2 \bigg \} \\
& = \sqrt{\gamma} \bigg \{-2 N K^2  + K  D_j \beta^j + K D_i \beta^i + 2 N K_{ij} K^{ij} - K^{ij} D_j \beta_i \\
& \qquad \qquad - K^{ij} D_i \beta_j -N R - N K_{ij} K^{ij} + N K^2 \bigg \} \\
& = \sqrt{\gamma} \bigg \{ -N K^2 + 2 K D_j \beta^j + N K_{ij} K^{ij} - 2 K^{ij} D_j \beta_i -N R \bigg \} \\
& = \sqrt{\gamma} \bigg \{ -N \Big [ R + K^2 - K_{ij} K^{ij} \Big ] + 2 K D_j \beta^j - 2 K^j {_i} D_j \beta^i \bigg \} \\
&= \sqrt{\gamma} \bigg \{ -N \Big [ R + K^2 - K_{ij} K^{ij} \Big ] + 2 \Big [ K\gamma^j {_i} - K^j {_i} \Big ] D_j \beta^i \bigg \} \\ 
&= \sqrt{\gamma} \bigg \{ -N \Big [ R + K^2 - K_{ij} K^{ij} \Big ] + 2 D_j \Big [ K\gamma^j {_i} \beta^i - K^j {_i} \beta^i \Big ] \\
& \quad -2 \beta^i \Big [\gamma^j {_i} D_j K - D_j K^j {_i} \Big ] \bigg \}  
\end{aligned}
\end{equation*}
\begin{equation}
\qquad \qquad \quad = \sqrt{\gamma} \bigg \{ -N \Big [ R + K^2 - K_{ij} K^{ij} \Big ] + 2 D_j \Big [ K \beta^j - K^j {_i} \beta^i \Big ] 
-2 \beta^i \Big [ D_i K - D_j K^j {_i} \Big ] \bigg \}  \, .
\label{ugth}
\end{equation}
With the aberrations of $ C_0 = R+ K^2 - K_{ij} K^{ij} $ and $  C_i = D_j K^j {_i} - D_i K $, the Hamiltonian density (\ref{ugth}) reduces to
\begin{equation}
 \mathcal{H} = \sqrt{\gamma} \bigg \{ -N C_0 + 2 \beta^i C_i + 2 D_j \Big [ K \beta^j - K^j {_i} \beta^i \Big ] \bigg \} \, .
\end{equation}
And the related Hamiltonian is obtained by
\begin{equation}
\begin{aligned}
 H &= \int_{\Sigma_t} \mathcal{H} d^3 x \\
& =-\int_{\Sigma_t} \bigg \{ N C_0 - 2 \beta^i C_i \bigg \} \sqrt{\gamma} d^3 x
+ 2 \int_{\Sigma_t} \sqrt{\gamma} D_j \Big [ K \beta^j - K^j {_i} \beta^i \Big ] d^3 x \, .
\label{sdf}
\end{aligned}
\end{equation}
Due to the boundary condition, the last integral of the (\ref{sdf}) vanishes and we get \emph{\textbf{the Hamiltonian of the gravitational vacuum field}} as
\begin{equation}
 H =-\int_{\Sigma_t} \bigg \{ N C_0 - 2 \beta^i C_i \bigg \} \sqrt{\gamma} d^3 x \, ,
\label{hami}
\end{equation}
where
\begin{equation}
 C_0 = R+ K^2 - K_{ij} K^{ij} \, ,
\label{cf}
\end{equation}
\begin{equation}
 C_i = D_j K^j {_i} - D_i K \, .
\label{fc}
\end{equation}
 \end{enumerate}
\emph{The crucial point is that the constraint equations (\ref{cf}) and (\ref{fc}) are nothing but exactly the constraint equations of the energy (\ref{gh}) and the momentum (\ref{mome}) [for vacuum] that we deduced during the 3+1 decomposition of the Einstein equation}. Moreover, \emph{\textbf{ for any spacetime to be a solution of the Einstein equation the related Hamiltonian equation (\ref{hami}) of it must be zero}}. 

\subsubsection{The General Gravitation Hamiltonian and The ADM Formalism} 

Contrary to the previous section, we will assume that the boundary term is not zero which with the 2+1 decomposition of the hypersurfaces that we deduced in the $ 1^{st} $ chapter will lead us to the well-known ADM formalism for the conserved quantities of a given hypersurface $ \Sigma_t $. Here, we will assume that \emph{the infinitesimal four dimensional volume element $ \partial \mathcal{V} $ is the union of two spacelike hypersurfaces $ \Sigma_{t_1} $ and  $ \Sigma_{t_2} $ at the lower and upper boundaries and a timelike hypersurface $ \mathcal{B} $ that covers the region between these two spacelike hypersurfaces} \cite{Poisson:2004ue}
\begin{equation}
 \partial \mathcal{V} = \Sigma_{t_2} \bigcup \,( - \Sigma_{t_1} ) \bigcup \mathcal{B} \, .
\end{equation}
Since the unit normal vector of $ \partial \mathcal{V} $ must be directed outward.However, the unit normal vector of the hypersurface $ \Sigma_{t_1} $ is \emph{future-directed} so it points inward. Then, with the help of the \emph{minus sign}, the unit normal vector of $ \Sigma_{t_1} $ will point outward, too. Now, in order to find the ADM formulas, we need to first deduce the Hamiltonian of this case: 

\begin{enumerate}
 \item The Gravitational Action and The Corresponding Hamiltonian when the Boundary Term is different than zero

The related  gravitational action is composed of the Einstein-Hilbert part, the matter (or boundary) part and a no dynamical part of $ S_0 $ that does not have any influence on the equation of motion \cite{Poisson:2004ue}:
\begin{equation}
 S_G[g] = S_H [g]+ S_B[g]-S_0 \, .
\label{sed}
\end{equation}
Here
\begin{equation}
 S_H[g] = \frac{1}{16 \pi} \int_\mathcal{V} {^4}R \sqrt{-g} d^4 x \, ,
\label{ac1}
\end{equation}
\begin{equation}
 S_B[g] = \frac{1}{8 \pi} \oint_{\partial \mathcal{V}} \varepsilon K \sqrt{|h|} d^3 y \, ,
\label{ac2}
\end{equation}
\begin{equation}
 S_0= \frac{1}{8 \pi} \oint_{\partial \mathcal{V}} \varepsilon K_0 \sqrt{|h|} d^3 y \, .
\label{ac3}
\end{equation}
By substituting the related actions (\ref{ac1}) and (\ref{ac2}) into (\ref{sed}), the gravitational action (\ref{sed}) becomes
\begin{equation}
 (16 \pi) S_G = \int_\mathcal{V} {^4}R \sqrt{-g} d^4 x + 2 \oint_{\partial \mathcal{V}} \varepsilon K \sqrt{|h|} d^3 y \, .
\label{dcs}
\end{equation}
Here, $ y^a $ are adapted coordinates of $ \partial \mathcal{V} $ , $ h_{ab} $ are the corresponding induced 3-metric, $ n^\alpha $ are the corresponding unit normal vector and $ K $ is the scalar extrinsic curvature. As we mentioned before, because the $ \partial \mathcal{V} $ is the union of two spacelike and one timelike hypersurfaces so $ \varepsilon = n^\alpha n_\alpha $ will be +1 or -1 depending on the type of hypersurfaces. The explicit form of (\ref{dcs}) is
\begin{equation}
\begin{aligned}
 (16 \pi) S_G &= \int_\mathcal{V} {^4}R \sqrt{-g}\, d^4 x + 2 \int_{\Sigma_{t_2}} n^\alpha n_\alpha K \sqrt{h}\, d^3 y 
+ 2 \int_{-\Sigma_{t_1}} n^\alpha n_\alpha K \sqrt{h}\, d^3 y \\
& \quad + 2 \int_\mathcal{B} r^\alpha r_\alpha \mathcal{K} \sqrt{- \gamma}\, d^3 y \\
& = \int_\mathcal{V} {^4}R \sqrt{-g}\, d^4 x - 2 \int_{\Sigma_{t_2}}  K \sqrt{h}\, d^3 y 
+2 \int_{\Sigma_{t_1}} K \sqrt{h}\, d^3 y \\
& \quad + 2 \int_\mathcal{B} \mathcal{K} \sqrt{- \gamma}\, d^3 y \, ,
\label{trt}
\end{aligned}
\end{equation}
where we have used $ n^\alpha n_\alpha = -1 $ of the spacelike hypersurface and $ r^\alpha r_\alpha = +1 $ of the timelike hypersurface. \emph{For convention, let us use the following form of the 3+1 decomposition of the spacetime Ricci scalar}, 
\begin{equation}
 {^4}R = R + K^{ab} K_{ab} -K^2 -2 \Big ( n^\alpha{ _{;\beta}}n^\beta - n^\alpha n^\beta{_{;\beta}}\Big )_{;\alpha} \, ,
\end{equation}
where $ '' {;} '' $ denotes the intrinsic covariant derivative. And, we know that the 3+1 decomposition of the spacetime metric is  
\begin{equation}
 \sqrt{-g}\, d^4 x = N \sqrt{h}\, dt\, d^3 y \, .
\end{equation}
Then, the 3+1 decomposition of the Einstein-Hilbert part is 
\begin{equation}
\begin{aligned}
 \int_\mathcal{V} {^4}R \sqrt{-g}\, d^4 x &= \int_{t_1}^{t_2} dt \bigg \{ \int_{\Sigma_t} R + K^{ab} K_{ab} -K^2 \bigg \} N \sqrt{h}\, d^3 y \\
& - 2 \int_\mathcal{V} \bigg \{ n^\alpha{ _{;\beta}}n^\beta - n^\alpha n^\beta{_{;\beta}}\bigg \}_{;\alpha} d^4 x \\
& = \int_{t_1}^{t_2} dt \bigg \{ \int_{\Sigma_t} R + K^{ab} K_{ab} -K^2 \bigg \} N \sqrt{h}\, d^3 y \\
&-  2 \oint_{\partial \mathcal{V}} \bigg \{ n^\alpha{ _{;\beta}}n^\beta - n^\alpha n^\beta{_{;\beta}}\bigg \} d \Sigma_\alpha \, .
\label{fsc}
\end{aligned}
\end{equation}
Because  of $ \partial \mathcal{V} = \Sigma_{t_2} \bigcup\, ( - \Sigma_{t_1} ) \bigcup \mathcal{B} $, we will decompose the closed integral of the equation (\ref{fsc}) into the corresponding integrals of $ \Sigma_{t_1} $, $ \Sigma_{t_2} $ and $ \mathcal{B} $. For simplicity, let us work on of  $ \Sigma_{t_1} $: The spacelike volume element is $ d \Sigma_\alpha = n_\alpha \sqrt{h}\, d^3 y $, then, we have  
\begin{equation}
\begin{aligned}
 -2 \int_{-\Sigma_{t_1}} \bigg \{ n^\alpha{ _{;\beta}} n^\beta - n^\alpha n^\beta{_{;\beta}}\bigg \} d \Sigma_\alpha
&=- 2 \int_{\Sigma_{t_1}} n^\beta {_{;\beta}} \sqrt{h}\, d^3 y \\ 
&= -2 \int_{\Sigma_{t_1}} K \sqrt{h}\, d^3 y \, ,
\label{trt1}
\end{aligned}
\end{equation}
where we used the fact that $  n^\alpha{ _{;\beta}} $ is an element of the hypersurface $ \Sigma_{t_1} $. Similarly, by evaluating the corresponding integral on $ \Sigma_{t_2} $ in the reverse direction, we will get  
\begin{equation}
2 \int_{\Sigma_{t_2}} K \sqrt{h}\, d^3 y \, .
\label{trt2}
\end{equation}
Observe that the results of (\ref{trt1}) and (\ref{trt2}) cancel out the $ 2^{nd} $ and $ 3^{th} $ terms on the right hand side of the gravitational action (\ref{trt2}). Thus, \emph{\textbf{the only contribution is coming from the integral over the timelike hypersurface $ \mathcal{B} $: Now, for timelike case 3-dimensional volume element is $ d \Sigma_\alpha = r_\alpha \sqrt{-\gamma}\, d^3 z $ and the unit spacelike normal $ r^\alpha $ of timelike hypersurface $ \mathcal{B} $ and the unit timelike vector normal vector $ n^\alpha $ of the spacelike hypersurface $ \Sigma_t $ are orthogonal to each other}} [that is, $ n^\alpha r_\alpha =0 $], then,
\begin{equation}
\begin{aligned}
  -2 \int_\mathcal{B} \bigg \{ n^\alpha{ _{;\beta}} n^\beta - n^\alpha n^\beta{_{;\beta}}\bigg \} d \Sigma_\alpha 
&= -2 \int_\mathcal{B} n^\alpha{_{;\beta}} n^\beta r_\alpha \sqrt{-\gamma}\, d^3 z \\
&= 2 \int_\mathcal{B} r_{\alpha;\beta} n^\alpha n^\beta \sqrt{-\gamma}\, d^3 z \, .
\label{trt3}
\end{aligned}
\end{equation}
With the help of (\ref{trt3}), the gravitational action (\ref{trt}) becomes,
\begin{equation}
\begin{aligned}
 (16 \pi) S_G &= \int_{t_1}^{t_2} dt \bigg \{ \int_{\Sigma_t}  \Big [R + K^{ab} K_{ab} -K^2 \Big ] \bigg \} N \sqrt{h}\,d^3 y \\
&+ 2 \int_\mathcal{B} \Big [ \mathcal{K} + r_{\alpha;\beta} n^\alpha n^\beta \Big ] \sqrt{-\gamma}\, d^3 z \, .
\label{trt4}
\end{aligned}
\end{equation}
Notice that the the gravitational action (\ref{trt4}) is composed of the 3+1 decomposition of the Einstein-Hilbert action and the integral over the timelike hypersurface  $ \mathcal{B} $. As we mentioned before the Einstein-Hilbert part results in the Hamiltonian and momentum constraints that ensure whether a given spacetime is a solution of the The Einstein equation or not. \emph{\textbf{The important point is that the integral over the timelike hypersurface $ \mathcal{B} $ will lead us to the conserved quantities of the hypersurfaces. That's, the boundary term of $ \mathcal{B} $ will give the ADM formalism. Therefore, we need to do the decompose the timelike hypersurface $ \mathcal{B} $ by assuming that $ \mathcal{B} $ is being foliated by the boundary of the spacelike hypersurface $ \Sigma_t $, $ S_t $, whose topology is supposed to be $ S^2 $}}. Now, the 2+1 decomposition of the 3-metric of $ \mathcal{B} $ is
\begin{equation}
 \sqrt{-\gamma}\, d^3 z = N \sqrt{\sigma}\,dt\, d^2 \theta \, .
\label{trt5}
\end{equation}
And scalar extrinsic curvature of $ \mathcal{B} $ is
\begin{equation}
 \mathcal{K} = \gamma^{ij} \mathcal{K}_{ij} = \gamma^{ij} \Big ( r_{\alpha;\beta} e^\alpha_i e^\beta_j \Big )
= r_{\alpha;\beta} \Big (\gamma^{ij} e^\alpha_i e^\beta_j \Big ) = r_{\alpha;\beta} \Big ( g^{\alpha\beta} - r^\alpha r^\beta \Big ) \, .
\label{trt6}
\end{equation}
So with the help of (\ref{trt5}) and (\ref{trt5}), the integrand of the related integral of (\ref{trt4}) that is over $ \mathcal{B} $  becomes
\begin{equation}
\begin{aligned}
 \mathcal{K} + r_{\alpha;\beta} n^\alpha n^\beta &= r_{\alpha;\beta} \Big ( g^{\alpha\beta} - r^\alpha r^\beta \Big )
 + r_{\alpha;\beta} n^\alpha n^\beta \\
&= r_{\alpha;\beta} \Big ( g^{\alpha\beta} - r^\alpha r^\beta + n^\alpha n^\beta \Big ) \\
&=r_{\alpha;\beta} \Big ( \sigma^{AB} e^\alpha_A e^\beta_B \Big ) \\
&= \sigma^{AB} \Big ( r_{\alpha;\beta} e^\alpha_A e^\beta_B \Big ) \\
&= \sigma^{AB} k_{AB} \\
&=k \, ,
\label{trt7}
\end{aligned}
\end{equation}
where $ k $ is the extrinsic curvature of $ S_t $. Thus, by substituting the result of (\ref{trt7}) into the gravitational action (\ref{trt4}), we get \emph{\textbf{the decomposition of the gravitational action when the boundary term is different than zero}}:
\begin{equation}
\begin{aligned}
 S_G &= \frac{1}{16 \pi} \int_{t_1}^{t_2} dt \bigg \{ \int_{\Sigma_t} \bigg ( R + K^{ab} K_{ab} -K^2 \bigg ) N \sqrt{h}\,d^3 y \\
& \qquad \qquad \qquad \quad + 2 \oint_{S_t} \Big ( k - k _0 \Big ) N \sqrt{\sigma} d^2 \theta \bigg \} \, .
\end{aligned}
\end{equation}
Here $ k_0 $ is \emph{the extrinsic curvature of } $ S_t $ \emph{embedded in flat space}. The $ k_0 $ is defined so that the gravitational action is zero for flat spacetime.

After construction of the action, it is time to find \emph{\textbf{the corresponding Hamiltonian of the system}}:In chapter 3, we have found that the relation between the extrinsic curvature and configuration variables  
\begin{equation}
 K_{ab}= \frac{1}{2N} \Big ( - \overset{.}h_{ab} + D_a \beta_b + D_b \beta_a \Big ) \, .
\label{trt9}
\end{equation}
And the corresponding canonical conjugate momentum is
\begin{equation}
 \pi^{ab} = \frac{\partial}{\partial \overset{.}h_{ab}} \Big ( \sqrt{-g} \mathcal{L}_G \Big ) \, .
\label{trt8}
\end{equation}
Due to the fact that the \emph{boundary part} is independent of $ \overset{.}h_{ab} $, the equation (\ref{trt8}) becomes
\begin{equation}
 (16 \pi) \pi^{ab} = \frac{\partial K_{mn}}{\partial \overset{.}h_{ab}} \frac{\partial}{K_{mn}} \Big ( 16 \pi \sqrt{-g} \mathcal{L}_G \Big ) \, ,
\end{equation}
where 
\begin{equation*}
 16 \pi \sqrt{-g} \mathcal{L}_G = \bigg [ R + \Big ( h^{ac} h^{bd} - h^{ab} h^{cd} \Big ) K_{ab} K_{cd} \bigg ] N \sqrt{h} \, .
\end{equation*}
Then,
\begin{equation*}
\begin{aligned}
 (16 \pi) \pi^{ab} &= - \frac{1}{2N} \delta_{ma} \delta_{nb} N \sqrt{h} \Big [ h^{ac} h^{bd} - h^{ab} h^{cd} \Big ]
\Big [\delta_{ma} \delta_{nb} K_{cd} + \delta_{mc} \delta_{nd} K_{ab} \Big ] \\
&= -\frac{\sqrt{h}}{2} \delta_{ma} \delta_{nb} \bigg \{ h^{mc} h^{nd} K_{cd} +h^{am} h^{bn} K_{ab}
 - h^{mn} h^{cd} K_{cd} \\
& \qquad  \qquad  \qquad -h^{ab} h^{mn} K_{ab} \bigg \} \\
&=- \sqrt{h} \delta_{ma} \delta_{nb} \Big [h^{am} h^{bn} K_{ab} -h^{ab} h^{mn} K_{ab} \Big ] \\
&= -\Big [ K^{ab} -K h^{ab} \Big ] \, .
\end{aligned}
\end{equation*}
so we get
\begin{equation}
 (16 \pi) \pi^{ab} = \sqrt{h} \Big ( K h^{ab} - K^{ab} \Big ) \, .
\label{trt10}
\end{equation}
In order to find the corresponding Hamiltonian, let us substitute the explicit forms of $ \pi^{ab} $ (\ref{trt9}) and $ \overset{.}h_{ab} $ (\ref{trt10}) in to the Hamiltonian density:
\begin{equation}
\begin{aligned}
 \mathcal{H}_G &= \pi^{ab} \overset{.}h_{ab} - \sqrt{-g} \mathcal{L}_G \\
 (16 \pi) \mathcal{H}_G &=(16 \pi) \pi^{ab} \overset{.}h_{ab} -(16 \pi) \sqrt{-g} \mathcal{L}_G \\
&=\sqrt{h} \Big ( K h^{ab} - K^{ab} \Big ) \Big (-2N K_{ab} + D_a \beta_b D_b \beta_a \Big ) \\
& \qquad - \Big ( R + K^{ab}K_{ab} -K^2 \Big ) N \sqrt{h} \\
&=\sqrt{h} \bigg \{ 2NK^{ab}K_{ab} - K^{ab}D_b \beta_a - K^{ab}D_a \beta_b - -2NKh^{ab}K_{ab} \\
&\qquad \quad + Kh^{ab}D_b \beta_a + Kh^{ab}D_a \beta_b -NR-NK^{ab}K_{ab} +NK^2\bigg \} \\
&=\sqrt{h} \bigg \{NK^{ab}K_{ab} -NK^2 -NR-2K^{ab}D_b\beta_a + 2Kh^{ab}D_b\beta_a \bigg \} \\
&=-N\sqrt{h} \underbrace{\Big [R+K^2 -K_{ab} K^{ab} \Big ] } -2 \sqrt{h} \Big [ K^{ab}-Kh^{ab} \Big ]D_b\beta_a \\
& \qquad \qquad \qquad \qquad \quad C_0 \\
&=-NC_0\sqrt{h} -2 \sqrt{h} \bigg \{ D_b \Big [ \beta_a \Big (  K^{ab}-Kh^{ab} \Big ) \Big ] - \beta_a \underbrace{\ D_b \Big [ K^{ab}-Kh^{ab} \Big ] } \bigg \} \, . \\
& \qquad \qquad \qquad \qquad \quad \qquad \qquad \qquad \qquad \quad \qquad \qquad C^a 
\label{11}
\end{aligned}
\end{equation}
Then, by integrating the Hamiltonian density (\ref{11}) over the spacelike hypersurface $ \Sigma_t $, we get the  Hamiltonian of the system as 
\begin{equation}
\begin{aligned}
 (16 \pi ) H_G &= \int_{\Sigma_t} 16 \pi \mathcal{H}_G \,d^3 y -2 \oint_{S_t} \Big ( k-k_0 \Big ) N \sqrt{\sigma}\, d^2 \theta \\
&= -\int_{\Sigma_t} \bigg \{ NC_0 -2 \beta_a C^a \bigg \} \sqrt{h}\, d^3 y \\
& \quad -2 \int_{\Sigma_t} D_b \Big [\beta_a \Big (K^{ab}-Kh^{ab} \Big ) \Big] \sqrt{h}\, d^3y-2 \oint_{S_t} \Big ( k-k_0 \Big ) N \sqrt{\sigma}\, d^2 \, .
\end{aligned}
\end{equation}
Thus, by using the general Stokes theorem, we reach our aim of \emph{\textbf{the gravitational Hamiltonian when the boundary term is different than zero}}
\begin{equation}
\begin{aligned}
 (16 \pi ) H_G &= -\int_{\Sigma_t} \Big \{ NC_0 - 2\beta_a C^a \Big \} \sqrt{h}\, d^3 y \\ 
& \quad-2 \oint_{S_t} \Big \{N(k-k_0) + \beta_a (K^{ab} -Kh^{ab}) r_b \Big \} \sqrt{\sigma} \,d^2 \theta \, .
\label{po}
\end{aligned}
\end{equation}
\item The ADM Formalism

\begin{enumerate}
 \item The ADM Mass

In the previous section, we have found the general gravitational Hamiltonian in the equation (\ref{po}). \emph{\textbf{Due to the convention that we follow, we should do the following changes in the gravitational Hamiltonian}} (\ref{po}):
\begin{equation}
\begin{aligned}
 & a \rightarrow i\,,\, b \rightarrow j \\
& h \rightarrow \gamma\,,\,y \rightarrow x \\
& k \rightarrow \kappa \,,\, k_o \rightarrow \kappa_o \\
& r \rightarrow s \,,\,\sigma \rightarrow q \\
& \theta \rightarrow y \, .
\end{aligned}
\end{equation}
Then, the gravitational Hamiltonian (\ref{po}) turns into
\begin{equation}
\begin{aligned}
 (16 \pi ) H_G &= -\int_{\Sigma_t} \Big \{ NC_0 - 2\beta^i C_i \Big \} \sqrt{\gamma}\, d^3 x \\ 
&  -2 \oint_{S_t} \Big \{N(\kappa-\kappa_0) + \beta^i (K_{ij} -K\gamma_{ij}) s^j \Big \} \sqrt{q} \,d^2 y \, ,
\end{aligned}
\end{equation}
where $ S_t $ is the boundary of $ \Sigma_t $ and has the topology of $ S^2 $; \textbf{x} is a well-defined coordinate system on the $ \Sigma_t $ and $ \gamma $ is the corresponding induced 3-metric on $ \Sigma_t $ ; $ \kappa $ is the scalar extrinsic curvature of $ S_t $ embedded in $ (\Sigma_t ,\gamma ) $; $ \kappa_0 $ is the scalar extrinsic curvature embedded in the flat spacetime $ (\Sigma_t ,f) $; $ \hat{\textbf{s}} $ is the spacelike unit vector that is normal to $ S_t $; \textbf{y} is a well-defined coordinate system on $ S_t $ and $ q $ is the corresponding 2-metric on $ S_t $.

Now, suppose that a given spacetime is a solution of the Einstein equation. Then, \\ the corresponding integral of constraints vanishes [ due to $C_0=0 $ and $ C_i=0 $] \cite{Gourgoulhon:2007ue},
\begin{equation}
 H_{sol.} = -\frac{1}{8 \pi} \oint_{S_t} \Big \{N(\kappa-\kappa_0) + \beta^i (K_{ij} -K\gamma_{ij}) s^j \Big \} \sqrt{q} \,d^2 y \, .
\end{equation}
The total mass of the $ \Sigma_t $ which is measured by an asymptotically inertial observer ($ N=1 $ and $ \beta=0 $ ) with a well-defined adapted coordinates of  $ (t,x^i ) $ is given by \emph{\textbf{the famous ADM energy formula}} \cite{Gourgoulhon:2007ue} of
\begin{equation}
 \mathcal{M}_{ADM}=-\frac{1}{8 \pi} \lim_{\underset{(r \rightarrow \infty)}{{S_t}\to\infty}} \oint_{S_t} (\kappa-\kappa_0 )\sqrt{q}\, d^2 y \, .
\end{equation}
$ \mathcal{M}_{ADM} $ is the conserved quantity associated the symmetry of the action under time translation \cite{Gourgoulhon:2007ue} . And, in terms of the intrinsic connection of $ \Sigma_t $ :
\begin{equation}
 \mathcal{M}_{ADM}=\frac{1}{16 \pi} \lim_{S_t\to\infty} \oint_{S_t} \Big [ D^j \gamma_{ij} - D_i (f^{kl}\gamma_{kl} ) \Big]s^i \, \sqrt{q}\, d^2 y \, .
\label{adm}
\end{equation}
\emph{Furthermore, as we mentioned before one of the conditions for a spacetime to be asymptotically flat is that there must be a coordinate system $ (x^i) $ in which the background metric \textbf{f} is diagonalized. Now, in this coordinate $ D_i = \frac{\partial}{\partial x^i} $ and $ f^{kl} = \delta^{kl} $. Therefore, the ADM energy formula (\ref{adm}) turns into a simpler one in this specific coordinate} \cite{Gourgoulhon:2007ue}:
\begin{equation}
 \mathcal{M}_{ ADM} =\frac{1}{16 \pi} \lim_{S_t\to\infty} \oint_{S_t} \Big ( \frac{\partial \gamma_{ij}}{\partial x^j } - \frac{\partial \gamma_{jj}}{\partial x^i} \Big )s^i \sqrt{q}\, d^2 y \, .
\end{equation}
Finally, \emph{the conformal form of the ADM energy} \cite{Gourgoulhon:2007ue} is
\begin{equation}
 \mathcal{M}_{ADM}=-\frac{1}{2\pi} \lim_{S_t\to\infty} \oint_{S_t} s^i \Big (D_i\Psi -\frac{1}{8} D^j \tilde{\gamma}_{ij} \Big ) \sqrt{q}\,d^2y \, .
\end{equation}

\begin{figure}[h]
\centering
\includegraphics[width=0.5\textwidth]{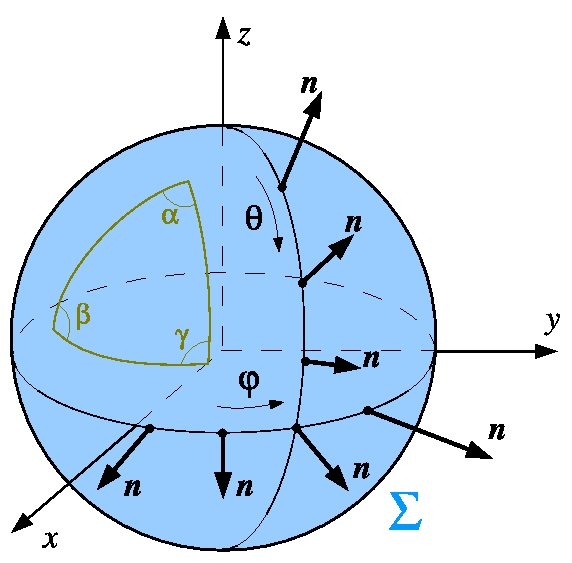}
\caption{The topology of $ S^2 $.Notice that $ \Sigma $ is equivalent to $ S_t $ and \textbf{n} to $ \hat{\textbf{r}} $ in the Schwarzschild case.}
\label{hyp_sphere}
\end{figure}

\textbf{Example:} The Schwarzschild spacetime in the adapted coordinates of $ (x^\alpha)= (t,r,\theta,\phi) $ is
\begin{equation}
 g_{\mu\nu} dx^\mu dx^\nu = -\Big (1-\frac{2m}{r} \Big )dt^2 + \Big (1-\frac{2m}{r} \Big )^{-1} dr^2 +r^2 \Big [d \theta^2 + sin^2\theta\, d \phi^2 \Big] \, .
\end{equation}
Now, $ (x^i) = (r,\theta,\phi ) $ can be taken as the spatial the coordinates on $ \Sigma_t $. Then, the induced 3-metric is
\begin{equation}
 \gamma_{ij} = diag \Big [ \Big (1-\frac{2m}{r} \Big )^{-1},r^2,r^2 sin^2 \theta \Big] \, .
\end{equation}
The components of the background metric become 
\begin{equation}
 f_{ij} = diag (1,r^2,r^2sin^2 \theta ) \, ,
\end{equation}
and their duals are
\begin{equation}
 f^{ij} = diag(1,r^{-2},r^{-2}sin^{-2} \theta ) \, .
\end{equation}
It is time to start to calculate $ \mathcal{M}_ {ADM} $: As we see in the figure (\ref{hyp_sphere}), r=constant corresponds $ S_t $, the corresponding coordinates on it are $ y^a=(\theta,\phi) $. Also, $ s^i \sqrt{q} d^2 y = r^2 sin\theta\, d\theta\, d\phi \, (\partial_r)^i $. Because, the unit spacelike vector $ \hat{\textbf{r}} $ is normal to the $ S_t $, then, $ (\partial_r)^i=(1,0,0) $. Therefore, the related integral (\ref{adm}) of this case becomes,
\begin{equation}
 \mathcal{M}_{ADM}=\frac{1}{16 \pi} \lim_{r\to\infty} \oint_{r=const.} \Big [D^j \gamma_{rj} -D_r (f^{kl}\gamma_{kl}) \Big ] r^2 sin\theta\, d\theta\, d\phi \, .
\label{adm1}
\end{equation}
In order to evaluate the corresponding $ \mathcal{M}_{ADM} $ that the hypersurface of the Schwarzschild holds, we have to first evaluate the integrands of the integral (\ref{adm1}): Let us start to calculate $ 2^{nd} $ integrand, 
\begin{equation}
 f^{kl} \gamma_{kl} =\gamma_{rr} + \frac{1}{r^2} \gamma_{\theta\theta} + \frac{1}{r^2 sin^2\theta} \gamma_{\phi\phi}
= \Big( 1-\frac{2m}{r} \Big )^{-1} + 2 \, .
\label{integ2}
\end{equation}
Since (\ref{integ2}) is a scalar field, we have
\begin{equation}
 D_r\Big(f^{kl}\gamma_{kl} \Big ) = \frac{\partial}{\partial r} \Big (f^{kl} \gamma_{kl} \Big ) = - \Big (1-\frac{2m}{r} \Big )^{-2} \frac{2m}{r^2} \, .
\label{adm2}
\end{equation}
Secondly, let us evaluate the $ 1^{st} $ integrand:
\begin{equation}
 D^j \gamma_{rj} = f^{jk} D_k \gamma_{rj} = D_r \gamma_{rr} + \frac{1}{r^2} D_\theta \gamma_{r\theta} + \frac{1}{r^2 sin^2\theta} D_\phi \gamma_{r\phi} \, .
\end{equation}
Now, the non-vanishing Christoffel symbols associated with $ D $ are 
\begin{equation}
 \tilde{\Gamma}^r_{\theta\theta} =-r \,\, \,and \,\,\, \tilde{\Gamma}^r_{\phi\phi}=-r sin^2\theta \, ,
\end{equation}
\begin{equation}
  \tilde{\Gamma}^\theta_{r \theta} = \tilde{\Gamma}^\phi_{\phi r} = \frac{1}{r} \,\,\,and \,\,\,\tilde{\Gamma}^\theta_{\phi\phi} = -cos\theta sin\theta \, ,
\end{equation}
\begin{equation}
  \tilde{\Gamma}^\phi_{r \phi} =  \tilde{\Gamma}^\phi_{\phi r} = \frac{1}{r} \,\,\, and \,\,\,  \tilde{\Gamma}^\phi_{\theta\phi}=\frac{1}{tan\theta} \, .
\end{equation}
With the associated covariant derivatives given by
\begin{equation}
 D_r \gamma_{rr} = \frac{\partial\gamma_{rr}}{\partial r} -2 \tilde{\Gamma}^i_{rr} \gamma_{ir} = \frac{\partial\gamma_{rr}}{\partial r} \, ,
\end{equation}
\begin{equation}
 D_\theta \gamma_{r\theta} =\frac{\partial\gamma_{r\theta}}{\partial\theta}- \tilde{\Gamma}^i_{\theta r}\gamma_{i\theta}- \tilde{\Gamma}^i_{\theta\theta}\gamma_{ri}= - \tilde{\Gamma}^\theta_{\theta r} \gamma_{\theta\theta}- \tilde{\Gamma}^r_{\theta\theta}\gamma_{r r} \, ,
\end{equation}
\begin{equation}
 D_\phi \gamma_{r\phi} = \frac{\partial\gamma_{r\theta}}{\partial\phi}- \tilde{\Gamma}^i_{\phi r}\gamma_{i\phi} \gamma_{r i}=- \tilde{\Gamma}^\phi_{\phi r} \gamma_{\phi\phi}- \tilde{\Gamma}^r_{\phi\phi}\gamma_{rr} \, ,
\end{equation}
the $ 1^{st} $ integrand is 
\begin{equation}
 \begin{aligned}
D^j\gamma_{rj} &= D_r \gamma_{rr} + \frac{1}{r^2} D_\theta \gamma_{r\theta} + \frac{1}{r^2 sin^2\theta} D_\phi \gamma_{r\phi}  \\
&=\frac{\partial\gamma_{rr}}{\partial r} -\frac{1}{r^2} \bigg \{ \tilde{\Gamma}^\theta_{\theta r}\gamma_{\theta\theta} +  \tilde{\Gamma}^r_{\theta\theta} \gamma_{rr} \bigg\}
-\frac{1}{r^2 sin^2\theta}\bigg \{  \tilde{\Gamma}^\phi_{\phi r} \gamma_{\phi\phi} +  \tilde{\Gamma}^r_{\phi\phi} \bigg \} \\
&=\frac{\partial}{\partial r}\Big [\Big(1-\frac{2m}{r} \Big )^{-1} \Big] \frac{-1}{r^2} \bigg \{\frac{1}{r}\times r^2 
- r \times \Big(1-\frac{2m}{r} \Big )^{-1} \bigg \} \\
&\quad -\frac{1}{r^2 sin^2\theta} \bigg \{\frac{1}{r}\times r^2 sin^2\theta-rsin^2\theta \times \Big(1-\frac{2m}{r} \Big )^{-1} \bigg \} \\
&= -\frac{2m}{r^2} \Big(1-\frac{2m}{r} \Big )^{-2} + \frac{4m}{r}\times\frac{1}{r-2m} \\
&=-\frac{2m}{r^2} \Big(1-\frac{2m}{r} \Big )^{-2} + \frac{4m}{r^2}\times \Big (1-\frac{2m}{r} \Big )^{-1} \\
&=\frac{2m}{r^2} \Big(1-\frac{2m}{r} \Big )^{-2} \Big (1- \frac{4m}{r} \Big) \, .
\label{adm3}
\end{aligned}
\end{equation}
With the help of the related results of (\ref{adm2}) and (\ref{adm3}), the integrand of the equation (\ref{adm1}) is obtained from
\begin{equation}
\begin{aligned}
 D^j\gamma_{rj}-D_r(f^{kl}\gamma_{kl}) &=\frac{2m}{r^2} \Big(1-\frac{2m}{r} \Big )^{-2} \Big (1- \frac{4m}{r} +1 \Big) \\
&=\frac{4m}{r^2}\Big(1-\frac{2m}{r} \Big )^{-1} \\
& \tilde{=} \frac{4m}{r^2} \,\,\, when \,\,\, r \rightarrow \infty \, .
\label{adm4}
\end{aligned}
\end{equation}
Let us substitute the result (\ref{adm4}) into the integral equation of $ \mathcal{M}_{ADM} $ (\ref{adm1})
\begin{equation}
\begin{aligned}
 M_{ADM}&=\frac{1}{16\pi} \lim_{r\to\infty} \oint_{r=cons.} \Big[D^j \gamma_{rj} - D_r(f^{kl}\gamma_{kl} )\Big ]r^2 sin\theta d\theta \, d\phi \\
&=\frac{1}{16\pi} \int_{0}^{2\pi} d\phi \int_{0}^{\pi} \frac{4m}{r^2} r^2 sin\theta\, d\theta \\
&=m \, ,
\end{aligned}
\end{equation}
which is \emph{\textbf{exactly the mass parameter of the Schwarzschild solution}}. 
\item The ADM Linear Momentum

We have seen that one of the condition for a spacetime to be asymptotically flat is that there must be a Cartesian coordinate system ($ x^i $) on each of Cauchy surface in which the background metric is diagonal and the each diagonal element must be 1. Therefore, the induced coordinates $ (\partial_i)_{i \in \{1,2,3\}} $ for the tangent spaces $ \mathcal{T}_p(\Sigma) $ provide three spatial direction for translation in coordinates. Now, the symmetry of the  action under spatial translations provide the $ 2^{nd} $ family of conserved quantities, $ \mathcal{P}_i $, of $ \Sigma_t $. Because of the symmetry under spatial translation, we choice an observer for which $ N=0 $ and $ \beta^i=1 $. Then, the components of related conserved quantities are given by the formula \cite{Gourgoulhon:2007ue} of
 \begin{equation}
 \mathcal{P}_i = \frac{1}{8 \pi} \underset{S_t \to \infty }{\lim} \oint_{S_t}\Big ( K_{jk} - K \gamma_{jk} \Big ) (\partial_i)^j s^k \sqrt{q}\,d^2 y \, ,
\end{equation}
where the index $ i $ can take the values of (1,2,3) and it shows which component of $ \mathcal{P}_i $ will be calculated. In another words, it determines the direction in which  the spatial translation will be done. Furthermore, the $ \mathcal{P}_i 's $ are known as the \emph{\textbf{ $ i^{th} $ component of the ADM linear momentum}} of the hypersurface $ \Sigma_t $ of \emph{the globally-hyperbolic asymptotically flat spacetime}. On the other hand, since the extrinsic curvature \textbf{K} of the $ \Sigma_t $ embedded in the Schwarzschild spacetime with the standard and isotropic coordinates vanishes, then, \emph{the corresponding ADM linear momentums $ \mathcal{P}_i $ of the hypersurface of the Schwarzschild spacetime vanishes}. 
\item The ADM 4-Momentum

 The ADM 4-Momentum \cite{Gourgoulhon:2007ue} is defined as
\begin{equation}
 \mathcal{P}^{ADM}_\alpha =\Big (-\mathcal{M}_{ADM},\mathcal{P}_i \Big) \, ,
\end{equation}
which  transform like the components of a 1-form under $ (x^\alpha) \rightarrow (x^{'\alpha}) $ during which \emph{the fundamental properties for a spacetime to be asymptotically flat are kept valid}.  

\item The ADM Angular Momentum

We suppose that the spacetime that we deal with receives the Killing vectors. Now, the angular momentum of the $ \Sigma_t $ of a globally-hyperbolic asymptotically flat spacetime which is related to the rotational symmetry of the action is obtained by using the rotational Killing vectors of the background metric $ (\phi_i)_{i\in \{1,2,3\}} $. In the Cartesian coordinates (x,y,z), the Killing vectors of the background metric about the x-axis,y-axis and z-axis are \cite{Gourgoulhon:2007ue}
\begin{equation}
 \phi_x =y\partial_z -z \partial_y\,,\,\phi_y =z\partial_x -x \partial_z \,,\,\phi_z =x\partial_y -y \partial_x \, .
\end{equation}
Then, the $ i^{th} $ \emph{\textbf{component of the angular momentum}} of the $ \Sigma_t $ can be defined as
\begin{equation}
 \mathcal{J}_i = \frac{1}{8\pi} \underset{S_t \to\infty }{lim} \oint_{S_t} \Big (K_{jk} -K \gamma_{jk} \Big )(\phi_i)^j s^k \sqrt{q}\, d^2y \, .
\end{equation}
Contrary to what we expect, $ \mathcal{J}_i $ do not transform like the 4-dimensional vectors under  $ (x^\alpha) \rightarrow (x^{'\alpha}) $ during which \emph{the fundamental properties for a spacetime to be asymptotically flat are kept valid} \cite{Gourgoulhon:2007ue}. That's, they are coordinate-dependent.  
\end{enumerate}
\end{enumerate}
Because of the coordinate-dependence of $ \mathcal{J}_i $, the scientists have tried to put the constraints on selecting coordinate systems such that $ \mathcal{J}_i $ is invariant under special subset of the related family of coordinate transformations. That's, they have being considered particular decays. For example, York \cite{York:1979cf} has considered the following decays of the $ \tilde{\gamma}_{ij} $ relative to the Cartesian coordinates for the background metric and the scalar extrinsic curvature
 \begin{equation}
\begin{aligned}
 \frac{\partial \tilde{\gamma}_{ij}}{\partial x^k} &= O[r^{-3}] \, , \\ 
K &=O[r^{-3}] \, .
\end{aligned}
\end{equation}
They are called the \emph{\textbf{quasi-isotropic gauge}} and \emph{\textbf{asymptotically maximal gauge}}, respectively. These \emph{\textbf{asymptotic gauge conditions}} are used to select the suitable coordinates. And it rejects some of well-known coordinates such as the standard Schwarzschild coordinates. Moreover, \\ York proposed that the angular momentum is carried by the $ O[r^{-3}] $ piece of \textbf{K} and is invariant under the change within this gauge \cite{Gourgoulhon:2007ue}.

\section{RELATION BETWEEN THE COTTON (CONFORMAL) SOLITON AND STATIC VACUUM SOLUTIONS}
\label{sec:Relation Between The Cotton(Conformal) Soliton And Static Vacuum Solutions}

This chapter is devoted to an application of the methods described in the previous chapters. The four-dimensional Einstein equation with a cosmological constant is given by 
\begin{equation}
{^{4}}R_{\mu\nu}-\frac{1}{2}Rg_{\mu\nu}+\Lambda g_{\mu\nu}=8\pi T_{\mu\nu} \, .
\end{equation}
 Alternatively, we could rewrite it as 
\begin{equation}
{^{4}}R_{\mu\nu}=8\pi\bigg(T_{\mu\nu}-\frac{1}{2}\Big(T-\frac{\Lambda}{4\pi}\Big)g_{\mu\nu}\bigg) \, .
\end{equation}
By following the same procedure as we did in chapter 4, the corresponding 3+1 Einstein system with $\Lambda\neq 0 $ becomes 
\begin{equation}
\bigg(\frac{\partial}{\partial t}-\mathcal{L}_{\beta}\bigg)\gamma_{ij}=-2NK_{ij} \, ,
\end{equation}
\begin{equation}
\begin{aligned}
\bigg(\frac{\partial}{\partial t}-\mathcal{L}_{\beta}\bigg)K_{ij}=-D_{i}D_{j}N+N\bigg\{ & R_{ij}+KK_{ij}-2K_{ik}K^{k}{_{j}}\\
 & +4\pi\bigg[\Big(S-E-\frac{\Lambda}{4\pi}\Big)\gamma_{ij}-2S_{ij}\bigg]\bigg\} \, ,
\end{aligned}
\end{equation}
 \begin{equation}
R+K^{2}-K_{ij}K^{ij}=2\Lambda+16\pi E \, ,
\end{equation}
 \begin{equation}
D_{j}K^{j}{_{i}}-D_{i}K=8\pi p_{i} \, .
\end{equation}
The \textbf{static vacuum equation} ($\overset{.}{\gamma}=0\,,\beta=0$) obtained from the previous system of equations is 
\begin{equation}
R_{ij}=N^{-1}D_{i}D_{j}N+\Lambda\gamma_{ij} \, ,
\end{equation}
where $ R=2 \Lambda $ and $ N^{-1}\bigtriangleup_{\gamma}N =-\Lambda $ \cite{Bartnik:2005hh}. From now, we will use $ \nabla $ as the intrinsic Levi-Civita connection and $ g_{ij} $ as the induced 3-metric (i.e. we are making a change of notation $ \gamma_{ij} \rightarrow g_{ij} $ and $ D \rightarrow \nabla $). The Cotton-York tensor \cite{York:1979cf}, \cite{York:1981bg}, \cite{Cotton:1899hj} is
\begin{equation}
C_{ij}=\epsilon_{j}{^{pq}}\Big(\nabla_{q}R_{ip}-\frac{1}{4}g_{ip}\nabla_{q}R\Big) \, .
\label{york-cot}
\end{equation}
Notice that the last term of the Cotton-York tensor (\ref{york-cot}) \{\cite{York:1979cf}, \cite{York:1981bg}, \cite{Cotton:1899hj}\} vanishes. Let us rewrite the Cotton-York tensor of the static vacuum field with $\Lambda\neq0$ in terms of the Ricci tensor: 
\begin{equation}
\begin{aligned}C_{ij} & =\epsilon_{j}{^{pq}}\nabla_{q}R_{ip}=\epsilon_{j}{^{pq}}\nabla_{q}\Big\{N^{-1}\nabla_{i}\nabla_{p}N\Big\}\\
 & =\epsilon_{j}{^{pq}}\Big\{-\frac{1}{N^{2}}\nabla_{q}N\nabla_{i}\nabla_{p}N+\frac{1}{N}\nabla_{q}\nabla_{i}\nabla_{p}\Big\} \, ,
\end{aligned}
\end{equation}
then
 \begin{equation}
\begin{aligned}NC_{ij} & =\epsilon_{j}{^{pq}}\Big\{-\nabla_{q}N(N^{-1}\nabla_{i}\nabla_{p}N)+\nabla_{q}\nabla_{i}\nabla_{p}N\Big\}\\
 & =\epsilon_{j}{^{pq}}\Big\{-[R_{ip}-\Lambda g_{ip}]\nabla_{q}N+\nabla_{q}\nabla_{i}\nabla_{p}N\Big\}\\
 & =-\epsilon_{j}{^{pq}}[R_{ip}-\Lambda g_{ip}]\nabla_{q}N+\epsilon_{j}{^{pq}}\nabla_{q}\nabla_{i}\nabla_{p}N\\
 & =-\epsilon_{j}{^{pq}}[R_{ip}-\Lambda g_{ip}]\nabla_{q}N+\epsilon_{j}{^{pq}}\Big\{[\nabla_{q},\nabla_{i}]\nabla_{p}N+\nabla_{i}\nabla_{q}\nabla_{p}N\Big\} \, .
\end{aligned}
\end{equation}
Because the multiplication between the anti-symmetric and symmetric tensor is zero, the last term vanishes yielding
 \begin{equation}
\begin{aligned}
NC_{ij} & =-\epsilon_{j}{^{pq}}[R_{ip}-\Lambda g_{ip}]\nabla_{q}N+\epsilon_{j}{^{pq}}[\nabla_{q},\nabla_{i}]\nabla_{p}N\\
 & =-\epsilon_{j}{^{pq}}[R_{ip}-\Lambda g_{ip}]\nabla_{q}N-\epsilon_{j}{^{pq}}R^{k}{_{pqi}}\nabla_{k}N\\
 & =-\epsilon_{j}{^{pq}}[R_{ip}-\Lambda g_{ip}]\nabla_{q}N-\epsilon_{j}{^{pq}}R_{kpqi}\nabla^{k}N \, .
\label{ziraat meco}
\end{aligned}
\end{equation}
In three dimensions the Weyl tensor is zero so the Riemann tensor can be written in terms of the Ricci tensor, the metric and the scalar curvature as  
\begin{equation}
R_{kpqi}=2g_{k[q}R_{i]p}+2g_{p[i}R_{q]k}-Rg_{k[q}g_{i]p} \, .
\end{equation}
By using this identity, the equation (\ref{ziraat meco}) becomes 
 \begin{equation}
C_{ij}=-\epsilon_{j}{^{pq}}\Big\{2R_{ip}\delta^{k}{_{q}}+g_{pi}R^{k}{_{q}}-2\Lambda g_{ip}\delta^{k}{_{q}}\Big\}N^{-1}\nabla_{k}N\label{gjfh} \, .
\end{equation}
With the definition of $ U_{k} \equiv N^{-1}\nabla_{k}N $, 
\begin{equation}
\begin{aligned}
C_{ij} & =-\epsilon_{j}{^{pq}}\Big\{2R_{ip}\delta^{k}{_{q}}+g_{pi}R^{k}{_{q}}-2\Lambda g_{ip}\delta^{k}{_{q}}\Big\}U_{k}\\
 & =-2\epsilon_{j}{^{pq}}R_{ip}U_{q}-\epsilon_{ji}{^{q}}\Big\{R^{k}{_{q}}-2\Lambda\,\delta^{k}{_{q}}\Big\}U_{k} \, .
\label{rfdg}
\end{aligned}
\end{equation}
We can get rid off the anti-symmetric part of (\ref{rfdg}) : Therefore, let us first do the interchange of the indices $ i\leftrightarrow j $ in (\ref{rfdg})
\begin{equation}
C_{ij}=-2\epsilon_{i}{^{pq}}R_{jp}U_{q}-\epsilon_{ij}{^{q}}\Big\{R^{k}{_{q}}-2\Lambda\,\delta^{k}{_{q}}\Big\}U_{k} \, .
\label{lkj4}
\end{equation}
The anti-symmetric part will drop by adding the equations (\ref{rfdg}) and (\ref{lkj4}), and we will have 
\begin{equation}
C_{ij}=-\epsilon_{i}{^{pq}}R_{jp}U_{q}-\epsilon_{j}{^{pq}}R_{ip}U_{q} \, ,
\end{equation}
 or
 \begin{equation}
C^{i}{_{j}}=-\epsilon^{ipq}R_{jp}U_{q}-\epsilon_{j}{^{pq}}R^{i}{_{p}}U_{q} \, .
\label{ezo}
\end{equation}
Let us define $ X^{ip} \equiv -\epsilon^{ipq}U_{q} $ and $ X_{j}{^{p}} \equiv -\epsilon_{j}{^{pq}}U_{q} $, then,
 \begin{equation}
C^{i}{_{j}}=X^{i}{_{p}}R^{p}{_{j}}+R^{i}{_{p}}X_{j}{^{p}} \, .
\end{equation}
This can be written as a matrix equation, let us rewrite it in the compact form
 \begin{equation}
\begin{aligned}
\textbf{C} &=\textbf{X}\textbf{R}+\textbf{R}\textbf{X}^{T}  \, ,\\
\textbf{C}^T &= \textbf{R}^T \textbf{X}^T + (\textbf{X}^T)^T \textbf{R}^T \, .
\label{dicle}
\end{aligned}
\end{equation}
Since $ \textbf{C}^T=\textbf{C} $ and $ \textbf{R}^T=\textbf{R} $, the equation (\ref{dicle}) can be written as
\begin{equation}
 \textbf{C} = \textbf{R}^T \textbf{X}^T + (\textbf{X}^T)^T \textbf{R} \, .
\end{equation}
By defining  $ \textbf{A}=\textbf{R} $ and $ \textbf{Y}=\textbf{X}^T $, we have
\begin{equation}
 \textbf{C}= \textbf{A}^T \textbf{Y} + \textbf{Y}^T \textbf{A} \, .
\label{alper}
\end{equation}
 Now, the matrix equation of the type 
\begin{equation}
\textbf{A}^{T}\textbf{X}\pm\textbf{X}^{T}\textbf{A}=\textbf{B} \, ,
\end{equation}
has the general solution of 
\begin{equation}
\textbf{X}=\frac{1}{2}\textbf{G}^{T}\textbf{B}\textbf{P}_{1}+\textbf{G}^{T}\textbf{B}(1-\textbf{P}_{1})+(1-\textbf{P}_{2}^{T})\textbf{Y}+(\textbf{P}_{2}^{T}\textbf{Z}\textbf{P}_{2}\textbf{A}) \, ,
\label{braden}
\end{equation}
where $ \textbf{Z} $ is a rank-2 antisymmetric tensor ; $ \textbf{P}_{1}=\textbf{G}\textbf{A} $ and $\textbf{P}_{2}=\textbf{A}\textbf{G}$ such that $\textbf{A}\textbf{P}_{1}=\textbf{P}_{2}\textbf{A}=\textbf{A}$; $\textbf{A}\textbf{G}\textbf{A}=\textbf{A}$ \cite{Braden:1998}.

By using the equation (\ref{braden}), the general solution for our equation (\ref{alper}) is 
\begin{equation}
 \textbf{Y}=\textbf{X}^T =\frac{1}{2} \textbf{R}^{-1} \textbf{C} + \textbf{Z} \textbf{R} \, ,
\end{equation}
or we have
 \begin{equation}
\textbf{X}=\frac{1}{2}\textbf{C}\textbf{R}^{-1}-\textbf{R}\textbf{Z} \, .
\label{mamisss}
\end{equation}
Let us examine this solution for the Cotton flow \cite{Kisisel:2008jx}: The equation of the gradient Cotton soliton \cite{Kisisel:2008jx} can be taken as
\begin{equation}
 C_{ij} + \nabla_i \nabla_j N =0 \, .
\end{equation}
By choosing an ansatz, $\nabla_{i}\nabla_{j} N=(R_{ij}-\Lambda\, g_{ij})N $, the general solution for the gradient Cotton soliton \cite{Kisisel:2008jx} becomes
\begin{equation}
\begin{aligned}
 X_i{^p} &=-\frac{1}{2} N \,\delta_i{^p} + \frac{1}{6}\,N\, R\, (R^{-1})_i{^p} -R_i{^m} Z_m{^p} \\
&=-\frac{1}{2} N \,\delta_i{^p} + \frac{\Lambda \,N}{3}\, (R^{-1})_i{^p} -R_i{^m} Z_m{^p} \, ,
\end{aligned}
\end{equation}
 since $ X_{j}{^{p}}=-\epsilon_{j}{^{pq}}U_{q}=-\epsilon_{j}{^{pq}} N^{-1} \nabla N $, we get
 \begin{equation}
\nabla^{m}N=\frac{1}{2}\epsilon^{ipm}R_{i}{^{n}}Z_{np}N \, .
\label{dkt1}
\end{equation}
And the corresponding constraint equation in which the rank-2 anti-symmetric tensor Z must satisfy is 
\begin{equation}
\frac{1}{2}\epsilon^{ipm}\nabla_{m}R_{i}{^{n}}Z_{np}+\frac{1}{2}\epsilon^{ipm}R_{i}{^{n}}\nabla_{m}Z_{np}+\frac{1}{4}R_{i}{^{n}}Z_{np}\Big(R^{ik}Z_{k}{^{p}}-R^{pk}Z_{k}{^{i}}\Big)=-\Lambda \, .
\label{dkt2}
\end{equation}
For the case of the \textbf{static vacuum solution with zero cosmological constant}, we found the general solution as 
\begin{equation}
\nabla^{m}N=\epsilon^{ipm}R_{j}{^{n}}Z_{np} \, .
\label{dkt3}
\end{equation}
And the corresponding constraint equation for Z is
 \begin{equation}
\epsilon^{jpi}R_{j}{^{n}}\nabla_{i}Z_{np}=0 \, .
\label{dkt4}
\end{equation}
It seems that the results (\ref{dkt1}), (\ref{dkt2}), (\ref{dkt3}) and (\ref{dkt4}) can be used to find which solution of the static field equation with $ \Lambda $ (or $ \Lambda =0 $) is also a solution of the gradient Cotton soliton \cite{Kisisel:2008jx}. Furthermore, there is only one 3-dimensional Ricci soliton \cite{Hamilton:1982} which is known as the Bryant soliton. However, the explicit metric is not known. We have not also been able to solve the constraint equations and have not found explicit metrics. But the formulation outlined above can be used to explore the gradient Cotton solitons \cite{Kisisel:2008jx} and the solutions of Topologically Massive Gravity (TMG) \cite{Deser:1981wh}, \cite{Deser:1982vy}, \cite{Chow:2009km} as well as the gradient Ricci solitons \cite{Hamilton:1982}.

\section{CONCLUSION}
\label{sec:Conclusion}

In this work, we have first learned how to foliate a globally hyperbolic four-dimensional spacetime by a continuous set of Cauchy surfaces $(\Sigma_{t})_{t\in\mathcal{R}} $ which, with particular numbers of the projection onto the hypersurface and along the unit normal vector, provide us the fundamental relations of the 3+1 formalism \cite{Hawking:1973pf}. Furthermore, with the help of these basic relations as well as the 3+1 decomposition of the stress-energy tensor, we have learned that the full projection of the Einstein equation onto $\Sigma $ gives the dynamical part and the other two projection give the constraint equations which are used to check whether a given spacetime is a solution of the Einstein equation or not. Moreover, with the help of the coordinate adapted to the flow and the shift vector $ \beta $, one can convert the 3+1 Einstein system to a set of PDEs \cite{Gourgoulhon:2007ue}.

Secondly, we have analyzed the flow of the hypersurfaces as if there is a conformal relation between a well-defined conformal background metric $ \tilde{\gamma} $ and the set of the induced 3-metrics associated with the hypersurfaces. Moreover, by constructing the fundamental conformal transformations of the intrinsic quantities of the hypersurface, the 3+1 conformal Einstein system is constructed. And we have emphasized that the trace and traceless parts of the 3+1 dynamical Einstein equation transform separately under the conformal transformation. Finally, we have reconstructed the 3+1 Einstein system for the  foliation which is maximally sliced ($ K=0 $). We have seen that the conformal background metric is nothing but a conformally flat background metric and the Cotton-York tensor \cite{Cotton:1899hj}, \cite{York:1979cf} vanishes in this case. We have also seen that this particular case leads to the 3+1 IWM system \cite{Isenberg:1978}, \cite{Wilson:1989} which is the conformal approximation to the general relativity.

Thirdly, we have reconstructed the Hamiltonian form \cite{Dirac:1958sc}, \cite{Dirac:1958sq} of the general relativity which provides us the ADM formalism for the conserved quantity of hypersurfaces of the globally-hyperbolic spacetimes which asymptotically approach to the well-defined spacetimes such as the Minkowski spacetime \cite{Arnowitt:1959ah}. Furthermore, we have seen that the quasi-isotropic gauge and the asymptotically maximal gauge force us to shrink the cluster of the coordinates in which the ADM angular momentum $ \mathcal{J}_i $ becomes invariant.

Finally, we have proposed a method in chapter 7  which we think will give the relation between the solutions of the gradient Cotton soliton \cite{Kisisel:2008jx} and of the static vacuum field equations. Furthermore, we think that this method can be used to find the relation between the solutions of the gradient Ricci \cite{Hamilton:1982} and the Cotton \cite{Kisisel:2008jx} solitons and the solutions of the Topologically Massive Gravity (TMG) \cite{Deser:1981wh}, \cite{Deser:1982vy}, \cite{Chow:2009km}.

\begin{acknowledgments}
I am heartily thankful to my supervisor Assoc. Prof. Dr. Bayram Tekin whose infinite energy, encouragement, guidance and support from the initial to the final level enabled me to develop an understanding of the subject. I would also like to thank the people who provided me an atmosphere that I dreamt to be in, Prof. Dr. Atalay Karasu, Prof. Dr. Metin \"{O}nder, Prof. Dr. Ay\c{s}e Karasu, Prof. Dr. B. \"{O}zg\"{u}r Sar{\i}o\u{g}lu, Prof. Dr. M\"{u}ge B. Evinay, Dr. Heeseung Zoe, Tahsin \c{C}. \c{S}i\c{s}man, \.{I}brahim  G\"{u}ll\"{u}, Deniz Olgu Devecio\u{g}lu, Mehmet Ali Olpak, \.{I}brahim Burak \.{I}lhan and G\"{o}khan Alka\c{c}.

I owe my deepest gratitude to Dr. Kubilay \.{I}nal (PETA\c{S}) for his supports. I would also like to dedicate this work to my family and to the people who always shared my load : Dr. Cengiz Burak, G\"{o}zde \c{C}i\c{c}ek, Zir. M\"{u}h. Mecit Demir, Servet Dengiz, \.{I}hsan Dengiz, Dr. Alper \"{O}zk\"{o}k, H. \c{S}irin Dengiz, Halil Dengiz, Nadire D. Co\c{s}ar, Dr. \,Engin Akg\"{u}nd\"{u}z, \.{I}brahim Kaya and \"{O}. Erdin\c{c} Da\u{g}deviren.

\underline{Disclaimer}: This Thesis closely follows the excellent lecture notes of Eric Gourgoulhon \emph{3+1 formalism and Bases of numerical relativity} arXiv: gr-qc/0703035v1. I would like to thank Prof. Gourgoulhon for sending me the figures and allowing me to reproduce them in my thesis. Save for the Cotton flow section, no originality in this thesis is claimed. I just expound on the computations given in the above mentioned work. For the 2+1 foliation of the spacelike and timelike hypersurfaces and the ADM energy part, I partially used Eric Poisson's book \emph{A Relativist's Toolkit, The Mathematics of Black-Hole Mechanics, Cambridge University, Cambridge (2004)}. 

The computations that I included in this thesis were presented as lecture series in front of some of the members of the gravity research group in METU.
\end{acknowledgments}.


\begin{thebibliography}{99}

\bibitem{Gourgoulhon:2007ue} E.~Gourgoulhon, ``3+1 formalism and bases of numerical relativity,'' [gr-qc/0703035 [GR-QC]].

\bibitem{Carroll:2004fv} S. M. ~Caroll, ``Spacetime and Geometry: An Introduction to General Relativity,'' Addison \ Wesley (\ Pearson \ Education), \ San \ Fransisco (2004) 

\bibitem{Straumann:2004} N. ~Straumann, ``General Relativity, with Applications to Astrophysics,'' Springer-\ Verlag, \ Berlin (2004) 

\bibitem{Poisson:2004ue} E.~Poisson, ``A Relativist's Toolkit , The Mathematics of Black-Hole Mechanics,'' Cambridge \ University \ Press,\ Cambridge (2004)

\bibitem{Darmois:1927vy} G.~Darmois, ``Les \'e quations de la gravitation einsteinienne,'' M\'emorial \ des \ Sciences \ Math\'ematiques {\bf 25}, \ Gauthier-Villars, \ Paris(1927)

\bibitem{Lichnerowicz:1939vy} A.~Lichnerowicz, ``Sur certains probl\'emes globaux relatifs au syst\'eme des \'equations d'Einstein,'' Hermann, \ Paris(1939); \ Actual. \ Sci. \ Ind. {\bf 833}.

\bibitem{Lichnerowicz:1982} A.~Lichnerowicz, ``L'int\'egration des \'equations de la gravitation relativiste et le probl\'eme des n corps,'' J. \ Math. \ Pures \ Appl. {\bf 23 },37 (1944)

\bibitem{Bruhat:1952} Y.~Four\'es-Bruhat, ``Th\'eor\'eme d'existence pour certains syst\'ems d'\'equations aux d\'eriv\'ees partielles non lin\'eaires,'' Acta Mathematica {\bf 88}, 141 (1952);

\bibitem{Dirac:1958sc} P.~A.~M.~Dirac,``The Theory of gravitation in Hamiltonian form,'' Proc.\ Roy.\ Soc.\ Lond.\  {\bf A246}, 333-343 (1958).

\bibitem{Dirac:1958sq} P.~A.~M.~Dirac, ``Generalized Hamiltonian dynamics,'' Proc.\ Roy.\ Soc.\ Lond.\  {\bf A246}, 326-332 (1958).

\bibitem{Wheeler:1964og} J.A. ~Wheeler, ``Geometrodynamics and the issue of the final state, in Relativity, Groups and Topology,'' edited by \ C. \ DeWitt, \ Gordon and \ Breach, \ New \ York (1964), p. 316

\bibitem{Hawking:1973pf} S. W. ~Hawking and G.F.R. ~Ellis, ``The large scale structure of space-time,'' Cambridge \ University \ Press, \ Cambridge (1973)

\bibitem{Cotton:1899hj} E. ~Cotton, ``Sur les vari\'et\'es \'a trois dimensions,'' Annales de la facult\'e des sciences de Toulouse S\'er. 2, {\bf 1}, 385 (1899)

\bibitem{York:1979cf} J. W. ~York,``Kinematics and dynamics of general relativity,in Sources of Gravitational Radiation,'' edited \ by \ L. \ L. \ Smarr, \ Cambridge \ University \ Press, \ Cambridge (1979), p. 83

\bibitem{York:1981bg} J. W. ~York, ``Gravitational Degrees of Freedom and the Initial-Value Problem,'' Phys.\ Rev.\ Lett.  {\bf 26}, 1656 (1971).

\bibitem{Arnowitt:1959ah} R.~L.~Arnowitt, S.~Deser and C.~W.~Misner, ``Dynamical Structure And Definition Of Energy In General Relativity,'' Phys.\ Rev.\  {\bf 116}, 1322 (1959).

\bibitem{Isenberg:1978} J. A. ~Isenberg,``Waveless Approximation Theories of Gravity,''preprint University of Maryland (1978)

\bibitem{Wilson:1989} J. R. ~Wilson and G.j. ~Mathews ``Relativistic hydrodynamics,'' edited by  C.R. Evans, L.S. Finn and D.W. Hobill,Cambridge University Press, Cambridge (1989),

\bibitem{Deruelle:2006gh} N.~Deruelle, ``General Relativity,'' a Primer,lectures at Institut Henri Poincar\'e, Paris (2006)

\bibitem{Bartnik:2005hh} R.~Bartnik and P.~Tod, ``A note on static metrics,'' Class.\ Quant.\ Grav.\  {\bf 23}, 569 (2006) [arXiv:gr-qc/0512097].

\bibitem{Kisisel:2008jx} A.~U.~O.~Kisisel, O.~Sarioglu and B.~Tekin, ``Cotton flow,'' Class.\ Quant.\ Grav.\  {\bf 25}, 165019 (2008) [arXiv:0803.1603 [hep-th]].

\bibitem{Braden:1998} H. W.~Braden, `` The equations $ A^T X \pm X^T A = B $ ,'' Siam J. Matrix Anal. Appl. Vol. 20, No. 2, pp.295-302 (1998) 

\bibitem{Hamilton:1982} R.~Hamilton, `` Three-manifolds with positive Ricci curvature,'' J. \ Diff. \ Geo.,17:255-306,1982

\bibitem{Deser:1981wh} S.~Deser, R.~Jackiw and S.~Templeton, ``Topologically massive gauge theories,'' Annals Phys.\  {\bf 140}, 372 (1982)

\bibitem{Deser:1982vy} S.~Deser, R.~Jackiw and S.~Templeton, ``Three-Dimensional Massive Gauge Theories,'' Phys.\ Rev.\ Lett.\  {\bf 48}, 975 (1982).

\bibitem{Chow:2009km} D.~D.~K.~Chow, C.~N.~Pope and E.~Sezgin, ``Classification of solutions in topologically massive gravity,'' Class.\ Quant.\ Grav.\  {\bf 27}, 105001 (2010) [arXiv:0906.3559 [hep-th]].

\end{thebibliography}
\end{document}